\documentclass[12pt]{article}
\usepackage{a4wide}
\usepackage{latexsym}
\usepackage{amsmath}
\usepackage{amssymb}
\usepackage{amsfonts}
\usepackage{amscd}
\usepackage{cite}
\usepackage{placeins}
\usepackage{tikz-cd}
\usepackage[export]{adjustbox}

\usepackage[pdfpagelabels=false]{hyperref}
\hypersetup{
	colorlinks,
	linkcolor={blue!50!black},
	citecolor={blue!50!black},
	urlcolor={blue!50!black}
}

\usepackage{pslatex}
\usepackage{graphicx}
\usepackage[latin1,utf8]{inputenc}
\usepackage[T1]{fontenc}

\RequirePackage{mathtools}
\usepackage{mathdots}
\usepackage{bm}
\usepackage{nomencl}

\usepackage{dcpic,pictex}

\numberwithin{equation}{section}

\allowdisplaybreaks

\newcommand{\bq}{\begin{eqnarray}}
\newcommand{\eq}{\end{eqnarray}}
\newcommand{\eps}{\varepsilon}

\newcommand{\qbar}{q}
\newcommand{\Yinvariant}{Y}
\newcommand{\Frobeniusbasis}{\psi}

\newcommand{\Period}{\Pi}
\newcommand{\normalisedPeriod}{\hat{\Pi}}

\newcommand{\branchpoint}{e}

\newcommand{\cicy}[2]{\begin{matrix} #1\end{matrix}\!\left[\begin{matrix}#2 \end{matrix}\right]}

\newcommand{\mirrorY}{Y^{\text{mirror}}_\tau}

\usepackage{pifont}
\newcommand{\cmark}{\ding{51}}%
\newcommand{\xmark}{\ding{55}}%

\newcommand\Ip{\rotatebox[origin=c]{180}{$\Pi$}}

\DeclareMathAlphabet\mathbfcal{OMS}{cmsy}{b}{n}

\makenomenclature

\makeatletter
\newcommand*{\centerfloat}{%
  \parindent \z@
  \leftskip \z@ \@plus 1fil \@minus \textwidth
  \rightskip\leftskip
  \parfillskip \z@skip}
\makeatother

\begin{document}

\thispagestyle{empty}

\begin{flushright}
  MITP/24-038 \\
  TUM-HEP-1503/24 
\end{flushright}

\vspace{1.5cm}

\begin{center}
{\Large\bf A Calabi--Yau-to-Curve Correspondence\\[1ex]
for Feynman Integrals}\\
  \vspace{1cm}
  {\large Hans~Jockers${}^{a}$, 
          S\"oren~Kotlewski${}^{a}$, 
          Pyry~Kuusela${}^{a}$, 
          Andrew~J.~McLeod${}^{b}$, 
          Sebastian~P\"ogel${}^{a}$, 
          Maik~Sarve${}^{a}$, 
          Xing~Wang${}^{c}$ and 
          Stefan~Weinzierl${}^{a}$ \\
  \vspace{1cm}
      {\small \em ${}^{a}$ PRISMA+ Cluster of Excellence,} 
      {\small \em Mainz Institute for Theoretical Physics,} \\
      {\small \em Johannes Gutenberg-Universit{\"a}t Mainz,}
      {\small \em D - 55099 Mainz, Germany}\\
  \vspace{2mm}
      {\small \em ${}^{b}$ Higgs Centre for Theoretical Physics,} 
      {\small \em School of Physics and Astronomy,} \\
      {\small \em The University of Edinburgh} 
      {\small \em Edinburgh EH9 3FD, Scotland, UK} \\
  \vspace{2mm}
      {\small \em ${}^{c}$ Physik Department,} 
      {\small \em TUM School of Natural Sciences,} \\
      {\small \em Technische Universit\"at M\"unchen,} 
      {\small \em D - 85748 Garching, Germany}
  } 
\end{center}

\vspace{2cm}

\begin{abstract}\noindent
{It has long been known that the maximal cut of the equal-mass four-loop banana integral is a period of a family of Calabi--Yau threefolds that depends on the kinematic variable $z=m^2/p^2$. We show that it can also be interpreted as a period of a family of genus-two curves. We do this by introducing a general Calabi--Yau-to-curve correspondence, which in this case locally relates the original period of the family of Calabi--Yau threefolds to a period of a family of genus-two curves that varies holomorphically with the kinematic variable $z$. In addition to working out the concrete details of this correspondence for the equal-mass four-loop banana integral, we outline when we expect a correspondence of this type to hold.}
\end{abstract}
\vspace*{\fill}

\newpage

\tableofcontents

\newpage

\section{Introduction}
\label{sect:intro}

Methods from algebraic geometry have come to play an increasingly important role in the study of perturbative quantum field theory.
This has been especially true in the calculation of Feynman integrals, where geometric objects such as
elliptic curves \cite{Sabry:1962,Broadhurst:1993mw,Laporta:2004rb,MullerStach:2011ru,Bloch:2013tra,Adams:2014vja,Adams:2017ejb,Adams:2018yfj,Ablinger:2017bjx,Broedel:2017kkb,Broedel:2017siw,Broedel:2018iwv,Broedel:2018qkq,Broedel:2019hyg,Bourjaily:2017bsb,Kristensson:2021ani,Bogner:2019lfa,Walden:2020odh,Weinzierl:2020fyx,McLeod:2023qdf},
higher genus curves \cite{Huang:2013kh,Georgoudis:2015hca,Doran:2023yzu,Marzucca:2023gto},
and Calabi--Yau manifolds \cite{Bloch:2014qca,Bloch:2016izu,Bourjaily:2018ycu,Bourjaily:2018yfy,Bourjaily:2019hmc,Broedel:2019kmn,Klemm:2019dbm,Vergu:2020uur,Bonisch:2020qmm,Bonisch:2021yfw,Pogel:2022yat,Pogel:2022ken,Pogel:2022vat,Duhr:2022pch,Duhr:2022dxb,Kreimer:2022fxm,Forum:2022lpz,Cao:2023tpx,McLeod:2023doa,Gorges:2023zgv,Bourjaily:2022bwx,Mishnyakov:2023wpd}
already begin to make an appearance at two loops.
Understanding the periods of these geometric objects turns out to be essential for developing both analytic and numerical control over the corresponding Feynman integrals.

One of the most powerful methods for computing Feynman integrals is the method of differential equations \cite{Kotikov:1990kg}.
In this approach, one first derives a system of differential equations for a set of master integrals, and then transforms the system of differential equations into an $\eps$-factorised form \cite{Henn:2013pwa}.
Once the system is in this form, it can be solved systematically in terms of iterated integrals \cite{Chen}.
The first and third steps of this procedure are algorithmic. The task of computing Feynman integrals is therefore reduced to finding an 
appropriate transformation that puts the system of differential equations into $\eps$-factorised form.

In this paper, we will focus on the example of the equal-mass four-loop banana integral, which is shown in fig.~\ref{fig_banana}.
The system of differential equations that describes this Feynman integral can be reduced to a basis of five master integrals, which depend on the kinematic variable $z=m^2/p^2$ and the dimensional regularisation parameter $\eps$. One of these masters is trivial, and can be chosen to be the product of four one-loop tadpole integrals. This leaves four non-trivial master integrals in the top sector.
The transformation to an $\eps$-factorised form is achieved by taking the first master integral in the top sector to be
\bq
 M_1\left(z,\eps\right) & = & \frac{\eps^4}{\Frobeniusbasis_0\left(z\right)} I_{11111}\left(z,\eps\right)~.
\eq
Here, $I_{11111}(z,\eps)$ denotes the Feynman integral with all propagators raised to the power one in $D=2-2\eps$ space-time dimensions.
The remaining master integrals for an $\eps$-factorised form are obtained from linear combinations of the previous master integrals 
and derivatives thereof.
The construction is known \cite{Pogel:2022ken} and we will not be concerned with these other master integrals in this paper.
Rather, we are interested in the function $\Frobeniusbasis_0(z)$.
It is known that this function is a period of a family of Calabi--Yau threefolds $Y_z$; that is, for any value of $z$, there is a Calabi--Yau threefold $Y_z$.
The Calabi--Yau threefold $Y_z$ possesses a holomorphic $(3,0)$-form $\Omega(z)$, and $\Frobeniusbasis_0(z)$ is obtained by integrating
$\Omega(z)$ along an integral cycle of $Y_z$.
In this way, periods of geometric objects enter the definition of master integrals when we put the system of differential equations 
into $\eps$-factorised form.

In this paper, we show that we may also view $\Frobeniusbasis_0(z)$ as a period of a family of genus-two curves $C_{z}$. We do this by explicit construction. That is, we show how to build a genus-two curve that has the property that a specific linear combination $\omega(z)$ of its holomorphic $(1,0)$-forms gives $\Frobeniusbasis_0(z)$ upon being integrated along a cycle of the genus-two curve.

The motivation for studying correspondences between Calabi--Yau geometries and higher genus curves for Feynman integrals is two-fold. From a conceptional perspective it offers insights into a motivic interpretation of Feynman integrals, which means that a given Feynman integral can possibly be represented by geometrically distinct but motivically equivalent realizations. From a practical point of view, the interpretation of Feynman integrals in terms of periods of (higher genus) curves, as opposed to Calabi--Yau periods, can be advantageous as the moduli spaces of higher genus curves and their modular properties have been studied more extensively in the mathematical literature.

Let us make this correspondence more precise.
On the Calabi--Yau side we may integrate the holomorphic $(3,0)$-form $\Omega$ along four independent integral three-cycles.
This yields four integral periods.
These four integral periods are annihilated by a Picard--Fuchs operator $L^{(0)}$ of degree four.
Alternatively, on the side of the genus-two curve, we may integrate the above mentioned holomorphic $(1,0)$-form $\omega$ along 
four one-cycles (two $a$-cycles and two $b$-cycles). 
We show that the periods so obtained are again annihilated by the same Picard--Fuchs operator $L^{(0)}$.

The guiding idea behind our construction is to identify the intermediate Jacobian (also known as the Picard variety) of the family of genus-two curves $C_z$ with a suitable intermediate Jacobian defined using the periods of the family of Calabi--Yau threefolds.
To explore whether such an identification can be done, we consider different types of intermediate Jacobians of Calabi--Yau threefolds.
For Calabi--Yau threefolds there are two known types of intermediate Jacobians --- due to Griffiths and Weil --- that differ by a choice of complex structure. 
These can be constructed globally over the entire parameter space of the Calabi--Yau threefolds $Y_z$, and they are independent of a choice of basis of integral periods. 
On the one hand, the family of Griffiths intermediate Jacobians varies holomorphically in the parameter $z$ and thus matches the expected holomorphic dependence of the Feynman integral on $z$. 
However, the Griffiths intermediate Jacobian of a (non-rigid) Calabi--Yau threefold is not an Abelian variety. This means that it does not correspond to a point in the Siegel upper half-space $\mathcal{H}_2$ that can be identified with the Picard variety of a curve. 
On the other hand, the family of Weil intermediate Jacobians of the Calabi--Yau threefolds does parametrise a family of Abelian varieties. Hence, they correspond to points in the Siegel upper half-space $\mathcal{H}_2$, and thus can be identified with a family of curves. 
However, this family of curves varies non-holomorphically with $z$, since the correspondence between the family of Calabi--Yau threefolds and the associated family of curves involves complex conjugation with respect to the parameter $z$. Thus, neither the Griffiths nor the Weil intermediate Jacobian allows us to associate a holomorphic family of genus-two curves, as we would like.

To remedy this situation, we construct yet another intermediate Jacobian by means of holomorphic extension, which we call the polarised holomorphic Jacobian of the Calabi--Yau threefold. 
We only define this novel intermediate Jacobian locally, near an explicit range of the real analytic kinematic variable $z$.
We show that it realises a family of Abelian varieties that give rise to a holomorphic family of curves. 
As these curves satisfy our desiderata, they allow us to realise the correspondence between Calabi--Yau periods and periods of curves that we are after in the computation of Feynman integrals.

We show how this construction works for the equal-mass four-loop banana graph in full detail, which we explicitly work out for a particular kinematic range. 
Namely, we observe that for $z \in ]0,\frac{1}{25}[$, 
the symmetric $2 \times 2$-matrix $\bm{N}$ that describes the point 
in the Siegel upper half-space ${\mathcal H}_2$ corresponding to the Weil intermediate Jacobian has a particular structure.
The diagonal entries of this matrix are purely imaginary, while the off-diagonal entries are real.
This allows us to determine the polarised holomorphic intermediate Jacobian --- which agrees with the Weil intermediate Jacobian on the interval $z \in ]0,\frac{1}{25}[$ --- explicitly.
We can thus show that this polarised holomorphic intermediate Jacobian is described in terms of a symmetric $2{\times}2$-matrix $\bm{H}$ which has the properties we want, namely it lives in the Siegel upper half-space $\mathcal{H}_2$ and varies holomorphically with $z$. For other kinematic ranges of $z$ a similar correspondence can be worked out, which can be obtained in practise by analytically continuing the Calabi--Yau periods to a different kinematic regime.

Given the polarised intermediate Jacobian, we can then construct a family of curves $C_z$ of genus two with the same intermediate Jacobian. We present the genus-two curve $C_z$ in the Rosenhain form
\bq
 C_z & : & 
 v^2 \; = \; u \left(u-\branchpoint_2\left(z\right)\right) \left(u-\branchpoint_3\left(z\right)\right) \left(u-\branchpoint_4\left(z\right)\right) \left(u-1\right)~,
\eq
where three branch points have been fixed at $0$, $1$ and $\infty$.
The other three branch points $\branchpoint_2(z)$, $\branchpoint_3(z)$ and $\branchpoint_4(z)$
are holomorphic functions of $z$. The two independent holomorphic $(1,0)$-forms $\omega_0$ and $\omega_1$ on this curve can be taken to be
\bq
 \omega_0\left(z\right) \; = \; \frac{du}{v\left(z\right)}~,
 & &
 \omega_1\left(z\right) \; = \; \frac{u \, du}{v\left(z\right)}~.
\eq
We show that we can construct a linear combination
\bq
\label{linear_combination}
 \omega\left(z\right) & = & c_0(z) \, \omega_0\left(z\right) + c_1(z) \, \omega_1\left(z\right)~,
\eq
where $c_0(z)$ and $c_1(z)$ are holomorphic functions, such that the four periods of $\omega$ are annihilated by 
the Picard--Fuchs operator $L^{(0)}$.

This paper is organised as follows.
In the next section we first introduce the equal-mass four-loop banana integral, define the Picard--Fuchs operator and introduce the Frobenius basis.
In section~\ref{sect:Calabi-Yau_to_Curve}, we introduce and give a detailed geometric account of the Calabi--Yau-to-curve correspondence, which in later sections is explicitly applied to the equal-mass four-loop banana integral. While this section spells out the broader picture of the proposed geometric correspondence, the readers focussed on the computational results for the explicit example of the four-loop banana integral may skip this section in a first read.
In section~\ref{sect:calabi_yau}, we discuss the geometry of the family of Calabi--Yau threefolds corresponding to the equal-mass four-loop banana integral, and show how data from mirror symmetry can be used to relate the periods in the Frobenius basis to the integral periods.
We then go on in section~\ref{sect:jacobian} to compute the Griffiths, Weil, and polarised holomorphic intermediate Jacobians that can be defined on this family of Calabi--Yau threefolds.
The last of these intermediate Jacobians provides the desired bridge to a genus-two curve. 
In section~\ref{sect:construction} we construct explicitly the genus-two curve whose Jacobian variety matches the polarised holomorphic intermediate Jacobian. 
Finally, section~\ref{sect:conclusions} contains our conclusions.
Some technical aspects are relegated to two appendices. 
In appendix~\ref{appendix:intermediate_Jacobians}, we give further technical details on intermediate Jacobians and their complex structures. 
In appendix~\ref{appendix:theta_constants}, we review relations among theta constants.

\begin{figure}
\centerfloat\includegraphics[width=1.1\textwidth]{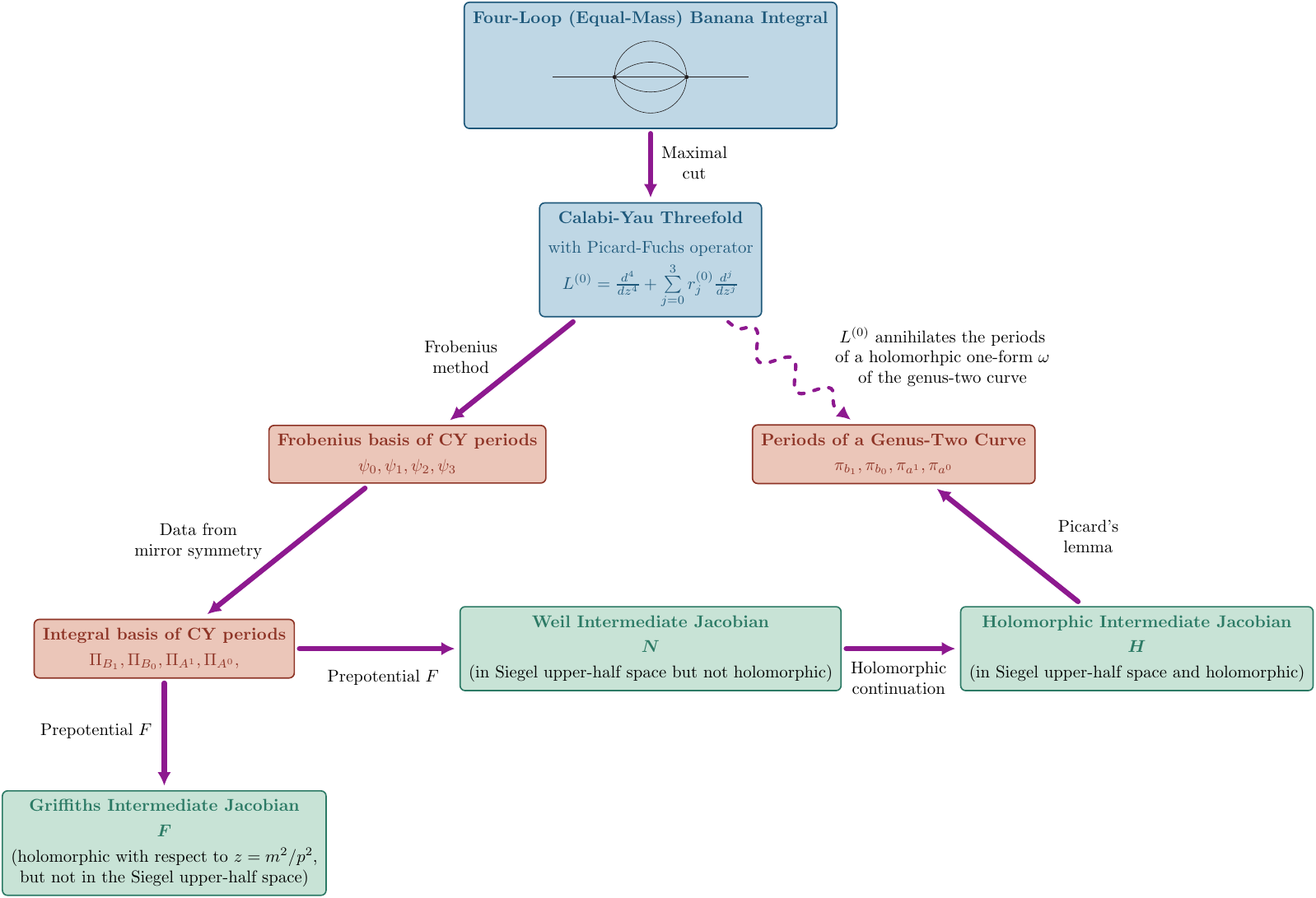}
\caption{
A flow chart visualising the construction of the Calabi--Yau-to-Curve correspondence presented in this paper.
}
\label{fig_flow_chart}
\end{figure}

For the convenience of the reader, we provide a graphical summary of the flow of this paper in fig.~\ref{fig_flow_chart}, as well as a summary of the most important symbols on the following pages.


\newpage
\printnomenclature
\nomenclature[s000]{\bf Feynman integrals:}{}
\nomenclature[s001]{$I_{11111}$}{The equal-mass four-loop banana integral with unit propagators}
\nomenclature[s002]{$z$}{The ratio $z=\frac{m^2}{p^2}$}
\nomenclature[s003]{$\eps$}{The dimensional regularisation parameter}
\nomenclature[s004]{$L^{(0)}$}{The Picard--Fuchs operator}
\nomenclature[s010]{\bf Calabi--Yau threefold:}{}
\nomenclature[s011]{$Y$}{The Calabi--Yau threefold}
\nomenclature[s012]{$\Omega$}{Holomorphic $(3,0)$-form}
\nomenclature[s013]{$\Frobeniusbasis_i$}{Frobenius basis ($0 \le i \le 3$)}
\nomenclature[s014]{$\tau$}{The image of the mirror map}
\nomenclature[s016]{$B_1, B_0, A^1, A^0$}{Symplectic basis of $H_3(Y,{\mathbb Z})$}
\nomenclature[s017]{$\Period_{B_1}, \Period_{B_0}, \Period_{A^1}, \Period_{A^0}$}{Basis of integral periods}
\nomenclature[s018]{$\normalisedPeriod_{B_1}, \normalisedPeriod_{B_0}, \normalisedPeriod_{A^1}, \normalisedPeriod_{A^0}$}{Normalised integral periods such that $\normalisedPeriod_{A^0}=1$}
\nomenclature[s030]{\bf Genus-two curve:}{}
\nomenclature[s031]{$C$}{The genus-two curve}
\nomenclature[s032]{$\omega_0, \omega_1$}{Holomorphic $(1,0)$-forms}
\nomenclature[s034]{$b_1, b_0, a^1, a^0$}{Symplectic basis of $H_1(C,{\mathbb Z})$}
\nomenclature[s035]{$\mathbfcal{X}$}{$2 \times 2$-matrix of $a$-cycle periods}
\nomenclature[s036]{$\mathbfcal{T}$}{$2 \times 2$-matrix of $b$-cycle periods}
\nomenclature[s043]{$\omega$}{Linear combination of $\omega_0$ and $\omega_1$}
\nomenclature[s047]{$\pi_{b_1}, \pi_{b_0}, \pi_{a^1}, \pi_{a^0}$}{Periods of $\omega$}
\nomenclature[s100]{\bf Calabi--Yau-to-curve correspondence:}{}

\nomenclature[s110]{$\Delta$}{An open disc in the Calabi--Yau complex structure moduli space $\mathcal{M}_{\text{cs}}$.}
\nomenclature[s110]{$\Delta_{\mathbb{R}}$}{A Lagrangian submanifold of $\Delta$.}
\nomenclature[s111]{$\mathcal{Y}_{\Delta}$}{The family of Calabi--Yau threefolds over $\Delta$.}
\nomenclature[s112]{$Y_{z}$}{A Calabi--Yau threefold with complex structure moduli $z$.}
\nomenclature[s113]{$F_J(z),X^I(z)$}{Alternative notation for the integral B- and A-periods of $Y_z$ in the context of projective special K\"ahler geometry.}
\nomenclature[s114]{$\alpha_I,\beta^J$}{An integral symplectic three-form cohomology basis of $Y_z$.}
\nomenclature[s115]{$F(X)$}{The prepotential of the complex structure moduli space $\mathcal{M}_{\text{cs}}$.}
\nomenclature[s116]{$\mathcal{C}_{g}$}{The family of stable genus-$g$ curves over $\Delta \subset \overline{\mathcal{M}}_{g}$.}
\nomenclature[s116]{$C_{g,z}$}{A stable genus-$g$ curve with complex structure moduli $z$.}

\nomenclature[s119]{$V_z$}{A half-dimensional subspace of $H^3(Y_z,\mathbb{C})$.}
\nomenclature[s120]{$J_2(Y_z,V_z)$}{The $2^{\text{nd}}$ intermediate Jacobian of $Y_z$ given by the subspace $V_z$.}
\nomenclature[s121]{$J_2^W(Y_z)$}{The $2^{\text{nd}}$ Weil intermediate Jacobian of $Y_z$.}
\nomenclature[s122]{$J_2^G(Y_z)$}{The $2^{\text{nd}}$ Griffiths intermediate Jacobian of $Y_z$.}
\nomenclature[s123]{$J_2^{\Delta_{\mathbb{R}}}(Y_z)$}{The $2^{\text{nd}}$ polarised holomorphic intermediate Jacobian of $Y_z$ constructed from $\Delta_{\mathbb{R}}$.}
\nomenclature[s124]{$J_1(C_{g,z})$}{The $1^{\text{st}}$ intermediate Jacobian of the curve $C_{g,z}$.}

\nomenclature[s130]{$\bm{F}$}{The symmetric matrix defining the $2^{\text{nd}}$ Griffiths intermediate Jacobian of $Y_z$}
\nomenclature[s131]{$\bm{N}$}{The symmetric matrix defining the $2^{\text{nd}}$ Weil intermediate Jacobian of $Y_z$}
\nomenclature[s132]{$\bm{H}$}{The symmetric matrix defining the $2^{\text{nd}}$ polarised holomorphic Jacobian of $Y_z$}
\nomenclature[s133]{$\bm{\tau}$}{The symmetric matrix defining the $1^{\text{st}}$ intermediate Jacobian of $C_{g,z}$}

\nomenclature[s140]{$\mathcal{H}_g$}{The Siegel upper half-space of degree $g$.}
\nomenclature[s141]{$\mathcal{A}_g$}{The moduli space of $g$-dimensional Abelian varieties.}
\nomenclature[s142]{$\overline{\mathcal{M}}_g$}{The moduli space of stable genus-$g$ curves.}
\nomenclature[s143]{$\overline{\mathcal{M}}_g^\mathcal{A}$}{The moduli space of stable genus-$g$ curves with non-degenerate intermediate Jacobian.}
\nomenclature[s144]{$\mathcal{S}_g$}{The Schottky locus of stable genus-$g$ curves.}

\nomenclature[s150]{$\Phi$}{The Calabi--Yau-to-curve correspondence map.}

\newpage
\clearpage

\newpage

\section{The Equal-Mass Four-Loop Banana Integral}
\label{sect:physics}

In this section we first introduce the family of the equal-mass four-loop banana integrals 
in subsection~\ref{sect:Feynman_integral_definition}.
We then focus on one particular member of this family, the integral $I_{11111}$ and 
discuss in subsection~\ref{sect:Picard_Fuchs} the differential equation satisfied by this integral.
This is an inhomogeneous differential equation of order four.
The $\eps^0$-part of the corresponding homogeneous differential equation 
defines a differential operator $L^{(0)}$.
This differential operator will play an important part in the rest of the paper.
In subsection~\ref{sect:Frobenius}, we study the solution space of this differential operator with the help
of the Frobenius method.
All content of this section is formulated without reference to geometry.


\subsection{Definition of the Feynman integral}
\label{sect:Feynman_integral_definition}

The family of the equal-mass four-loop banana integrals is defined by
\bq
\label{def_four_loop_banana_integral}
 I_{\nu_1 \nu_2 \nu_3 \nu_4 \nu_5}
 = e^{4 \gamma_E \eps}
 \left(p^2\right)^{\nu-2D}
 \int \left( \prod\limits_{a=1}^{5} \frac{d^Dk_a}{i \pi^{\frac{D}{2}}} \right)
 i \pi^{\frac{D}{2}} \delta^D\left(p-\sum\limits_{b=1}^{5} k_b \right)
 \left( \prod\limits_{c=1}^{5} \frac{1}{\left(-k_c^2+m^2\right)^{\nu_c}} \right),
\eq
where $D$ denotes the number of space-time dimensions,
$\eps$ the dimensional regularisation parameter,
$\gamma_E$ the Euler--Mascheroni constant
and 
\bq
 \nu & = & \sum\limits_{j=1}^5 \nu_j~.
\eq 
We consider these integrals in $D=2-2\eps$ space-time dimensions.
\begin{figure}
\begin{center}
\includegraphics[scale=0.8]{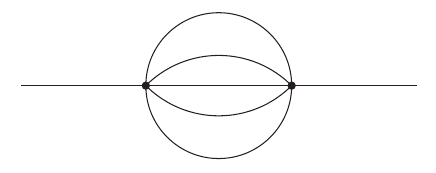}
\end{center}
\caption{
The four-loop banana graph.
}
\label{fig_banana}
\end{figure}
The corresponding Feynman graph is shown in fig.~\ref{fig_banana}. It is convenient to introduce the dimensionless variable
\bq
\label{def_y}
 z & = & \frac{m^2}{p^2}~.
\eq
Feynman's $i\delta$-prescription translates to the fact that $z$ has a small negative imaginary part:
$z \rightarrow z - i \delta$.

The Feynman parameter representation is given by
\bq
 I_{\nu_1 \nu_2 \nu_3 \nu_4 \nu_5}
 & = &
 \frac{e^{4 \gamma_E \eps}\Gamma\left(\nu-2D\right)}{\prod\limits_{j=1}^{5}\Gamma(\nu_j)}
 \int\limits_{a_j \ge 0} d^{5}a \; \delta\left(1-\sum\limits_{j=1}^{5} a_j \right) \; 
 \left( \prod\limits_{j=1}^{5} a_j^{\nu_j-1} \right)
 \frac{\left[ {\mathcal U}\left(a\right) \right]^{\nu-\frac{5D}{2}}}{\left[ {\mathcal F}\left(a\right) \right]^{\nu-2D}}~,
\eq
where the two graph polynomials are given by
\bq
\label{def_U_F}
 {\mathcal U} 
 & = & 
 a_1 a_2 a_3 a_4 a_5 \left( \frac{1}{a_1} + \frac{1}{a_2} + \frac{1}{a_3} + \frac{1}{a_4} + \frac{1}{a_5} \right)~,
 \nonumber \\
 {\mathcal F} 
 & = & 
 a_1 a_2 a_3 a_4 a_5 
 \left[
  z \left( a_1+a_2+a_3+a_4+a_5\right) \left( \frac{1}{a_1} + \frac{1}{a_2} + \frac{1}{a_3} + \frac{1}{a_4} + \frac{1}{a_5} \right) -  1
 \right]~.
\eq
We are interested in the case where the indices $\nu_1, \nu_2, \dots, \nu_5$ are integers.
Integration-by-parts identities \cite{Chetyrkin:1981qh} and symmetries allow us to express any Feynman integral 
from the countable set $(\nu_1,\nu_2,\nu_3,\nu_4,\nu_5)\in{\mathbb Z}^5$
as a finite linear combination of a subset of this family.
The integrals of this subset are called master integrals, and define a basis of a vector space.
For the case at hand, a possible basis is given by
\bq
 I_{11110}~, \quad \; I_{11111}~, \quad \; I_{11112}~, \quad \; I_{11113}~, \quad \; I_{11114}~.
\eq


\subsection{The Picard--Fuchs operator}
\label{sect:Picard_Fuchs}

We are in particular interested in the integral $I_{11111}$,
i.e. the integral where all propagators occur to the power one.
The integral $I_{11111}$ satisfies a linear inhomogeneous differential equation of order $4$:

\bq
\label{full_Picard_Fuchs_dgl}
 L \; I_{11111}
 & = &
 - \frac{5! }{z^{4} \left(1-z\right)\left(1-9z\right)\left(1-25z\right)}
 \eps^4 I_{11110}~,
\eq
with
\bq
 L 
 & = &
 \frac{d^4}{dz^4} + \sum\limits_{j=0}^{3} r_j \frac{d^j}{dz^j}~.
\eq

The differential operator $L$ is called the Picard--Fuchs operator for the integral $I_{11111}$.
The coefficients $r_j$ are rational functions in $z$ and polynomials in $\eps$.
We denote by $L^{(0)}$ the $\eps^0$-part of $L$ which is given by

\bq
\label{def_Picard_Fuchs_eps_0}
 L^{(0)} 
 & = &
 \frac{d^4}{dz^4} 
 + \left[ \frac{6}{z} - 2 \frac{1}{\left(1-z\right)} - 2 \frac{9}{\left(1-9z\right)}  - 2 \frac{25}{\left(1-25z\right)} \right] \frac{d^3}{dz^3}
 \nonumber \\
 & &
 + \frac{7-518z+6501z^2-8550z^3}{z^2\left(1-z\right)\left(1-9z\right)\left(1-25z\right)} \frac{d^2}{dz^2}
 + \frac{1-196z+3963z^2-7200z^3}{z^3\left(1-z\right)\left(1-9z\right)\left(1-25z\right)} \frac{d}{dz}
 \nonumber \\
 & &
 - \frac{5-285z+900z^2}{z^3\left(1-z\right)\left(1-9z\right)\left(1-25z\right)}~.
\eq

The operator $L^{(0)}$ will play an important role in the rest of the paper.
Note that $L^{(0)}$ appears as operator $34$ in the list of \cite{Almkvist:2005}.
This can be seen using the operator identity

\bq
\lefteqn{
 z^4 \left(1-z\right)\left(1-9z\right)\left(1-25z\right) \; L^{(0)} 
 = } & &
\nonumber \\
 & &
 \theta_z^4
 - z \left(35\theta_z^4+70\theta_z^3+63\theta_z^2+28\theta_z+5\right)
 + z^2\left(\theta_z+1\right)^2\left(259\theta_z^2+518\theta_z+285\right)
 \nonumber \\
 & &
 - 225z^3\left(\theta_z+1\right)^2\left(\theta_z+2\right)^2~.
\eq

Here, $\theta_z=z \frac{d}{dz}$ denotes the Euler operator (also known as the logarithmic derivative) in the variable $z$. The Riemann symbol of the differential operator $L^{(0)}$ is given by
\begin{align}
	\mathcal{P}\left\{
	\begin{tabular}{ccccc}
		0& $\frac{1}{25}$ & $\frac{1}{9}$ &~1~& \hskip-4pt $\infty$ \hskip-4pt{}\\[2pt]
		\hline
		0&0&0&0&1\\
		0&1&1&1&1\\
		0&1&1&1&2\\
		0&2&2&2&2
	\end{tabular}
	\hskip5pt z ~\right\},
\end{align}
so we see in particular that the point $z=0$ is a point of maximal unipotent monodromy (a MUM-point).

\subsection{The Frobenius basis of periods near the MUM-point}
\label{sect:Frobenius}
Let us now turn to the study of solutions to the differential equation
\bq
\label{homogeneous_diff_eq}
L^{(0)} \psi & = & 0
\eq
near the point $z=0$. Using the method of Frobenius, we obtain a basis $\Frobeniusbasis_0$, $\Frobeniusbasis_1$, $\Frobeniusbasis_2$, $\Frobeniusbasis_3$ of solutions of \eqref{homogeneous_diff_eq}. We may write the Frobenius basis of solutions~as
\bq
\label{Frobenius_series}
\Frobeniusbasis_0
& = & \phantom{\frac{1}{\left(2\pi i\right)^2}}
\sum\limits_{n=0}^\infty 
a_{0,n} z^{n}~,
\nonumber \\
\Frobeniusbasis_1
& = &
\frac{1}{\left(2\pi i\right)}\phantom{{}^2}
\sum\limits_{n=0}^\infty 
\left[
a_{1,n} 
+
a_{0,n} 
\ln z
\right]
z^{n}~,
\nonumber \\
\Frobeniusbasis_2
& = &
\frac{1}{\left(2\pi i\right)^2}
\sum\limits_{n=0}^\infty 
\left[
a_{2,n} 
+
a_{1,n} 
\ln z
+
\frac{1}{2}
a_{0,n} 
\ln^2 z
\right]
z^{n}~,
\nonumber \\
\Frobeniusbasis_3
& = &
\frac{1}{\left(2\pi i\right)^3}
\sum\limits_{n=0}^\infty 
\left[
a_{3,n} 
+
a_{2,n} 
\ln z
+
\frac{1}{2}
a_{1,n} 
\ln^2 z
+
\frac{1}{6}
a_{0,n} 
\ln^3 z
\right]
z^{n}~.
\eq
We choose a normalisation so that $a_{i,0}= \delta_{i0}$. The coefficients are then given by
\bq
\label{Frobenius_coeffs}
a_{0,n}
& = & \!\!
\sum\limits_{n_1+n_2+n_3+n_4+n_5=n}
\left( \frac{n!}{n_1! n_2! n_3! n_4! n_5!} \right)^2~,
\nonumber \\
a_{1,n}
& = & \!\!
\sum\limits_{n_1+n_2+n_3+n_4+n_5=n}
\left( \frac{n!}{n_1! n_2! n_3! n_4! n_5!} \right)^2
2 \left[ S_1\left(n\right) - S_1\left(n_1\right) \right]~,
\nonumber \\
a_{2,n}
& = & \!\!
\sum\limits_{n_1+n_2+n_3+n_4+n_5=n}
\left( \frac{n!}{n_1! n_2! n_3! n_4! n_5!} \right)^2
\left\{ 2 \left[ S_1\left(n\right) - S_1\left(n_1\right) \right]
\left[ S_1\left(n\right) - S_1\left(n_2\right) \right]
- S_2\left(n\right)
\right\}~,
\nonumber \\
a_{3,n}
& = & \!\!
\sum\limits_{n_1+n_2+n_3+n_4+n_5=n}
\left( \frac{n!}{n_1! n_2! n_3! n_4! n_5!} \right)^2
\left\{
\frac{2}{3} S_3\left(n\right)
- 2 S_2\left(n\right) \left[ S_1\left(n\right) - S_1\left(n_2\right) \right]
\right. \nonumber \\
& & \left.
+ \frac{4}{3} \left[ S_1\left(n\right) - S_1\left(n_1\right) \right]
\left[ S_1\left(n\right) - S_1\left(n_2\right) \right]
\left[ S_1\left(n\right) - S_1\left(n_3\right) \right]
\right\}~,
\eq
where $S_m(n)$ denotes the harmonic sum
\bq
S_m\left(n\right) & = &  \sum\limits_{j=1}^n \frac{1}{j^m}~.
\eq
The first few coefficients $a_{k,n}$ are given 
\begin{table}
	\begin{center}
		{\renewcommand{\arraystretch}{1.4} 
			\begin{tabular}{c|rrrrrrrr}
				$a_{k,n}$ & $n=0$ & $n=1$ & $n=2$ & $n=3$ & $n=4$ & $n=5$ & $n=6$ & $n=7$ \\
				\hline
				$k=0$ & $1$ & $5$  & $45$            & $545$                & $7885$                & $127905$                  & $2241225$                 & $41467725$ \\
				$k=1$ & $0$ & $8$  & $100$           & $\frac{4148}{3}$     & $\frac{64198}{3}$     & $\frac{1804058}{5}$       & $\frac{19434106}{3}$      & $\frac{2559002582}{21}$ \\
				$k=2$ & $0$ & $1$  & $\frac{197}{4}$ & $\frac{33637}{36}$   & $\frac{2402477}{144}$ & $\frac{121787041}{400}$   & $\frac{4134822577}{720}$  & $\frac{3940929461909}{35280}$ \\
				$k=3$ & $0$ & $-2$ & $-\frac{89}{4}$ & $-\frac{19295}{108}$ & $-\frac{933155}{864}$ & $\frac{114928799}{36000}$ & $\frac{1490204749}{4320}$ & $\frac{78690073506979}{7408800}$ \\
		\end{tabular}}
	\end{center}
	\caption{
		The first few coefficients $a_{k,n}$.
	}
	\label{table_coeff_a}
\end{table}
in table~\ref{table_coeff_a}.

The series solutions in eq.~(\ref{Frobenius_series}) are convergent for $|z| < \frac{1}{25}$. The radius of convergence is determined by the next singularity of the Picard--Fuchs operator, which is at $z=\frac{1}{25}$. In terms of Feynman integrals this is related to the physical threshold located at $z=\frac{1}{25}$.

\subsection{Overview of geometric interpretations of the Picard--Fuchs operator}
The previous discussion uses only the general theory of differential equations. 
However, the Picard--Fuchs operator $L^{(0)}$ has also interesting geometric interpretations: 
it is intimately related to a family of Calabi--Yau manifolds known as ($\mathbb{Z}_5$-quotients of) Hulek--Verrill manifolds \cite{Hulek:2005} (see also \cite{Candelas:2021lkc}), describing the variation of the periods of the family with the variable $z$. 
This relation allows us to express many of the interesting properties of the banana integral in terms of the geometric properties of the Hulek--Verrill Calabi--Yau manifolds (see for example \cite{Bonisch:2020qmm}). The correspondence between Hulek--Verrill manifolds and banana integrals is by now well-known, dating back to at least \cite{Vanhove:2018mto,Candelas:2019llw}.

However, the Hulek--Verrill Calabi--Yau geometries are not the only ones related to the differential operator $L^{(0)}$. 
In this work, we construct a family of genus-two curves whose periods are governed by the Picard--Fuchs operator $L^{(0)}$, and single out a holomorphic one-form such that the periods of the form are also annihilated by $L^{(0)}$. We first present the general formalism that allows us to carry out this construction, which can be applied to any Calabi--Yau threefold with one or two complex structure parameters. Part of the construction even carries over to Calabi--Yau threefolds with more than two complex structure parameters. However, in these cases with higher dimensional complex structure parameter spaces the Schottky problem poses an obstruction to formulate a Calabi--Yau-to-curve correspondence for generic values of the complex structure parameters. The requirements for spelling out such a correspondence --- in particular the dependence on the number of complex structure parameters of the Calabi--Yau threefold --- are discussed in detail in section~\ref{sect:an_corr}. We work out the specific example for the banana integral in section~\ref{sect:jacobian}.


\section{The Calabi--Yau-to-Curve Correspondence} \label{sect:Calabi-Yau_to_Curve}
In this section, we describe in detail the geometric origin of the constructed correspondence between Calabi--Yau threefolds and complex genus-$g$ curves. For the underlying mathematical principles, see for instance the textbooks \cite{MR1288523,MR1967689,Cox1999a,Hori:2003ic}. Firstly, for Calabi--Yau threefolds, we introduce and review their integral periods, projective special K\"ahler manifolds for their complex structure moduli spaces, and the notion of their Griffiths and Weil intermediate Jacobians. We develop the notion of what we call polarised holomorphic intermediate Jacobians, which is tailored to define a useful correspondence between Calabi--Yau threefolds and genus-$g$ curves in the context of Feynman integral calculations.  Secondly, we review the complex structure moduli space of genus-$g$ curves and define their period matrices and their associated Griffiths intermediate Jacobians, which for genus-$g$ curves coincide with their Weil intermediate Jacobians. Thirdly, for Calabi--Yau threefolds with a single complex structure parameter on the one hand and for genus-two curves on the other hand, we spell out the Calabi--Yau-to-curve correspondence explicitly. It is this particular correspondence that is applied in the following sections.

This section is meant for the reader interested in the conceptual origin of the presented Calabi--Yau-to-curve correspondence. For those readers who are mainly interested in the presented correspondence on a computational level, this section can be skipped on first reading.


\subsection{The complex structure moduli space of Calabi--Yau threefolds} \label{sect:CY2curve}
Let us consider a family of three-dimensional Calabi--Yau varieties $\mathcal{Y}_{\mathcal{M}_\text{cs}}$ over its complex structure moduli space~$\mathcal{M}_\text{cs}$ of complex dimension $m$. In this work we are not so much concerned with the global structure of the moduli space $\mathcal{M}_\text{cs}$. Instead, for a (small) complex $m$-dimensional open disc $\Delta\subset\mathcal{M}_\text{cs}$, we focus on the local family $\mathcal{Y}_{\Delta} \subset\mathcal{Y}_{\mathcal{M}_\text{cs}}$ of Calabi--Yau varieties
\begin{equation}
	\begin{CD}
		Y_z @>>> \mathcal{Y}_{\Delta} \\
		@. @V\pi VV \\
		@. \Delta
	\end{CD}
\end{equation}
where the fibre $Y_z$ denotes the Calabi--Yau variety associated to the complex structure modulus $z\in \Delta$ and the map $\pi$ projects a member $Y_z$ of the family $\mathcal{Y}_\Delta$ to the corresponding parameter $z\in \Delta$ of the (local) complex structure moduli space $\Delta$.

The three-form cohomology groups $H^3(Y_z,\mathbb{R})$ of the Calabi--Yau manifolds $Y_z$ of dimension $2(m+1)$ of a smooth local family $\mathcal{Y}_{\Delta}$ over the contractable disc $\Delta$ are constant because these cohomology classes are topological invariants of smooth manifolds.\footnote{For smooth families over non-contractable open subsets of $\mathcal{M}_\text{cs}$ or for families with singular Calabi--Yau varieties $Y_z$ which correspond to lower-dimensional loci in moduli space, the cohomology groups $H^3(Y_z,\mathbb{R})$ still furnish a local system.}  
The constant complex cohomology classes $H^3(Y_z,\mathbb{C}) = H^3(Y_z,\mathbb{R})\otimes\mathbb{C}$ decompose into Dolbeault cohomology classes $H^{(3-n,n)}(Y_z)$, $n=0,\ldots,3$, as
\begin{equation}
	H^3(Y_z,\mathbb{C}) = \bigoplus_{n=0}^3 H^{(3-n,n)}(Y_z) \ ,
\end{equation}  
which vary over the complex structure moduli space $\Delta$. In particular, the nowhere vanishing $(3,0)$-form $\Omega(z) \in H^{(3,0)}(Y_z)$ varies holomorphically. That is to say, the $(3,0)$-forms $\Omega(z)$ furnish a holomorphic section of the holomorphic line bundle over $\Delta$, the fibres of which are the complex one-dimensional Dolbeault cohomology classes $H^{(3,0)}(Y_z)$. Moreover, the $(3,0)$-form $\Omega(z)$ unambiguously determines the complex structure of the Calabi--Yau threefold $Y_z$. This result is often referred to as Torelli's theorem.

The Dolbeault cohomology groups $H^{(3-n,n)}(Y_z)$ for $n\ge 1$ do not furnish holomorphic bundles over the moduli space $\Delta$. However, the Dolbeault cohomology classes give rise to the Hodge filtration
\begin{equation} \label{eq:filtration}
	H^{(3,0)}(Y_z) \simeq F^3H^3_\mathbb{C}\subset F^2H^3_\mathbb{C} \subset \ldots \subset
	F^0H^3_\mathbb{C}\simeq H^3(Y_z,\mathbb{C}) \ ,
\end{equation}
with
\begin{equation}
	F^{3-p} H^3_\mathbb{C} \simeq \bigoplus_{j=0}^p H^{(3-j,j)}(Y_z) \ , \quad p=0,1,2,3 \ .
\end{equation} 
The important point is that the Hodge filtration varies holomorphically over $\Delta$ \cite{MR0229641,MR0233825,MR0258824}. This means that --- as opposed to the Dolbeault cohomology groups $H^{(3-n,n)}(Y_z)$ for $n\ge 1$ --- the  filter pieces $F^{3-p}H^3_\mathbb{C}$ give rise to holomorphic vector bundles over $\Delta$. 

As discussed, the holomorphic $(3,0)$-form $\Omega(z)$ encodes the complex structure of the Calabi--Yau manifold $Y_z$. Infinitesimal complex structure deformations are given in terms of a three-form $\theta(z)$ by deformations $\Omega(z) \mapsto \Omega(z) + \lambda \theta(z) + O(\lambda^2)$. A non-trivial first order deformation $\theta(z)$ must be a cohomology class in $H^{(2,1)}(Y_z)$ \cite{MR0112154}. Conversely, the Bogomolov--Tian--Todorov theorem ensures that any infinitesimal deformation by a cohomology class in $H^{(2,1)}(Y_z)$ extends to a finite deformation \cite{MR0514769,MR0915841,MR1027500}. Therefore the dimension of the complex structure moduli space is given by
\begin{equation}
	m = \dim_\mathbb{C} \Delta = h^{2,1}(Y_z)  \ ,
\end{equation}  
in terms of the Hodge numbers $h^{p,q}(Y_z) = \dim_\mathbb{C} H^{(p,q)}(Y_z)$. 

\subsection{The integral periods of Calabi--Yau threefolds} \label{sect:integral_periods}
Instead of describing the complex structure moduli space in terms of $\Omega(z)$ directly, it is convenient to parametrise the complex structure moduli space by the periods of $\Omega(z)$, which are defined as follows: Let us consider the homology groups $H_3(Y_z,\mathbb{Z})$, which --- as their dual cohomology groups --- are topological and thus constant over $\Delta$ (or at least furnish a local system over $\Delta$).  As a result, each integral homology class $\gamma \in H_3(Y_z,\mathbb{Z})$ defines an integral period
\begin{equation} \label{eq:integral_periods_definition}
	\Pi_\gamma(z) = \int_\gamma \Omega(z) \ ,
\end{equation}  
which is a holomorphic function over the (local) moduli space $\Delta$. We often choose an integral symplectic basis of three-cycles $(B_J, A_I)$, $I=0,\ldots,m$, which obeys the canonical (oriented) skew-symmetric intersection relations
\begin{equation} \label{eq:AB_intersections}
	A^I \cap B_J = \delta^I_J \ , \quad
	A^I \cap A^J = B_I \cap B_J = 0 \ , \quad I,J = 0,\ldots, m \ .
\end{equation}
These relations do not uniquely fix a basis. Rather, they define a basis of symplectic three-cycles only up to an integral symplectic transformation $\operatorname{Sp}(2(m+1),\mathbb{Z})$. With respect to a chosen basis, we define the integral period vector
\begin{equation} \label{eq:DefABPeriods}
	\Pi(z) = \left( \Pi_{B_m}(z), \ldots,    \Pi_{B_0}(z),    \Pi_{A^m}(z)  , \ldots ,  \Pi_{A^0}(z)   \right) ^ T  \ .
\end{equation}
The integral periods with respect to the three-cycles $A^I$ and $B_J$, $I,J=0,\ldots,m$, are called integral A- and B-periods, respectively. We often denote them in the following by
\begin{equation}  \label{eq:XIFJ}
	X^I(z) = \int_{A^I} \Omega(z) \ , \qquad
	F_J(z) = \int_{B_J} \Omega(z) \ , \qquad I,J = 0,\ldots, m \ .
\end{equation}
Note that any integral period $\Pi_\Gamma(z)$ is uniquely expressed as an integral linear combination of A- and B-periods. Furthermore, due to Torelli's theorem, the period vector $\Pi(z)$ fully determines the complex structure of the Calabi--Yau manifold $Y_z$ --- analogously as the holomorphic $(3,0)$-form $\Omega(z)$. In fact, we recover the holomorphic $(3,0)$-form $\Omega(z)$  from the integral periods as follows: Let $(\alpha_I, \beta^J)$, $I,J=0,\ldots,m$, be an integral basis of $H^3(Y_z,\mathbb{Z})$ that is dual to the chosen symplectic homology basis $(B_J, A^I)$, i.e. 
\begin{equation} \label{eq:cohomology_basis}
	\int_{A^I} \alpha_J = -\int_{B_J} \beta^I = \delta^{I}_J \ , \qquad
	\int_{A^I} \beta^J = \int_{B_I} \alpha_J = 0 \ , \qquad I, J = 0,\ldots, m \ .
\end{equation}
Then the holomorphic $(3,0)$-form $\Omega(z)$ is given by\footnote{As is often customary in the literature, here and in the following we denote by $(\alpha_I\ , \beta^J)$ and $\begin{pmatrix} F_J \\ X^I \end{pmatrix}$ the $2(m+1)$-dimensional row and column vectors formed by the cohomology elements $\alpha_I$ and $\beta^J$, $I,J=0,\ldots,m$, and the A- and B-periods $X^I$ and $F_J$, $I,J=0,\ldots,m$, respectively.  Furthermore, after the second equal sign, we use for ease of notation the Einstein sum convention for the indices of the periods, which is also used in following.}
\begin{equation} \label{eq:OmegaExpand}
	\Omega(z)  = (\alpha_I\ , \beta^J) \ \bm{\Sigma} \  \begin{pmatrix}
		F_J \\ X^I
	\end{pmatrix} =   \alpha_I\,  X^I(z) -   \beta^J F_J(z)    \ , \quad \bm{\Sigma} =\begin{pmatrix*}[r]
	\bm{0}&\bm{1} \\ -\bm{1}&\bm{0}
\end{pmatrix*} \ .
\end{equation}

Finally, let us discuss how the A- and B-periods defined above depend on the chosen symplectic homology basis  $(B_J,A^I)$ and the normalisation of the holomorphic $(3,0)$-form $\Omega(z)$. Firstly, changing the integral symplectic basis  $(B_J, A^I)$ by an integral symplectic transformation
\begin{equation} \label{eq:SympTrans}
	\begin{pmatrix}
		B_J \\ A^I
	\end{pmatrix} \rightarrow \bm \Gamma \begin{pmatrix}
	B_J \\ A^I
\end{pmatrix} \ ,  \quad\bm \Gamma = \begin{pmatrix}   
		\bm{A} & \bm{B} \\ \bm{C} & \bm{D}
	\end{pmatrix} \in \operatorname{Sp}(2(m+1),\mathbb{Z}) \ ,
\end{equation}
 obeying
 \begin{equation}
 	\label{eq:modtrafo}
 	\bm {\Gamma}^T \bm{\Sigma} \bm{\Gamma}=\bm{\Sigma}
 \end{equation}
transforms the period vector as
\begin{equation} \label{eq:PeriodTrans}
	\begin{pmatrix} F_J(z)  \\ X^I(z) \end{pmatrix} \mapsto
	\bm{\Gamma} \begin{pmatrix} F_J(z)  \\ X^I(z) \end{pmatrix} = 
	\begin{pmatrix} 
		  \bm{A}^{\,\,K}_{J} F_K(z)+ \bm{B}_{JK} X^K(z)  \\ 
		{\bm{C}}^{\,IK} F_K(z)+\bm{D}^{\,I}_{\,K} X^K(z) 
	\end{pmatrix} \ .
\end{equation}
In terms of the integral $(m+1)\times(m+1)$ block-matrices $\bm{A}$, $\bm{B}$,  $\bm{C}$, and  $\bm{D}$ equation~(\ref{eq:modtrafo}) translates to
\begin{equation}
	\bm{D}^T \bm{A} - \bm{B}^T \bm{C} = \bm{1} \ , \qquad
	\bm{C}^T \bm{A} - \bm{A}^T \bm{C} =  \bm{0} \ , \qquad
	\bm{D}^T \bm{B} - \bm{B}^T \bm{D} = \bm{0} \ .
\end{equation}
Such transformations are also referred to as modular transformations. Secondly, rescaling the holomorphic $(3,0)$-form $\Omega(z)$ with a non-zero holomorphic function $f: \Delta \to \mathbb{C}^*$ changes the periods $\Pi_\Gamma(z)$ to $f(z) \Pi_\Gamma(z)$. Therefore, we say that the periods $\Pi_\Gamma(z)$ are homogeneous of degree one. 

\subsection{Griffiths intermediate Jacobians of Calabi--Yau threefolds} \label{sect:Griffiths_Jacobian_general}

For our work, an important tool to describe the complex structure moduli space of the Calabi--Yau manifolds $Y_z$ is the second Griffiths intermediate Jacobian $J_2^G(Y_z)$, which is given by \cite{MR0229641,MR0233825,MR1288523}\footnote{By the following fairly common notation we mean the double quotient $ (H^3(Y_z,\mathbb{C})/F^2H_{\mathbb{C}}^3)/H^3(Y_z, \mathbb{Z})$.}
\begin{equation} \label{eq:DefJ2G}
	J_2^G(Y_z) \simeq H^3(Y_z,\mathbb{C}) / \left( F^2H_{\mathbb{C}}^3 \oplus H^3(Y_z, \mathbb{Z}) \right) \ .
\end{equation}   
The Griffiths intermediate Jacobians $J_2^G(Y_z)$ vary holomorphically over $\Delta$ because the subspaces $F^2H_{\mathbb{C}}^3$ of the filtrations~\eqref{eq:filtration} furnish a holomorphic bundle over the moduli space $\Delta$. Moreover, since $F^2H_{\mathbb{C}}^3$ and $H^3(Y_z,\mathbb{C})$ are of complex dimension $(m+1)$ and $2(m+1)$, respectively, $H^3(Y_z,\mathbb{C}) /  F^2H_{\mathbb{C}}^3$ is a complex vector space of dimension $(m+1)$, which in $J_2^G(Y_z)$ is divided by the $2(m+1)$ dimensional lattice $H^3(Y_z, \mathbb{Z}) \subset H^3(Y_z,\mathbb{C}) / F^2H_{\mathbb{C}}^3$. Hence, altogether the Griffiths intermediate Jacobian is a complex torus of the form
\begin{equation}
	J_2^G(Y_z) \simeq \mathbb{C}^{m+1}/\Lambda \ ,
\end{equation} 
where the $2(m+1)$ dimensional lattice $\Lambda$ is determined by the complex structure of the Calabi--Yau threefold $Y_z$. 

Let us now determine the lattice $\Lambda$ of the intermediate Jacobian $J_2^G(Y_z)$ explicitly. First, we use Griffiths transversality \cite{MR0229641,MR0233825}, which asserts that the subspace $F^2H_{\mathbb{C}}^3$ is generated by $\Omega(z)$ and its $m$-dimensional gradient with respect to the complex structure moduli $z$, i.e.
\begin{equation} \label{eq:F2H3}
	F^2H^3_\mathbb{C} = \langle\!\langle \Omega(z), \partial_z \Omega(z) \rangle\!\rangle \ .
\end{equation}  
Thus, with the explicit form~\eqref{eq:OmegaExpand} for $\Omega(z)$ the quotient $H^3(Y_z,\mathbb{C}) /  F^2H_{\mathbb{C}}^3 \simeq \mathbb{C}^{m+1}$ is determined by $(m+1)$ equivalence relations, which with respect to the symplectic integral basis $(\alpha_I, \beta^J)$ of $H^3(Y_z,\mathbb{C})$ become in matrix form
\begin{equation} \label{eq:EQRel}
	0 \sim \begin{pmatrix} \alpha_I,  \beta^J\end{pmatrix} \bm{\Sigma} \begin{pmatrix} \mathbfcal{F} \\  \mathbfcal{X} \end{pmatrix} \ ,
\end{equation}
with the $(m+1)\times(m+1)$ matrices
\begin{equation} \label{eq:XFmat}
	\mathbfcal{X} = \begin{pmatrix} X^I , \partial_z X^I \end{pmatrix} \ , \qquad
	\mathbfcal{F} = \begin{pmatrix} F_J , \partial_z F_J \end{pmatrix} \ ,
\end{equation}   
and where $\begin{pmatrix} \alpha_I,  \beta^J\end{pmatrix}$ denotes the $2(m+1)$-dimensional row vector of symplectic cohomology classes. 
As the matrix $\mathbfcal{X}$ is invertible for smooth Calabi--Yau threefolds $Y_z$ this equivalence relation relates the cohomology classes $\alpha_I$ to $\beta^J$ as
\begin{equation}
	\alpha_I \sim   \beta^J \, \left(\mathbfcal{F} \mathbfcal{X}^{-1}  \right)_{JI} \ , \qquad I=0,\ldots,m \ .
\end{equation}  
These relations allow us to explicitly divide out the integral cohomology classes $H^3(Y_z,\mathbb{Z})$, such that we readily arrive at the lattice $\Lambda$ of the form
\begin{equation}
	\Lambda = \mathbb{Z}^{m+1} +  \mathbb{Z}^{m+1}\left(  \mathbfcal{F} \mathbfcal{X}^{-1} \right) \ .
\end{equation}  
In summary, we find that the second Griffiths intermediate Jacobian $J_2^G(Y_z)$ of the Calabi--Yau threefold $Y_z$ reads
\begin{equation} \label{eq:J2Yz}
	J_2^G(Y_z) \simeq \mathbb{C}^{m+1}/( \mathbb{Z}^{m+1} + \mathbb{Z}^{m+1}\left(  \mathbfcal{F} \mathbfcal{X}^{-1} \right) )
	\simeq \mathbb{C}^{m+1}/(  \mathbb{Z}^{m+1}  \mathbfcal{X} + \mathbb{Z}^{m+1} \mathbfcal{F}) \ .
\end{equation}
where the $(m+1)\times(m+1)$ matrices $\mathbfcal{X}$ and $\mathbfcal{F}$ are calculated from the A- and B-periods according to eq.~\eqref{eq:XFmat}. 

Many of the properties of the family of Calabi--Yau threefolds $\mathcal{Y}_\Delta$ described so far generalise to families of Calabi--Yau manifolds of higher odd complex dimension. However, what is specific to families of three-dimensional Calabi--Yau manifolds is the fact that their (local) complex structure moduli spaces $\mathcal{M}_\text{cs}$ are projective special K\"ahler manifolds of complex dimension $m$ \cite{Ceresole:1992su,Bershadsky:1993cx}.\footnote{For an introduction to projective special K\"ahler manifolds, see for instance ref.~\cite{Freed:1997dp}} Such projective special K\"ahler manifolds are locally entirely given by their holomorphic prepotential $F(X)$, which is a holomorphic function of homogeneous degree two in terms of the A-periods $X^I$, $I=0,\ldots,m$, which are of homogeneous degree one. Furthermore, the B-periods $F_I$ of homogeneous degree one are the derivatives of the prepotential $F(X)$ with respect to the A-periods $X^I$, i.e.
\begin{equation} \label{eq:FI}
	F_I(X) = \frac{\partial F(X)}{\partial X^I} \  .
\end{equation}  
The K\"ahler potential $K$ of the projective special K\"ahler manifolds is given in terms of the prepotential $F(X)$ as
\begin{equation} \label{eq:Kdef}
	K = - \log \, \operatorname{Im} \left( 2  \, X^I \, \overline{F}_I(X) \right) \  .
\end{equation}
The K\"ahler potential $K$ should be viewed as a real function of a choice of affine coordinates $z^\alpha$, $\alpha=1,\ldots,m$, that are constructed from the projective coordinates $X^I$ for instance as $z^\alpha = \frac{X^\alpha}{X^0}$.\footnote{Rescaling the holomorphic projective coordinates $X^I(z)$ with a holomorphic function $f(z)$ changes the K\"ahler potential by a K\"ahler transformation $K \mapsto K - f(z) - \overline{f(z)}$. Similarly, K\"ahler potentials calculated with respect to different affine coordinates differ by a suitable K\"ahler transformation as well.} In the given context, these affine coordinates $z^\alpha$ coincide with the coordinates of the complex structure moduli space $\Delta$.

Considering the complex structure moduli space $\Delta$ from the perspective of the projective special K\"ahler manifold, the holomorphic $(3,0)$-form $\Omega$ can be interpreted as a function of the projective coordinates $X^I$ (which in turn depend on the affine complex structure moduli $z$). Then the filtered subspace $F^2H^3_\mathbb{C}$ calculated in eq.~\eqref{eq:F2H3} is generated by the gradient of $\Omega(X^I)$ with respect to the projective coordinates $X^I$, namely
\begin{equation} \label{eq:F2H3two}
	F^2H^3_\mathbb{C} = \langle\!\langle  \partial_I \Omega(X) \rangle\!\rangle_{I=0,\ldots,m} \ .
\end{equation}
Note that as the periods $X^I$ and $F_J$ are of homogenous degree one, $\Omega$ obeys the identity $\Omega = \sum_I X^I  \partial_{X^I} \, \Omega$, and hence the gradient with respect to the projective coordinates $X^I$ also contains the filtered subspace $F^3H^3_\mathbb{C}$. Thus, with respect to the generators in eq.~\eqref{eq:F2H3two} the equivalence relation~\eqref{eq:EQRel} becomes
\begin{equation}
	0 \sim \begin{pmatrix}   \alpha^I , \beta_J \end{pmatrix}  \bm{\Sigma}
	\begin{pmatrix}  \bm{F} \\ \bm{1} \end{pmatrix} \ ,
\end{equation} 
where $\bm{1}$ is the $(m+1)\times(m+1)$ identity matrix and the $(m+1)\times(m+1)$ matrix $\bm{F}$ of homogeneous degree zero reads in terms of the prepotential $F(X)$
\begin{equation} \label{eq:FIJ}
	\bm{F} =(F_{IJ}) \ , \qquad F_{IJ} = \frac{\partial^2 F}{\partial X^I \partial X^J} \ .
\end{equation}   
Therefore, we arrive at the convenient expression for the second Griffiths intermediate Jacobian~\eqref{eq:J2Yz} of the Calabi--Yau threefold $Y_z$: 
\begin{equation} \label{eq:J2GYzF}
	J_2^G(Y_z) \simeq \mathbb{C}^{m+1}/( \mathbb{Z}^{m+1} +  \mathbb{Z}^{m+1}  \bm{F}) \ .
\end{equation}  

Let us examine the Griffiths intermediate Jacobian $J_2^G(Y_z)$ for smooth Calabi--Yau threefolds~$Y_z$ further. First, we consider the Hermitian intersection pairing $Q$ for smooth Calabi--Yau threefolds $Y_z$ given by
\begin{equation}  \label{eq:DefQ}
	Q: H^3(Y_z,\mathbb{C}) \times H^3(Y_z,\mathbb{C}) \to \mathbb{C}, \ (\theta,\chi) \mapsto i \int \theta\wedge \overline\chi  \ .
\end{equation}   
It has signature $(m+1,m+1)$  because it is induced from the symplectic intersection pairing on $H^3(Y_z,\mathbb{R})$. Restricting $Q$ to the Dolbeault cohomology groups $H^{(3-n,n)}$, we have that\footnote{This can be verified by noting that in suitable local holomorphic coordinates $u$ of the Calabi--Yau threefold $Y_z$ a (positive) volume form takes the form $i (du^1 \wedge du^2 \wedge du^3) \wedge (d\overline u^1 \wedge d\overline u^2 \wedge d\overline u^3)$.} 
\begin{equation}
	\left\{
	\begin{matrix} 
		Q|_{H^{(3-n,n)}} \text{ is positive definite for $n$ even} \\[0.5ex]
		Q|_{H^{(3-n,n)}} \text{ is negative definite for $n$ odd} 
	\end{matrix} \right\} \ .
\end{equation}
Therefore, the Hermitian form $Q$ restricted to $F^2H^3_\mathbb{C} = H^{(3,0)}(Y_z) \oplus H^{(2,1)}(Y_z)$ has signature $(1,m)$, and in the basis~\eqref{eq:F2H3two} the restriction to $F^2H^3_\mathbb{C}$ becomes
\begin{equation}
	i \int_{Y_z} \partial_I\Omega \wedge \overline{\partial_J \Omega} = 2\, \operatorname{Im}(F_{IJ}) \ .
\end{equation}
Thus the imaginary part of the matrix $\bm{F}$ has signature $(1,m)$. This demonstrates the well-known result that the complex torus~\eqref{eq:J2GYzF} of the Griffiths intermediate Jacobian $J_2^G(Y_z)$ for Calabi--Yau threefolds with $m>0$ is not an Abelian variety,\footnote{A complex torus $\mathbb{C}^g/(\mathbb{Z}^g + \mathbb{Z}^g\bm{\tau})$ with non-degenerate matrix~$\operatorname{Im}(\bm{\tau})$ is an Abelian variety if it admits an embedding into some complex projective space $\mathbb{CP}^N$. Such an embedding can be constructed if and only if the complex torus $\mathbb{C}^g/(\mathbb{Z}^g + \mathbb{Z}^g\bm{\tau})$ possesses an ample line bundle $\mathcal{L}$. For $\operatorname{Im}(\tau)$ positive definite, the genus-$g$ theta function $\theta: \mathcal{H}_g \times \mathbb{C}^g \to \mathbb{C}$ given by $\theta(\bm{\tau},z) =\sum_{n\in \mathbb{Z}^g} e^{i \pi (n^T \bm{\tau} n +2 n^T z)}$ is well-defined because the sum over $n \in \mathbb{Z}^g$ is convergent. Then the theta function $\theta(\bm{\tau},z)$ realises a holomorphic section of an ample line bundle $\mathcal{L}$. Conversely, for $\operatorname{Im}(\tau)$ with indefinite signature there is no such holomorphic theta function and no ample line bundle $\mathcal{L}$. Therefore, the complex torus $\mathbb{C}^g/(\mathbb{Z}^g + \mathbb{Z}^g\bm{\tau})$  is not an Abelian variety in this case.}
because neither the symmetric matrix $\bm{F}$ nor its negative $-\bm{F}$ reside in the Siegel upper half-space $\mathcal{H}_{m+1}$,\footnote{\label{footnote_minus_sign} The matrices $\bm{F}$ and $-\bm{F}$ give rise to identical complex tori because their defining lattices $\mathbb{Z}^{m+1} + \mathbb{Z}^{m+1}\bm{F}$ and  $\mathbb{Z}^{m+1} +\mathbb{Z}^{m+1}(- \bm{F})$ are identical.} where $\mathcal{H}_{g}$, for any positive integer $g$, is defined as
\begin{equation} \label{eq:SiegelHalf}
	\mathcal{H}_{g} = \left\{ \ \bm{\tau} \in \operatorname{Sym}(g,\mathbb{C}) \, \middle| \, \operatorname{Im}(\bm{\tau}) \text{ is positive definite } \right\} \ .
\end{equation}  

\subsection{Weil intermediate Jacobians of Calabi--Yau threefolds} \label{sec:WeilJac}
The Weil intermediate Jacobian $J_2^W(Y_z)$ of the Calabi--Yau manifold $Y_z$ is an Abelian variety. It arises from the quotient
\begin{equation} \label{eq:DefJ2W}
	\begin{aligned}
		J_2^W(Y_z) &\simeq H^3(Y_z,\mathbb{C}) / \left( H^{(3,0)}(Y_z) \oplus \overline{H^{(2,1)}(Y_z)} \oplus H^3(Y_z, \mathbb{Z}) \right) \\
		&\simeq H^3(Y_z,\mathbb{C}) / \left( H^{(3,0)}(Y_z) \oplus H^{(1,2)}(Y_z) \oplus H^3(Y_z, \mathbb{Z}) \right)  \ .
	\end{aligned}   
\end{equation}  
Here we divide out the complex $(m+1)$-dimensional subspace $H^{(3,0)}(Y_z) \oplus H^{(1,2)}(Y_z)$, on which the Hermitian intersection pairing $Q$ of eq.~\eqref{eq:DefQ} restricts to a positive definite Hermitian form. Following similar arguments to those for the Griffiths intermediate Jacobian, the Weil intermediate Jacobian $J_2^W(Y_z)$ becomes the quotient
\begin{equation} \label{eq:DefJ2Wtwo}
	J_2^W(Y_z) \simeq \mathbb{C}^{m+1} / (\mathbb{Z}^{m+1} + \mathbb{Z}^{m+1} \bm{N} ) \ .
\end{equation}
The definiteness of the restricted intersection pairing implies that
$\bm{N}$ is a symmetric $(m+1) \times (m+1)$ matrix in the Siegel upper half-space $\mathcal{H}_{m+1}$.

Let us now explicitly construct the defining matrix $\bm{N}$ of the Weil intermediate Jacobian. We introduce the projective K\"ahler covariant derivative 
\begin{equation} \label{eq:Kahler_covariant_derivative}
	\nabla_I = \frac{\partial}{\partial X^I} + K_I \ , \qquad
	K_I = \frac{\overline F_I - \overline X^L F_{L I}}{\overline X^L F_L - X^L \overline F_L}=-\frac{\overline{X}^L\operatorname{Im}(F_{IL})}{\operatorname{Im}(\overline{X}^LF_L)} \ ,
\end{equation}
defined in terms of the K\"ahler potential~\eqref{eq:Kdef}.  Using eq.~\eqref{eq:OmegaExpand} and the definition of the K\"ahler covariant derivative, we find
\begin{equation}
	0 = \int_{Y_z} \nabla_I \Omega \wedge \overline\Omega  \ .
\end{equation}
Since $i \Omega \wedge \overline \Omega$ is a volume form of the Calabi--Yau threefold $Y_z$, the three-form $\nabla_I\Omega$ cannot contain a $(3,0)$ part, which --- combined with Griffiths transversality --- implies that
\begin{equation}
	\nabla_I \Omega \in H^{(2,1)}(Y_z) \quad \text{for all} \quad I=0,\ldots,m \ .
\end{equation}
Due to the homogeneity identity $\Omega = X^I \partial_I\Omega$, the holomorphic $(3,0)$-form $\Omega$ is in the kernel of the differential operator $X^I \nabla_I$, i.e. $X^I \nabla_I \Omega = 0$ and $\overline X^I \overline{\nabla}_I \overline\Omega =0$. Moreover, it is trivially true that $\overline\nabla_I \Omega \in H^{(3,0)}(Y_z)$ and $\overline X^I \overline\nabla_I \Omega =\Omega$. As a result, the subspace $H^{(3,0)}(Y_z) \oplus H^{(1,2)}(Y_z)$ is generated by
\begin{equation} \label{eq:Weil_intermediate_Jacobian_subspace_V}
	\begin{split}
	H^{(3,0)}(Y_z) \oplus H^{(1,2)}(Y_z) &= \langle\!\langle \overline{\nabla}_I ( \Omega + \overline\Omega)  \rangle\!\rangle_{I=0,\ldots, m}\\&=\langle\!\langle \alpha_J\overline{\nabla}_I ( X^J+\overline{X}^J)-\beta^I \overline{\nabla}_I (F_J+\overline{F}_J)  \rangle\!\rangle_{I=0,\ldots, m} \ .
	\end{split}
\end{equation}
Analogously to the derivation of the relations~\eqref{eq:EQRel} for the Griffiths intermediate Jacobian, these basis elements yield the coset relations of the Weil intermediate Jacobian which are of the form
\begin{equation}
	0 \sim \begin{pmatrix} \alpha_I  ,  \beta^J \end{pmatrix} \bm{\Sigma} \begin{pmatrix}  \mathbfcal{N} \\ \mathbfcal{Z} \end{pmatrix} \ , \qquad
	\mathbfcal{Z} = \big(\delta^I_J +  (X^I +\overline X^I)\overline K_J \big) \ , \qquad
	\mathbfcal{N} = \big(\overline F_{IJ} + (F_I +\overline F_I) \overline K_J  \big) \ ,
\end{equation}
so that the matrix $\bm{N} \in \mathcal{H}_{m+1}$ defining the Weil intermediate Jacobian~\eqref{eq:DefJ2Wtwo} becomes\footnote{Similarly to the Griffiths intermediate Jacobian, $\mathbfcal{N}\mathbfcal{Z}^{-1}$ and $-\mathbfcal{N}\mathbfcal{Z}^{-1}$ give rise to the same complex torus. The latter convention is chosen in equation \eqref{eq:Nmatrix} in order to obtain $\bm{N}\in \mathcal{H}_{m+1}$.}
\begin{equation}
	\label{eq:Nmatrix}
	\bm{N} = -  \mathbfcal{N}\mathbfcal{Z}^{-1} \ .
\end{equation}
In terms of the inverse matrix
\begin{equation}
	\mathbfcal{Z}^{-1} = \left( \delta^I_J - \frac{  (X^I +\overline X^I)\overline K_J }{ \overline K_L X^L } \right)  
	= \left( \delta^I_J-\frac{ (X^I+\overline X^I)\operatorname{Im}(F_{JM})X^M}{\operatorname{Im}(F_{KL})X^K X^L } \right) \ , 
\end{equation}
we arrive for the matrix $\bm{N}$ at the explicit expression 
\begin{equation} \label{eq:DefN}
	N_{IJ} = - \overline F_{IJ} - 2 i \frac{\operatorname{Im}(F_{I K}) X^K \operatorname{Im}(F_{J L}) X^L}{\operatorname{Im}(F_{KL})X^K X^L} \ . 
\end{equation}  

The matrix $\bm{N}$ of the Weil intermediate Jacobian $J^W_2(Y_z)$ also appears in the context of four-dimensional $N=2$ supergravity theories \cite{deWit:1984rvr,deWit:1984wbb,Andrianopoli:1996cm,Moore:1998pn}.\footnote{In $N=2$ supergravity literature, the matrix $\bm{N}$ is usually defined with the opposite sign such that the imaginary part of $\bm{N}$ is negative definite.} Namely, for an $N=2$ supergravity theory with $m$ vector multiplets, the target space manifold of the complex scalar fields in the vector multiplet sector is a projective special K\"ahler manifold of complex dimension $m$. In such a theory, the matrix $\bm{N}$ appears (up to an overall conventional sign) as the gauge kinetic coupling function for the $m$ vector fields from the vector multiplets combined with the graviphoton vector field from the gravity multiplet. Furthermore, upon compactifying type~IIB string theory on a smooth Calabi--Yau threefold $Y_z$, the resulting low energy effective $N=2$ supergravity theory consists of $m$ vector multiplets, whose projective special K\"ahler target space is the complex structure moduli space of the family of Calabi--Yau threefolds $\mathcal{Y}_\Delta$ \cite{Bodner:1989cg,Bohm:1999uk}.

\subsection{Griffiths versus Weil Jacobians of Calabi--Yau threefolds} \label{sec:GvsWJac}
Let us first discuss the modular properties of the Griffiths and Weil intermediate Jacobians $J^G_2(Y_z)$ and $J^W_2(Y_z)$. These intermediate Jacobians --- as defined in eqs.~\eqref{eq:DefJ2G} and \eqref{eq:DefJ2W}, respectively --- are constructed from the Dolbeault cohomology groups $H^{(3-n,n)}(Y_z)$ and the filtration~\eqref{eq:filtration}. As both the Dolbeault cohomology groups $H^{(3-n,n)}(Y_z)$ and the filtration~\eqref{eq:filtration} do not rely on a choice of a symplectic homology basis   $(B_J, A^I)$, the Griffiths and Weil intermediate Jacobians $J^G_2(Y_z)$ and $J^W_2(Y_z)$ do not depend on   $(B_J, A^I)$ either. Therefore, they are both invariant under the modular transformations~\eqref{eq:SympTrans}. However, the explicit form of the $(m+1)\times(m+1)$ matrices $\bm{F}$ and $\bm{N}$ that define the lattices $\Lambda$ of the intermediate Jacobians given in eqs.~\eqref{eq:J2GYzF} and \eqref{eq:DefJ2Wtwo} certainly depend on a choice of symplectic basis   $(B_J, A^I)$. Differentiating the B-periods~\eqref{eq:PeriodTrans} with respect to $X^I$, one finds that the matrix $\bm{F}$ transforms under a symplectic transformation~\eqref{eq:SympTrans} as
\begin{equation}
	\bm{F} \mapsto (\bm A \, \bm F + \bm B )(\bm C \, \bm F + \bm D)^{-1} \ .
\end{equation} 
It is straightforward to check that the matrix $\bm{N}$ defined in eq.~\eqref{eq:DefN} transforms according to \cite{Andrianopoli:1996cm}
\begin{equation} \label{eq:transN}
	(-\bm{N}) \mapsto (\bm A \, (-\bm N) + \bm B )(\bm C \, (-\bm N) + \bm D)^{-1} \ .
\end{equation}
We may write eq.~(\ref{eq:transN}) as
\begin{equation} \label{eq:transN_v2}
	\bm{N} \mapsto (\tilde{\bm A} \, \bm N + \tilde{\bm B} )(\tilde{\bm C} \, \bm N + \tilde{\bm D})^{-1} 
\end{equation}
with $\tilde{\bm \Gamma} =\rho({\bm \Gamma})$ and $\rho$ is the group automorphism
\bq
 \rho & : & \operatorname{Sp}(2(m+1),\mathbb{Z}) \; \to \; \operatorname{Sp}(2(m+1),\mathbb{Z})~, \; 
 \begin{pmatrix}   
		\bm{A} & \bm{B} \\ \bm{C} & \bm{D}
	\end{pmatrix}
 \; \mapsto \;
\begin{pmatrix*}[r]   
		\bm{A} & -\bm{B} \\ -\bm{C} & \bm{D}
	\end{pmatrix*} \ .
\eq
Note that these transformations preserve the Siegel upper half-space $\mathcal{H}_{m+1}$, which means that all modular images of the matrix $\bm{N} \in \mathcal{H}_{m+1}$ remain in the Siegel upper half-space $\mathcal{H}_{m+1}$ as well. In particular, this property ensures that the signatures of the imaginary parts of both matrices $\bm{F}$ and $\bm{N}$ are preserved under these transformations. 

As a consequence of the modular invariance of the Griffiths and Weil intermediate Jacobians $J_2^G(Y_z)$ and $J_2^W(Y_z)$ on the one hand, and the transformation behaviour of the matrices $\bm{F}$ and $\bm{N}$ on the other hand, all lattices constructed from modular images of $\bm{F}$ and $\bm{N}$ must respectively be equivalent. This can actually be checked explicitly by spelling out the equivalences among lattices $\Lambda$.

Apart from the distinct signatures (for $m\ge 1$) of the imaginary parts of the matrices~$\bm{F}$ and $\bm{N}$, which imply that only the Weil intermediate Jacobian $J^W_2(Y_z)$ is an Abelian variety --- in contrast to the Griffiths intermediate Jacobian $J^G_2(Y_z)$ --- there is another important difference: Namely, the Griffiths intermediate Jacobian $J^G_2(Y_z)$ varies holomorphically over the complex structure moduli space $\Delta$, whereas the Weil intermediate Jacobian $J^W_2(Y_z)$ does not. 
This is a consequence of the fact that the filtration~\eqref{eq:filtration} varies holomorphically whereas the Dolbeault cohomology groups $H^{(3-n,n)}(Y_z)$ for $n=1,\ldots,3$ do not vary holomorphically over the complex structure moduli space $\Delta$. The former enter in the definition of the Griffiths intermediate Jacobian~\eqref{eq:DefJ2G}, while the latter appear in the definition of the Weil intermediate Jacobian~\eqref{eq:DefJ2W}. 
More explicitly, we see that the defining matrix $\bm{F}$ of the Griffiths intermediate Jacobian is a holomorphic function of the holomorphic A-periods $X^I$ and the defining matrix $\bm{N}$ of the Weil intermediate Jacobian is not holomorphic (c.f., eq.~\eqref{eq:DefN}).

\subsection{Polarised holomorphic Jacobians of Calabi--Yau threefolds} \label{sect:polholJac}
The definitions of the Griffiths and Weil intermediate Jacobians raise the question about more general definitions of intermediate Jacobians. More generally, we can construct a complex torus of dimension $m+1$ given any half-dimensional subspace $V_z \subset H^3(Y_z,\mathbb{C})$ such that 
\begin{align}
	V_z \oplus \overline{V_z} = H^3(Y_z,\mathbb{C})~.
\end{align}
To be explicit, any such vector space $V_z$ can be given in terms of $(m+1)$-linear independent basis vectors, that is
\begin{equation} \label{eq:half-dimensional_subspace_V}
	V_z = \langle\!\langle \,  \alpha_I f_L^I- \beta^J g_{J,L} \,  \rangle\!\rangle_{L=0,\ldots,m} \ .
\end{equation}
Then, following arguments completely analogous to the cases of the Griffiths and Weil intermediate Jacobians, this subspace gives rise to the following complex torus that we call here an intermediate Jacobian $J_2(Y_z,V_z)$: 
\begin{equation} \label{eq:intermediate_Jacobian_J2_definition}
	J_2(Y_z,V_z) \simeq H^3(Y_z,\mathbb{C}) / \left( V_z \oplus H^3(Y_z,\mathbb{Z}) \right) \simeq \mathbb{C}^{m+1} / (\mathbb{Z}^{m+1} +  \mathbb{Z}^{m+1} \bm{M}) \ ,
\end{equation}
where the matrix $\bm{M}$ is given by
\begin{align} \label{eq:DefM}
	\bm{M} = -\bm{g}\ \bm{f}^{-1} ~,\qquad \bm{f}=(f_L^I(X))~, \quad \bm{g}=(g_{J,L}(X))~.
\end{align}
Due to the negative sign in this definition for $\bm M$, the signature of the imaginary part of the matrix~$\bm{M}$ conforms with the signature of the Hermitian form \eqref{eq:DefQ} restricted to $V_z$. From eq.~\eqref{eq:intermediate_Jacobian_J2_definition}, it is immediately clear that the intermediate Jacobian $J_2(Y_z,V_z)$ depends only on the subspace $V_z$ and not on the particular expressions $\bm{f}$ and $\bm{g}$ in the chosen basis $(\alpha_I,\beta^J)$ of $H^3(Y_z,\mathbb{Z})$. Indeed, a simple computation shows that under a symplectic change of basis \eqref{eq:SympTrans}, the matrix $\bm{M}$ transforms as 
\begin{align}
	(-\bm{M}) \mapsto (\bm{A}\, (-\bm{M}) + \bm{B})(\bm{C}\, (-\bm{M}) +  \bm{D})^{-1} \ , 
\end{align}
which shows that the quotient in eq.~\eqref{eq:intermediate_Jacobian_J2_definition} after the change-of-basis is isomorphic to the original. Note also that if the restriction of the Hermitian intersection pairing $Q$ to $V_z$ is positive definite, the resulting intermediate Jacobian is an Abelian variety. 

Like the Griffiths and Weil intermediate Jacobians, this construction can be extended for families of half-dimensional spaces $V_z \subset H^3(Y_z,\mathbb{C})$. To be exact, let us denote by $\mathcal{E}_U$ the Hodge bundle over $U \subseteq \mathcal{M}_{\text{cs}}$\footnote{This construction can be applied to the whole complex structure moduli space $\mathcal{M}_{\text{cs}}$. However, in this work we are mainly interested in the case where $U = \Delta$ is a simply-contractible subspace.} whose fibres are the middle cohomology spaces $H^3(Y_z,\mathbb{C})$. Then let $\mathcal{V}_U \subset \mathcal{E}_U$ be a complex subbundle\footnote{Recall that a complex vector bundle is one whose fibres are complex vector spaces, and the transition functions are $C^\infty$-functions. This should be contrasted with a holomorphic vector bundle whose fibres are also complex vector spaces but whose transition functions are required to be holomorphic.} of $\mathcal{E}_U$ over $U$ whose fibres are the vector spaces $V_z \subset H^3(Y_z,\mathbb{C})$: 
\begin{align}
\begindc{\commdiag}[300]
\obj(-1,3)[Vz]{$V_z$}
\obj(1,3)[calV]{$\mathcal{V}_U$}
\obj(3,3)[calE]{$\mathcal{E}_U$}
\obj(3,1){$U$}
\mor{calV}{$U$}{$\pi|_{\mathcal{V}_U}$}[\atright,\solidarrow]
\mor{calE}{$U$}{$\pi$}[\atleft,\solidarrow]
\mor(-1,3)(1,3){}[\atright,\solidarrow]
\mor(1,3)(3,3){}[\atright,\injectionarrow]
\enddc~.
\end{align}
Then applying the map $J_2$ of eq.~\eqref{eq:intermediate_Jacobian_J2_definition} fibrewise on $\mathcal{V}_U$ gives a corresponding family $\mathcal{J}_2(\mathcal{Y}_U,\mathcal{V}_U)$ of intermediate Jacobians of Calabi--Yau manifolds $Y_z$ belonging to a family $\mathcal{Y}_U$ over the base $U \subset \mathcal{M}_{cs}$:
\begin{equation}
	\begin{CD}
		J_2(Y_z,V_z) @>>> \mathcal{J}_2(\mathcal{Y}_U,\mathcal{V}_U) \\
		@. @V\pi VV \\
		@. U
	\end{CD}
\end{equation}
One easily sees that this construction can be used to obtain the families of Griffiths (respectively Weil) Jacobians. To construct these, one simply starts with the vector bundles $\mathcal{G}_{\mathcal{M}_{\text{cs}}}$ (resp. $\mathcal{W}_{\mathcal{M}_{\text{cs}}}$) over the complex structure moduli space $\mathcal{M}_{\text{cs}}$ whose fibres are given by the eq.~\eqref{eq:F2H3two} (resp. \eqref{eq:Weil_intermediate_Jacobian_subspace_V}). These vector bundles are defined over the whole complex structure moduli space $\mathcal{M}_{\text{cs}}(Y_z)$. Then applying the map $J_2$ of eq.~\eqref{eq:intermediate_Jacobian_J2_definition} gives families of intermediate Jacobians, whose fibres are exactly the Griffiths (resp. Weil) Jacobians defined in eq.~\eqref{eq:DefJ2G} (resp. eq.~\eqref{eq:DefJ2W}):
\begin{equation}
	\begin{CD}
		J_2^G(Y_z) @>>> \mathcal{J}_2^G \\
		@. @V\pi VV \\
		@. U
	\end{CD} 
\qquad \qquad
	\begin{CD}
	J_2^W(Y_z) @>>> \mathcal{J}_2^W \\
	@. @V\pi VV \\
	@. U
\end{CD}~,
\end{equation}
where we have introduced the convenient shorthand notation 
\begin{align}
	\mathcal{J}_2^G := \mathcal{J}_2(\mathcal{Y}_{\mathcal{M}_{\text{cs}}},\mathcal{G}_{\mathcal{M}_{\text{cs}}})~, \qquad 	\mathcal{J}_2^W := \mathcal{J}_2(\mathcal{Y}_{\mathcal{M}_{\text{cs}}},\mathcal{W}_{\mathcal{M}_{\text{cs}}})
\end{align}
to avoid proliferation of indices.

For the purposes of the present work we are in particular interested in the cases where one starts with, instead of only a complex vector bundle, a holomorphic vector bundle $\mathcal{V}_U$ so that the fibres $V_z$ are locally given by
\begin{equation} \label{eq:half-dimensional_subspace_V_families}
	V_z = \langle\!\langle \,  \alpha_I f_L^I(X) -  \beta^J g_{J,L}(X) \,  \rangle\!\rangle_{L=0,\ldots,m} \ ,
\end{equation} 
where the defining functions $f_L^I(X)$ and $g_{J,L}(X)$ are holomorphic functions of the A-periods $X^I$ (and hence holomorphic functions over $U$). Then we arrive at a holomorphic family $\mathcal{J}_2(\mathcal{Y}_U,\mathcal{V}_U)$ of intermediate Jacobians. If, in addition, the restriction of the Hermitian intersection pairing $Q$ to $V_z$ is positive definite for all $z \in U$, we refer to such families of intermediate Jacobians as polarised holomorphic intermediate Jacobians, or often polarised holomorphic Jacobians for short.

While the construction of polarised holomorphic intermediate Jacobians is somewhat unusual from a geometric perspective, these Jacobians arise in this work via the following construction: Let us consider an analytic Lagrangian submanifold $\Delta_\mathbb{R}$ embedded in the projective special K\"ahler complex structure moduli space~$\Delta$.\footnote{The submanifold $\Delta_\mathbb{R}$ embedded in the K\"ahler manifold $\Delta$ of complex dimension $m$ is Lagrangian, if the submanifold $\Delta_\mathbb{R}$ has real dimension $m$ and if the K\"ahler form of $\Delta$ pulled back to $\Delta_\mathbb{R}$ vanishes.} The complex (but not holomorphic) vector bundle $\mathcal{W}_{\mathcal{M}_{\text{cs}}}$, on which the family $\mathcal{J}_2^W$ of Weil intermediate Jacobians is based, can be restricted to a vector bundle $\mathcal{W}_{\Delta_{\mathbb{R}}} := \mathcal{W}_{\mathcal{M}_{\text{cs}}}|_{\Delta_{\mathbb{R}}}$ over the base $\Delta_{\mathbb{R}}$. In this case there exists an open subset $U \subset \Delta$ with $\Delta_\mathbb{R} \subset U$ such that the complex vector bundle uniquely extends to a holomorphic vector subbundle $\mathcal{V}_U \subset \mathcal{H}_U$, by which we mean that $\mathcal{V}_{\Delta_\mathbb{R}} := \mathcal{V}_U|_{\Delta_\mathbb{R}} = \mathcal{W}_{\Delta_{\mathbb{R}}}$.\footnote{Since we are working locally, we can use a local trivialisation of $\mathcal{H}_\Delta \simeq \mathbb{C}^{2(m+1)} \times \Delta$ where the subbundle $\mathcal{W}_\Delta$ is defined by taking the fibres to be given by the subspaces \eqref{eq:half-dimensional_subspace_V_families} where the non-holomorphic functions $f_L^I(X)$ and $g_{L,J}(X)$ are defined via \eqref{eq:Weil_intermediate_Jacobian_subspace_V}. On $\Delta_\mathbb{R}$ $f_L^I(X)$ and $g_{L,J}(X)$ restrict to real analytic functions, which can be holomorphically continued to some neighbourhood $U \subset \Delta$ of $\Delta_{\mathbb{R}}$. This defines the holomorphic subbundle $\mathcal{V}_U \subset \mathcal{E}_U$ that extends $\mathcal{W}_{\Delta_\mathbb{R}}$.} 

Restricting the neighbourhood $U$ further if needed, we can obtain a vector bundle $\mathcal{V}_U$ such that the restriction $Q|_{V_z}$ of the Hermitian intersection pairing to the fibre $V_z$ is positive definite for any fibre $V_z$.\footnote{By construction $Q|_{V_z}=Q|_{W_z}$ if $z \in \Delta_{\mathbb{R}}$, and we know that $Q|_{W_z}$ is positive definite since the fibre $W_z$ is associated to a Weil intermediate Jacobian. The signature of $Q|_{V_z}$ varies continuously along the base $U$, and positive definiteness is an open condition. Therefore we can always find a neighbourhood $\widetilde{U} \subset U$ of $\Delta_{\mathbb{R}}$ such that $Q|_{W_z}$ is positive definite for $z \in \widetilde{U}$.} Applying the map $J_2$ fibrewise we obtain a family $\mathcal{J}_2^{\Delta_{\mathbb{R}}} := J_2(\mathcal{Y}_U,\mathcal{V}_U)$ of polarised holomorphic intermediate Jacobians over the open subset $U \subset \Delta$ that holomorphically extends the family $\mathcal{J}_2^W|_{\Delta_{\mathbb{R}}}$ of Weil intermediate Jacobians restricted to $\Delta_\mathbb{R}$. The merit of this construction is that we can apply tools of complex analysis to the holomorphic family $\mathcal{J}_2^{\Delta_\mathbb{R}}$, in order to compute physical quantities intrinsic to the submanifold $\Delta_\mathbb{R}$. We denote the Jacobians by $J^{\Delta_\mathbb{R}}_2(Y_z)$ for brevity, and the corresponding family of matrices $\bm{M}$ is denoted by $\bm H(z)$. We summarise the relations between the Jacobians, vector bundles, and matrices, as well as some of their properties in table~\ref{tab:Jacobians}.

\begin{table}
	\begin{center}
		\renewcommand{\arraystretch}{1.4}
		\begin{tabular}{c|cccccc}
			Jacobian & Notation & Matrix & Vector bundle & Defined on & in $\mathcal{H}_{m+1}$ & Holomorphic  \\
			\hline
			Griffiths & $J^G_2(Y_z)$ & $\bm{F}$ & $\mathcal{G}_{\mathcal{M}_{\text{cs}}}$ & $\mathcal{M}_\text{cs}$ & \xmark & \cmark \\
			Weil & $J^W_2(Y_z)$ & $\bm{N}$ & $\mathcal{W}_{\mathcal{M}_{\text{cs}}}$ & $\mathcal{M}_\text{cs}$& \cmark & \xmark \\
			Holomorphic & $J^{\Delta_{\mathbb{R}}}_2(Y_z)$ & $\bm{H}$ & $\mathcal{V}_{U}$ & $U \supset \Delta_{\mathbb{R}}$ & \cmark & \cmark			
		\end{tabular}
	\end{center}
	\caption{Summary and comparison of the Griffiths, Weil, and polarised holomorphic intermediate Jacobians.}
	\label{tab:Jacobians}
\end{table}

\subsection{The complex structure moduli space of stable \texorpdfstring{genus-$g$}{genus-g} curves} \label{sect:genus-g_curves}
In this subsection, we review some properties of stable genus-$g$ curves. For more details on genus-$g$ curves, see for instance refs.~\cite{MR1139765,MR1288523}. In algebraic geometry, an algebraic curve $C$ of (arithmetic) genus $g$ is a Riemann surface with a complex structure that is almost everywhere smooth, except for possibly a finite number of double points. Such a curve is called stable if its automorphism group is finite, which is automatically true if the (arithmetic) genus $g$ of the curve is bigger than one.\footnote{For $g=0$ or $g=1$ an algebraic curve becomes stable if we include three marked points or one marked point, respectively. We do not consider marked points of algebraic curves in this work, as the focus is on stable curves of genus $g\ge 2$ and in particular on stable curves of genus $2$.} 

The complex structure moduli space of curves of (arithmetic) genus $g$ is denoted by $\overline{\mathcal{M}}_g$. It has the complex dimension
\begin{equation}
	\dim_\mathbb{C} \overline{\mathcal{M}}_g = 3g -3 \ , \qquad g\ge 2 \ .
\end{equation}
As for families of Calabi--Yau threefolds, we focus in this work on the local structure of the complex structure moduli space $\overline{\mathcal{M}}_g$. To this end, we consider a family of genus-$g$ curves $\mathcal{C}_g$ over a $(3g-3)$-dimensional complex disc $\Delta \subset \overline{\mathcal{M}}_g$, i.e. 
\begin{equation}
	\begin{CD}
		C_{g,z} @>>> \mathcal{C}_g \\
		@. @V\pi VV \\
		@. \Delta
	\end{CD}
\end{equation}
where $\pi$ projects a member $C_{g,z}$ of the family $\mathcal{C}_g$ to its complex structure modulus $z\in\Delta$. For a stable curve $C_{g,z}$, we associate the Hodge filtration
\begin{equation}
	H^{(1,0)}(C_{g,z}) \simeq F^1H^1_\mathbb{C} \subset F^0H^1_\mathbb{C} \simeq H^1(C_{g,z},\mathbb{C}) \ ,
\end{equation}
which varies holomorphically over the disc $\Delta$. Note that
\begin{equation}
	\dim_\mathbb{C}  H^{(1,0)}(C_{g,z})= \dim_\mathbb{C} F^1H^1_\mathbb{C} = g \ , \qquad 
	\dim_\mathbb{C}  H^1(C_{g,z},\mathbb{C}) = \dim_\mathbb{C} F^0 H^1_\mathbb{C} = 2g \ .
\end{equation}  
The Hodge filtration allows to define the first Griffiths intermediate Jacobian $J_1(C_{g,z})$ as
\begin{equation} \label{eq:J1C}
	J_1(C_{g,z}) \simeq H^1(C_{g,z},\mathbb{C})/ ( F^1H^1_\mathbb{C} \oplus H^1(C_{g,z},\mathbb{Z})) 
	\simeq \mathbb{C}^{g}/(\mathbb{Z}^g +  \mathbb{Z}^g \bm{\tau}) \ .
\end{equation}  
We define the Hermitian intersection pairing $Q$ by
\begin{equation}
	Q: H^1(C_{g,z},\mathbb{C}) \times H^1(C_{g,z},\mathbb{C}) \to \mathbb{C}, \ (\theta,\chi) \mapsto i \int \theta\wedge \overline\chi  \ .
\end{equation}
For any stable smooth curve $C_{g,z}$ the restriction of $Q$ to $F^1H^1_\mathbb{C}$ is positive definite.
Hence, the Griffiths intermediate Jacobian $J_1(C_{g,z})$ of such a curve is an Abelian variety 
described by a modular matrix $\bm{\tau}$ in the Siegel upper half-space $\mathcal{H}_g$ defined in eq.~\eqref{eq:SiegelHalf}. 
This implies that the Griffiths intermediate Jacobian coincides with the Weil intermediate Jacobian, 
and we often refer to both of them simply as the intermediate Jacobian $J_1(C_{g,z})$ 
of the curve $C_{g,z}$.\footnote{The first Griffiths intermediate Jacobian of a variety (and hence also for a curve $C_{g,z}$) is also called the Picard variety in the literature, see for instance ref.~\cite{MR1288523}.} 

While for any stable smooth curve $C_{g,z}$ the intermediate Jacobian $J_1(C_{g,z})$ is a non-degenerate Abelian variety, 
the intermediate Jacobian $J_1(C_{g,z})$ of a stable but singular curve may or may not degenerate. 
Therefore, we consider in the following the submoduli space of stable curves $\overline{\mathcal{M}}_g^\mathcal{A}$ with a non-degenerate intermediate Jacobian $J_1(C_{g,z})$, i.e.,
\begin{equation} \label{eq:sMA}
  \overline{\mathcal{M}}_g^\mathcal{A} = \left\{ C_{g,z} \in \overline{\mathcal{M}}_g \, \middle| \ J_1(C_{g,z}) \text{ is a non-degenerate Abelian variety} \ \right\}
  \subset \overline{\mathcal{M}}_g \ ,
\end{equation}
and the disc $\Delta$ for the families of curves $\mathcal{C}_g$ is taken in the following to lie in the moduli space $\overline{\mathcal{M}}_g^\mathcal{A}$. Then over such a disc $\Delta\subset \overline{\mathcal{M}}_g^\mathcal{A}$the intermediate Jacobians $J_1(C_{g,z})$ furnish by construction a family of Abelian varieties that vary holomorphically over $\Delta$. Moreover, Torelli's theorem asserts that the Abelian variety that arises from the intermediate Jacobian of a genus-$g$ curve, unambiguously encodes the complex structure of the genus-$g$ curve. In other words, the map $C_{g,z} \mapsto J_1(C_{g,z})$ is injective.

The defining matrix $\bm{\tau}$ of the intermediate Jacobian can explicitly be computed from the periods of the curve $C_{g,z}$. Let us introduce a symplectic basis $(b^j,a_i)$, $i,j=0,\dots,g-1$, of homology one-cycles in $H_1(C_{g,z},\mathbb{Z})$ with the canonical oriented intersections 
\begin{equation} \label{eq:symplectic_homology_basis_C}
	a^i \cap b_j = \delta^i_j \ , \qquad a^i \cap a^j = b_i \cap b_j = 0 \ , \qquad i,j = 0,\ldots,g-1 \ ,
\end{equation}  
together with a basis $\omega_n$, $n=0,\ldots, g-1$, of holomorphic $(1,0)$-forms generating the Dolbeault cohomology group $H^{(1,0)}(C_{g,z})$. Then the $2g\times g$ period matrix $\bm{P}$ of the curve $C_{g,z}$ is defined as
\begin{equation} \label{eq:genus-g_period_matrix_definition}
	\bm{P} = \begin{pmatrix}
	\mathbfcal{T} \\ \mathbfcal{X}
	\end{pmatrix}    \ ,
\end{equation}
in terms of the $g\times g$ matrices of A-periods and B-periods $\mathbfcal{X}=({\mathcal{X}^j}_n)$ and $\mathbfcal{T}=(\mathcal{T}_{jn})$ given by
\begin{equation}
	{\mathcal{X}^j}_n =  \int_{a^j} \omega_n \ , \qquad
	\mathcal{T}_{jn} =  \int_{b_j} \omega_n \ , \qquad n = 0, \dots, g-1~.
\end{equation}
Carrying out the same steps as for the derivation of the intermediate Jacobians for Calabi--Yau threefolds, we arrive at the following expression for the $g\times g$ matrix $\bm{\tau}$ in eq.~\eqref{eq:J1C}:
\begin{equation} \label{eq:period_matrix_tau_definition}
	\bm{\tau} = \mathbfcal{T} \mathbfcal{X}^{-1} \ .
\end{equation}
Note that the matrices $\bm{X}$ and $\mathbfcal{T}$ depend on the chosen basis elements $\omega_n$, $n=0,\ldots,g-1$, and on the choice of symplectic basis $(b_j, a^i)$, whereas the matrix $\bm{\tau}$ depends only on the symplectic basis $(b_j, a^i)$. With respect to a symplectic transformation $\operatorname{Sp}(2g,\mathbb{Z})$ acting on the basis $(b_j, a^i)$, we can readily deduce that the matrix $\bm{\tau}$ transforms as
\begin{equation} \label{eq:tauaction}
	\bm{\tau} \mapsto (\bm A\, \bm{\tau} + \bm B)(\bm C\, \bm{\tau} + \bm D)^{-1} \quad \text{for} \quad
	\begin{pmatrix}   
		\bm{A} & \bm{B} \\ \bm{C} & \bm{D}
	\end{pmatrix} \in \operatorname{Sp}(2g,\mathbb{Z}) \ .
\end{equation}

Let us close the discussion of stable curves with the Riemann--Schottky problem. To each stable curve $C$ the intermediate Jacobian $J_1(C)$ assigns a unique Abelian variety. That is to say, we have an injective map
\begin{equation} \label{eq:ABJmap}
	J_1:\ \overline{\mathcal{M}}_g^\mathcal{A} \to \mathcal{A}_g, \ C \mapsto J_1(C) \ ,
\end{equation} 
where $\overline{\mathcal{M}}_g^\mathcal{A}$ is the moduli space of stable curves with non-degnerate intermediate Jacobians as defined in eq.~\eqref{eq:sMA}. Moreover, $\mathcal{A}_g$ is the moduli space of Abelian varieties of complex dimension $g$, which is isomorphic to the quotient of $\mathcal{H}_g$ by the action of $\operatorname{Sp}(2g,\mathbb{Z})$:
\begin{equation}
	\mathcal{A}_g  \simeq  \mathcal{H}_g / \operatorname{Sp}(2g,\mathbb{Z}) \ , \qquad \dim_\mathbb{C} \mathcal{A}_g =  \dim_\mathbb{C} \mathcal{H}_g = \frac12 g(g+1) \ ,
\end{equation}  
where the modular group $\operatorname{Sp}(2g,\mathbb{Z})$ acts on a point $\bm{\tau}$ in the Siegel upper half-space by the transformation~\eqref{eq:tauaction}. Note that for genus $g=2$ and $g=3$ the dimensions of the moduli spaces $\overline{\mathcal{M}}_g^\mathcal{A}$ and $\mathcal{A}_g$ are the same. As a matter of fact, for $g=2$ and $g=3$ the moduli spaces $\overline{\mathcal{M}}_g^\mathcal{A}$ and $\mathcal{A}_g$ get identified.\footnote{For $g=1$, the stability condition requires consideration of one-dimensional complex tori with a marked point, i.e. elliptic curves, and their complex one-dimensional moduli space is denoted by $\overline{\mathcal{M}}_{1,1}^\mathcal{A}$. For elliptic curves the intermediate Jacobian map $J_1: \overline{\mathcal{M}}_{1,1}^\mathcal{A} \to \mathcal{A}_1$ is also a bijection.} However, for curves of higher genera $g>3$ the map $J_1$ is no longer surjective. The image of the map $J_1$ in $\mathcal{A}_g$ is known as the Schottky locus $\mathcal{S}_g$. Of particular significance for us is the trivial but useful observation that on $\mathcal{S}_g$ one can define the inverse map
\begin{align} \label{eq:inverse_Jacobian_map}
	J_1^{-1}: \mathcal{S}_g \to \overline{\mathcal{M}}_g^\mathcal{A}  \ .
\end{align}
Identifying the non-trivial Schottky locus for stable curves of genus $g>3$ is known as the Riemann--Schottky problem. For a survey article on this interesting subject, see for instance ref.~\cite{MR2931868}.

\subsection{The real analytic Calabi--Yau-to-curve correspondence} \label{sect:an_corr}

To construct the real analytic Calabi--Yau-to-curve correspondence, let us consider a family of Calabi--Yau threefolds $\mathcal{Y}_\Delta$ parametrised by the (local) $m$-dimensional complex structure moduli space~$\Delta$. As discussed in section~\ref{sec:WeilJac}, the Weil intermediate Jacobian $J^W_2$ assigns to each point in the complex structure moduli space an Abelian variety of complex dimension $m+1$, which defines the map\footnote{By a slight abuse of notation we denote this map by the same symbol $J_2^W$ as the Weil intermediate Jacobian --- and use analogous notation for other Jacobians as well.}
\begin{equation} 
	J^W_2: \Delta \to \mathcal{A}_{m+1} \ .
\end{equation} 
The Abelian variety in the image is explicitly represented in terms of the lattice~\eqref{eq:DefJ2Wtwo} with the matrix $\bm{N}$ of eq.~\eqref{eq:DefN}, which --- cf., section~\ref{sec:GvsWJac} --- demonstrates that, for $m>0$, the map $J^W_2$ is not holomorphic but only real analytic. Since the map $J^W_2$ is injective, the image of $J^W_2$ defines in the moduli space of Abelian varieties $\mathcal{A}_{m+1}$ a real analytic subspace $\mathcal{R}_{\Delta}$ of real dimension $2m$. We call the subspace $\mathcal{R}_\Delta$ the real analytic Calabi--Yau--Schottky locus of the family of Calabi--Yau threefolds $\mathcal{Y}_\Delta$. Furthermore, intersecting the Calabi--Yau--Schottky locus $\mathcal{R}_\Delta$ with the Schottky locus $\mathcal{S}_{m+1}$ of genus $m+1$ curves, we call the pre-image of this intersection with respect to the map $J^W_2$ the curve locus $\Delta_\mathcal{C}$ in the complex structure moduli space $\Delta$. Hence, restricting to this curve locus, we obtain the injective real analytic map
\begin{equation} \label{eq:J1_inverse}
	J^W_2: \Delta_\mathcal{C} \to \mathcal{S}_{m+1} \ ,
\end{equation}
which induces --- using Torelli's theorem --- a real analytic map $\Phi = J_1^{-1} \circ J_2^W$ from $\Delta_{\mathcal{C}}$ into the moduli space of genus $m+1$ stable curves
\begin{equation} \label{eq:RACtoCY}
	\Phi: \ \Delta_\mathcal{C} \to \overline{\mathcal{M}}_{m+1}^\mathcal{A} \ .
\end{equation}
We refer to this map as the real analytic Calabi--Yau-to-curve correspondence, as it assigns to a Calabi--Yau threefold $Y_z$ in the family $\mathcal{Y}_\Delta$ on the curve locus $\Delta_\mathcal{C} \subset \Delta$ a stable curve of genus $m+1$.

In a similar way, we can construct the inverse to the Calabi--Yau-to-curve correspondence. On the intersection $\mathcal{R}_\Delta \cap \mathcal{S}_{m+1}$ of the Calabi--Yau--Schottky locus $\mathcal{R}_\Delta$ and the Schottky locus $\mathcal{S}_{m+1}$ in the moduli space of Abelian varieties $\mathcal{A}_{m+1}$, the inverse of the map \eqref{eq:ABJmap} is defined and we refer to the pre-image of the intersection $\mathcal{R}_\Delta \cap \mathcal{S}_{m+1}$ as the Calabi--Yau locus $\overline{\mathcal{M}}_{m+1, \mathcal{Y}_{\Delta}}$, so that we arrive at the inverse real analytic Calabi--Yau-to-curve correspondence
\begin{equation} \label{eq:InverseRACtoCY} 
	\Phi^{-1}: \ \overline{\mathcal{M}}_{m+1, \mathcal{Y}_{\Delta}} \to \Delta \ .
\end{equation}

Note that, depending on the details of the geometric setup and the choice of the local family of Calabi--Yau threefolds $\mathcal{Y}_\Delta$, the curve locus $\Delta_\mathcal{C}$ arising from the described intersection may be empty. Then the constructed injective real analytic map $J_2^W$ and the resulting real analytic Calabi--Yau-to-curve correspondence $\Phi$ becomes empty. It would be interesting to study non-trivial examples of this type explicitly for Calabi--Yau families with $m\ge3$, which is, however, beyond the scope of this work. 

\begin{figure}
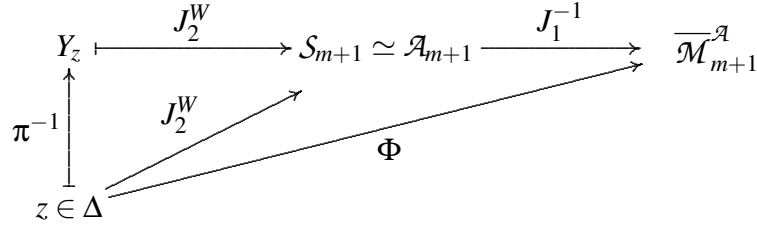

	\begin{align} \notag
		\begindc{\commdiag}[300]
		\obj(-1,1)[U]{$z \in \Delta$}
		\obj(-1,3)[Yz]{$Y_z$}
		\obj(3,3)[calS]{$\mathcal{S}_{m+1} \simeq \mathcal{A}_{m+1}$}
		\obj(7,3)[calM]{$\hskip10pt\overline{\mathcal{M}}_{m+1}^\mathcal{A}$}
		\mor{U}{calS}{$J_2^W$}[\atleft,\solidarrow]
		\mor{U}{Yz}{$\pi^{-1}$}[\atleft,\aplicationarrow]
		\mor{Yz}{calS}{$J_2^W$}[\atleft,\aplicationarrow]
		\mor{U}{calM}{$\Phi$}[\atright,\solidarrow]		
		\mor{calS}{calM}{$J_1^{-1}$}[\atleft,\solidarrow]	
		\enddc
	\end{align}
	\caption{A diagram summarising, for $m=1,2$ the construction of the map $\Phi$ realising the Calabi--Yau-to-curve correspondence. The map $\pi^{-1}$ gives a Calabi--Yau manifold $Y_z$ given a point $z$ in the local moduli space $\Delta$. Here $J_2^W$ denotes both the map that, given a Calabi--Yau manifold $Y_z$, gives the Weil intermediate Jacobian $J_2^W(Y_z)$, but also the composition $J_2^W \circ \pi^{-1} : \Delta \to \mathcal{S}_{m+1}$. $J_1$ denotes the map that maps a stable curve of genus-$(m{+}1)$ to its intermediate Jacobian.}
	\label{fig:Map_Phi}
\end{figure}

For Calabi--Yau families with one-dimensional or two-dimensional complex structure moduli spaces the associated curves are of genus two or three, respectively. As discussed in section~\ref{sect:genus-g_curves} for genus two and three the Riemann--Schottky problem is trivial in the sense that the Schottky locus $\mathcal{S}_g$ coincides with the moduli space of Abelian varieties $\mathcal{A}_g$ for $g=2,3$. Then the curve locus $\Delta_\mathcal{C}$ becomes the entire moduli space $\Delta$, and the real analytic Calabi--Yau-to-curve correspondence~\eqref{eq:RACtoCY} is realised on the entire moduli space $\Delta$, i.e. 
\begin{equation} \label{eq:RACtoCYlow}
	\Phi: \Delta \to \overline{\mathcal{M}}_{m+1}^\mathcal{A} \quad \text{for} \quad m=1,2 \ .
\end{equation}
We summarise the definition of the above map in figure~\ref{fig:Map_Phi}. For the inverse real analytic Calabi--Yau-to-curve correspondence~\eqref{eq:InverseRACtoCY} such a simplification does not occur in any dimension, because the dimension of the Calabi--Yau--Schottky locus $\mathcal{R}_\Delta$ of an $m$-dimensional family $\mathcal{Y}_{\Delta}$ of Calabi--Yau threefolds is always smaller than the dimension of the associated moduli space of Abelian varieties $\mathcal{A}_{m+1}$.

\subsection{The local holomorphic Calabi--Yau-to-curve correspondence} \label{sect:hol_corr}
Let us consider for a family of Calabi--Yau threefolds $\mathcal{Y}_\Delta$ over the local moduli space $\Delta$ a (local) Lagrangian submanifold $\Delta_\mathbb{R}$ with the associated Lagrangian family of Calabi--Yau threefolds~$\mathcal{Y}_{\Delta_\mathbb{R}}$. In the following, we restrict to one- and two-dimensional families of Calabi--Yau threefolds, i.e.
\begin{equation}
	\dim_\mathbb{C} \Delta = \dim_\mathbb{R} \Delta_\mathbb{R} = m \quad \text{with} \quad m\in\{ 1,2 \} \ ,
\end{equation}
In these dimensions the real analytic Calabi--Yau-to-curve correspondence is described by the map~\eqref{eq:RACtoCYlow}, which restricted to the Lagrangian sublocus $\Delta_\mathbb{R}$ becomes
\begin{equation} \label{eq:CYtoCurveLag}
	\Phi_{\mathbb{R} }:\ \Delta_\mathbb{R} \to {\overline{\mathcal{M}}_{m+1}^\mathcal{A} }\quad \text{for} \quad m=1,2 \ .
\end{equation}  
This restricted correspondence can be holomorphically extended using the results of section \ref{sect:polholJac}. As discussed there, given a family $\mathcal{J}^W_2$ of Weil intermediate Jacobians and a Lagrangian sublocus $\Delta_{\mathbb{R}}$, we are able to obtain a family $\mathcal{J}_2^{\Delta_{\mathbb{R}}}$ of polarised holomorphic intermediate Jacobians on an open subset $U\subset \Delta$ such that $\Delta_\mathbb{R} \subset U$. The benefit of considering the family $\mathcal{J}_2^{\Delta_{\mathbb{R}}}$ is that it is based on a holomorphic vector bundle $\mathcal{V}$ as opposed to $\mathcal{J}^W_2$ which is only based on a complex vector bundle $\mathcal{W}$. The family $\mathcal{J}_2^{\Delta_{\mathbb{R}}}$ is constructed so that its members are Abelian varieties, and for $g=2,3$, that is for $m=1,2$, the moduli space $\mathcal{A}_g$ of Abelian varieties gets identified with the Schottky locus $\mathcal{S}_g$. Therefore we can apply the map $J_1^{-1}$ \eqref{eq:J1_inverse} fibrewise to obtain a family of genus-$(m+1)$ curves. In this way, the family $\mathcal{J}_2^{\Delta_\mathbb{R}}$ of intermediate Jacobians extends the Calabi--Yau-to-curve correspondence~\eqref{eq:CYtoCurveLag} as well, namely we obtain a map
\begin{equation}
	\Phi_{U}^{\Delta_\mathbb{R}}:\ U \to {\overline{\mathcal{M}}_{m+1}^\mathcal{A}} \quad \text{for} \quad m=1,2 \ ,
\end{equation} 
where the superscript $\Delta_\mathbb{R}$ is to remind us that this construction depends on the Lagrangian sublocus $\Delta_\mathbb{R}$ to which the vector bundle $\mathcal{W}$ was restricted. We refer to this as the local holomorphic Calabi--Yau-to-curve correspondence between the local moduli space $U$ of the family of Calabi--Yau threefolds and the associated family of genus $m+1$ curves in the image of~$\Phi_{U}^{\Delta_\mathbb{R}}$.

A few comments in order: To construct the correspondence explicitly, it is useful to recall that given a symplectic integral homology basis $(B_J, A^I)$ of $H_3(X,\mathbb{Z})$ and its dual basis $(\alpha_I,\beta^J)$ of $H^3(X,\mathbb{Z})$, the Weil intermediate Jacobian $J_2^W(Y_z)$ \eqref{eq:DefJ2Wtwo} can be represented in terms of a matrix $\bm{N}(z)$, which depends on the point $z \in \Delta$, defined in eq.~\eqref{eq:DefN}. Let $x$ denote a particular choice of coordinates on $\Delta_{\mathbb{R}}$, then, restricting $\bm{N}(z)$ to $\Delta_{\mathbb{R}}$, we obtain a real analytic family $\bm{N}(x)|_{\Delta_{\mathbb{R}}}$ of matrices in $\mathcal{H}_{m+1}$. Then these are holomorphically continued to a holomorphic family of matrices $\bm H(w)$ on a tubular neighbourhood $U$ of $\Delta_{\mathbb{R}}$, by formally substituting $x \to w$ and interpreting $w$ as the holomorphic coordinate on $U$. 

To explicitly verify that the analytic continuation is well-defined, we need to ensure that the above construction is independent of the choice of a symplectic basis $(B_J, A^I)$ of $H_3(Y_z,\mathbb{Z})$: based on earlier discussions, we know that under a symplectic change of basis \eqref{eq:SympTrans}, the matrix $\bm{N}(z)$ transforms according to eq.~\eqref{eq:transN}. This implies, in particular, that on $\Delta_{\mathbb{R}}$ also the matrix $\bm{N}(x)$ transforms as
\begin{align}
	(-\bm{N}(x)) \mapsto (\bm{A}\, (-\bm{N}(x)) + \bm{B})(\bm{C}\, (-\bm{N}(x)) + \bm{D})^{-1} =: \bm \Gamma(-\bm{N}(x))\ .
\end{align}
We can analytically continue $\bm \Gamma(-\bm{N}(x))$ simply by substituting the holomorphic continuation $\bm{H}(w)$ of $\bm{N}(x)$ for $\bm{N}(x)$ in the formula above. Since formal power series form a field, this is certainly holomorphic in a suitable neighbourhood $U$ of $\Delta_{\mathbb{R}}$. Further, this agrees with $\bm \Gamma(-\bm{N}(x))$ on $\Delta_{\mathbb{R}}$. Therefore, this is the unique analytic continuation of $\bm \Gamma(-\bm{N}(x))$. In other words, we have verified that the following diagram commutes:
\begin{equation}
	\begin{CD}
		-\bm{N}(x)|_{\Delta_{\mathbb{R}}} @>\text{hol.}>\text{cont.}> -\bm{H}(z) \\
		@V\bm \Gamma VV @VV \bm \Gamma V \\
		\bm \Gamma(-\bm{N}(x)|_{\Delta_{\mathbb{R}}}) @>\text{hol.}>\text{cont.}> \bm \Gamma(-\bm{H}(z))
	\end{CD}
\end{equation}
Recall that for any matrix $\bm{N}$, $\bm{N}$ and $\bm{\Gamma}(- \bm{N}(x))$ define the same Jacobian $J_2(Y_z,V_z)$ differing only by a modular transformation corresponding to a symplectic change of the basis $(b_j, a^i)$ of homology one-cycles of $J_2(Y_z,V_z)$. Then the above argument shows that the analytic continuation gives a well-defined modular Calabi--Yau-to-curve correspondence $\Phi_{U}^{\Delta_{\mathbb{R}}}$ that depends only on the chosen Lagrangian subspace $\Delta_{\mathbb{R}}$. 


\section{From the Banana Integral to a Calabi--Yau Threefold}
\label{sect:calabi_yau}

In this section, we review the salient features of the Calabi--Yau geometries associated to the four-loop banana graph, which we use to construct an interesting instance of the Calabi--Yau-to-curve correspondence discussed above. We begin in subsection~\ref{sect:calabi_yau_threefold} by introducing the Calabi--Yau threefold $Y_z$ whose periods $\Pi_i(z)$ (see eq.~\eqref{eq:integral_periods_definition}) satisfy the Picard--Fuchs differential equation $L^{(0)} \Pi_i(z) = 0$ for the operator $L^{(0)}$ of eq.~\eqref{def_Picard_Fuchs_eps_0} associated to the integral $I_{11111}$. This identification goes back to refs.~\cite{Vanhove:2018mto,Candelas:2019llw}.

In addition to the manifold $Y_z$, we will need to study its mirror manifold $Y^{\text{mirror}}_\tau$, associated to $Y_z$ by a correspondence known as mirror symmetry (for a detailed review, see for example \cite{Hori:2003ic,Cox1999a}). In the string theory context, mirror symmetry may be understood as a conjecture that for every Calabi--Yau manifold $X$, there exists a mirror manifold $X^{\text{mirror}}$ such that the type IIA string theory compactified on $X$ is dual to the IIB string theory compactified on $X^{\text{mirror}}$. This duality gives a correspondence between the complex structure moduli space $\mathcal{M}_{\text{cs}}(Y_z)$ of $Y_z$, with the complexified quantum-corrected K\"ahler structure moduli space $\mathcal{M}_{cK}(Y^{\text{mirror}}_\tau)$ of its mirror. In particular one can obtain a map between the complex structure parameter $z \in \mathcal{M}_{\text{cs}}(Y_z)$ and the complexified K\"ahler parameter $\tau \in \mathcal{M}_{cK}(Y^{\text{mirror}}_\tau)$.  In subsection~\ref{sect:mirror_map} we construct this mirror map explicitly.

For us, the main importance is the fact that mirror symmetry relates the middle cohomology $H^3(Y_z,\mathbb{C})$ with the even cohomology $H^{2*}(Y^{\text{mirror}}_\tau,\mathbb{C})$ of its mirror. Since it is easy to construct an integral basis of the even cohomology $H^{2*}(Y^{\text{mirror}}_\tau,\mathbb{Z})$, we can use mirror symmetry to find a basis of integral periods for $Y$ in terms of the Frobenius basis of periods (which are the solutions already discussed in subsection~\ref{sect:Frobenius}). The integral periods are, in turn, needed to construct the Griffiths and Weil intermediate Jacobians. In subsection~\ref{sect:topological_data} we collect topological data of these manifolds with the help of which we work out this basis of integral periods explicitly in subsection~\ref{sect:integral_basis}.


\subsection{The Hulek--Verrill Calabi--Yau threefold and its mirror}
\label{sect:calabi_yau_threefold}

To find the Calabi--Yau threefold $Y_z$ associated to the equal-mass four-loop banana integral, it turns out to be natural to start with the geometry associated to the unequal-mass banana with pairwise distinct parameters $z_i = m_i^2/p^2$ for $i=1,\dots,5$. 
Geometrically, the parameters $z_i$ furnish coordinates on the complex structure moduli space $\mathcal{M}_{cs}$ of the associated manifold.

To start, we consider the variety defined in ${\mathbb C}{\mathbb P}^4$ by the vanishing of the second graph polynomial ${\mathcal F}$:
\begin{equation}
	Y^{\mathrm{sing}} =
	\left\{ \left[a_1:a_2:a_3:a_4:a_5\right] \in {\mathbb C}{\mathbb P}^4 \; | \; {\mathcal F}\left(a\right)=0 \right\}~,
\end{equation}
where, in the unequal-mass case,
\begin{equation}
	\mathcal{F} =  
	a_1 a_2 a_3 a_4 a_5 
	\left[
	\left( a_1+a_2+a_3+a_4+a_5\right) \left( \frac{z_1}{a_1} + \frac{z_2}{a_2} + \frac{z_3}{a_3} + \frac{z_4}{a_4} + \frac{z_5}{a_5} \right) - 1
	\right].
\end{equation}
Note that in the equal-mass case $z_1=z_2=\dots=z_5:=z$ this reduces to $\mathcal{F}(a)$ appearing in eq.~\eqref{def_U_F}.

This construction of $Y^{\text{sing}}$ gives a family of singular quintic Calabi--Yau varieties parametrised by $(z_1,\dots,z_5)$. However, we wish to find a family of smooth compact Calabi--Yau manifolds, so we consider a series of birational transformations. First, we take the intersection $\widehat Y := Y^{\text{sing}} \cap \mathbb{T}^4$ of the singular variety $Y^{\mathrm{sing}}$ with the projective torus 
${\mathbb T}^4={\mathbb C}{\mathbb P}^4 \backslash \left\{ a_1 \cdot a_2 \cdot a_3 \cdot a_4 \cdot a_5 = 0 \right\}$. The resulting family of manifold is given as the vanishing locus of (see eq.~(\ref{def_U_F}))
\bq \label{eq:non-compact_HV}
\left( a_1+a_2+a_3+a_4+a_5\right) \left( \frac{z_1}{a_1} + \frac{z_2}{a_2} + \frac{z_3}{a_3} + \frac{z_4}{a_4} + \frac{z_5}{a_5} \right) 
& = & 1
\eq
and is smooth outside of the conifold locus \cite{Hulek:2005,Candelas:2021lkc}
\begin{equation}
	\mathcal{D} = \prod_{\epsilon_i = \pm 1}\left( 1 + \epsilon_1 \sqrt{z_1} + \epsilon_2 \sqrt{z_2} + \epsilon_3 \sqrt{z_3} + \epsilon_4 \sqrt{z_4} + \epsilon_5 \sqrt{z_5} \right) = 0~.
\end{equation}
This, in turn, is birational to
\begin{align} \label{eq:HV_definition}
	a_1+a_2+a_3+a_4+a_5+a_6
	\; = \; 
	\frac{z_1}{a_1} + \frac{z_2}{a_2} + \frac{z_3}{a_3} + \frac{z_4}{a_4} + \frac{z_5}{a_5} + \frac{1}{a_6}
	\; = \; 0~,
\end{align}
which is easily seen by eliminating $a_6=-a_1-a_2-a_3-a_4-a_5$.\footnote{We introduce this family of geometries birational to $Y^{\text{sing}}$ as it turns out to be easier to obtain their toric compactifications and mirror manifolds than those of \eqref{eq:non-compact_HV} \cite{Candelas:2021lkc}. However, in principle one could directly consider the toric compactification of \cite{Hulek:2005}. The price one has to pay is that these manifolds are singular, and one needs to resolve the singularities to obtain a smooth Calabi--Yau threefold.}

However, the resulting geometry is non-compact, so to obtain compact manifolds, one can use techniques of toric geometry to compactify $\widehat Y$ to obtain a five-parameter family of smooth Calabi--Yau threefold $\text{HV}_{(z_1,\dots,z_5)}$ birational to $\widehat Y$ and thus to $Y^{\text{sing}}$. The Calabi--Yau threefolds $\text{HV}_{(z_1,\dots,z_5)}$ are so-called Hulek--Verrill Calabi--Yau threefolds, first studied in detail by Hulek and Verrill in \cite{Hulek:2005}, after which these manifolds have appeared prominently in physics in both string theory (see for example \cite{Candelas:2019llw,Candelas:2021lkc,Candelas:2021mwz,Candelas:2023yrg}) and scattering amplitudes (e.g. \cite{Vanhove:2018mto,Bonisch:2020qmm,delaCruz:2024xit,Lairez:2022zkj,Pogel:2022vat}) contexts. 

We have obtained a smooth family of Calabi--Yau threefolds, so we are now ready to start making contact with the differential operator $L^{(0)}$. Namely, we can verify that the periods of $\text{HV}$ satisfy the Picard--Fuchs equation $L^{(0)} \Pi_i = 0$ along the one-parameter subfamily $\text{HV}_{(z,\dots,z)}$, corresponding to the equal-mass case $z_1=\cdots=z_5=:z$. To perform this verification, one could use the Griffiths--Dwork method \cite{Dwork:1962,Griffiths:1969} (for a review, see for example \cite{Cox1999a}) to obtain the Picard--Fuchs differential ideal. However, in this case a more practical approach is to directly compute the fundamental period $\Pi_0(z)$ by using the residue map \cite{Cox1999a}:
\begin{equation} \label{eq:HV_fundamental_period}
	\begin{aligned}
		\Pi_{0}(z) &= - \frac{1}{(2\pi i)^5} \oint_0 \frac{d a_1}{a_1} \cdots \oint_0 \frac{d a_5}{a_5} 
		\frac1{\left[(a_1+\ldots+a_5)\left(\frac{z}{a_1} + \ldots + \frac{z}{a_5} \right) - 1 \right]} \\
		&= \sum_{n_1+n_2+n_3+n_4+n_5=n}
		\left( \frac{n!}{n_1! n_2! n_3! n_4! n_5!} \right)^2 z^{n} = \Frobeniusbasis_0(z)~.
	\end{aligned}
\end{equation}
By studying the symmetries of the manifolds $\text{HV}_{(z,\dots,z)}$, it can be shown \cite{Candelas:2021lkc} that in the present case this is enough to guarantee that every period $\Pi_i(z)$ satisfies the equation $L^{(0)} \Pi_i(z) = 0$.  

However, the family $\text{HV}_{(z,\dots,z)}$ is not yet the family uniquely associated to the operator $L^{(0)}$, as we had to impose the condition $z_1=\cdots=z_5=:z$ by hand to obtain periods $\Pi_i(z)$ that satisfy the differential equation $L^{(0)}\Pi(z) = 0$. In addition, as discussed in section \ref{sect:genus-g_curves}, to obtain a genus-$g$ curve from the Calabi--Yau geometry, we wish to restrict to geometries with $g \leq 3$, implying that we need $h^{1,2} \leq 2$. However, by studying the toric compactification carefully \cite{Hulek:2005,Candelas:2021lkc}, one can show that $h^{1,2}(\text{HV}) = 5$.\footnote{This is just a reflection of the fact that the family $\text{HV}_{(z_1,\dots,z_5)}$ has five complex structure parameters~$z_i$.}

To obtain an honest one-parameter family of manifolds whose periods are governed by the Picard--Fuchs differential operator $L^{(0)}$ given in eq.~\eqref{def_Picard_Fuchs_eps_0}, we can consider the $\mathbb{Z}_5$-action that cyclically permutes the coordinates $a_i$. This acts as a symmetry of the manifold $\text{HV}_{(z_1,\dots,z_5)}$ exactly when $z_1 = \cdots = z_5 =: z$, thus automatically imposes the condition we needed to obtain periods annihilated by $L^{(0)}$. In addition, this action is free outside the singular points $z=0,1/25,1/9,1,\infty$ of $L^{(0)}$ where the manifold $\text{HV}_{(z,\dots,z)}$ is also singular. Therefore, for such manifolds, we can take the quotient by the $\mathbb{Z}_5$-action to obtain a family $Y_z := \text{HV}_{(z,\dots,z)}/\mathbb{Z}_5$ of smooth Calabi--Yau threefolds. This family has $h^{1,2}(Y_z)=1$, so the middle cohomology $H^3(Y_z,\mathbb{C})$ is four-dimensional. This corresponds to the order of the differential operator $L^{(0)}$ discussed in section~\ref{sect:Picard_Fuchs}
and to the number of master integrals in the top sector of the four-loop equal-mass banana family in section~\ref{sect:Feynman_integral_definition}.

It can be easily shown that the periods of $Y_z$ are, up to combinatorial constants, exactly the periods of the family $\text{HV}_{(z,\dots,z)}$, and therefore the computation in eq.~\eqref{eq:HV_fundamental_period} is enough to show that they also satisfy the differential equation $L^{(0)} \Pi_i(z) = 0$~\cite{Candelas:2021lkc}.\footnote{There is also a freely-acting $\mathbb{Z}_2$ corresponding to $a_i \mapsto 1/a_i$. Taking a quotient with respect to $\mathbb{Z}_5\times\mathbb{Z}_2$ gives also a one-parameter family of smooth Calabi--Yau threefolds whose periods are governed by $L^{(0)}$. The only tangible difference between the two quotients is some of the topological data such as the triple intersection numbers and the second Chern class. In this paper we will concentrate for simplicity only on the $\mathbb{Z}_5$-quotient. However, all of the discussion applies for the $\mathbb{Z}_5 \times \mathbb{Z}_2$-quotient \textit{mutatis mutandis}. A detailed discussion on the quotients and their geometry can be found in \cite{Candelas:2021lkc}.}

There is a standard technique for obtaining the mirror manifold of a hypersurface or a complete intersection in a toric variety \cite{Batyrev:1993oya,Batyrev:1995ca}. We will not describe these techniques here, referring the interested reader to \cite{Candelas:2021lkc} for details, and just note that these can be applied in the present case to identify the mirror manifold $Y^{\text{mirror}}$. The identification proceeds by first constructing the mirror $\text{H}\Lambda_{(\tau_1,\dots,\tau_5)}$ of the full family $\text{HV}_{(z_1,\dots,z_5)}$ of Hulek--Verrill manifolds with five complex structure parameters. These are given by the simple complete intersection Calabi--Yau manifolds\footnote{See for example \cite{Candelas:1987kf} and \cite{Hubsch:1992nu} for accessible introduction and explanation of the notation used here.}
\begin{align}
	\text{H}\Lambda_{(\tau_1,\dots,\tau_5)} = \cicy{\mathbb{P}^1\\\mathbb{P}^1\\\mathbb{P}^1\\\mathbb{P}^1\\\mathbb{P}^1}{1 & 1\\1 & 1\\1 & 1\\ 1 & 1\\ 1 & 1}.
\end{align}
Then the family of mirrors $\mirrorY$ of the one-parameter quotient manifolds $Y_z = \text{HV}_{(z,\dots,z)}/\mathbb{Z}_5$ is given by taking the $\mathbb{Z}_5$-quotient of the complete intersection $\text{H}\Lambda_{(\tau,\dots,\tau)}$ with all K\"ahler parameters $\tau_1 = \cdots = \tau_5 =: \tau$ equal. The complex structure parameters can be chosen so that a $\mathbb{Z}_5$-action which cyclically permutes the coordinates of the projective spaces $\mathbb{P}^1$ acts as a symmetry. Then the quotient with respect to this symmetry gives the family of mirror manifolds $\mirrorY$:
\begin{align}
	\mirrorY = \text{H}\Lambda_{(\tau,\dots,\tau)} / \mathbb{Z}_5~.
\end{align}

\subsection{The mirror map}
\label{sect:mirror_map}
It can be shown \cite{Candelas:1990rm} that the mirror map, which maps the coordinate $z$ on $\mathcal{M}_{\text{cs}}(Y_z)$ to the coordinate $\tau$ on $\mathcal{M}_{cK}(Y^{\text{mirror}}_\tau)$, can be given near the point of maximal unipotent monodromy $z=0$ of $L^{(0)}$ in terms of the holomorphic solution $\Frobeniusbasis_0$ and the single-logarithmic solution $\Frobeniusbasis_1$ by
\bq
\label{mirror_map}
\tau = \frac{\Frobeniusbasis_1}{\Frobeniusbasis_0},
& &
\qbar = e^{2\pi i \tau}.
\eq
We may express $\qbar$ as a power series in $z$:
\bq
\label{examples_qbar_z_expansion}
\qbar
& = & 
z + 8 z^2 + 92 z^3 + 1288 z^4 + 20398 z^5 + 350968 z^6
+ {\mathcal O}\left(z^7\right).
\eq
The map from $z$ to $\qbar$ can also be inverted, yielding $z$ as a power series in $\qbar$:
\bq
\label{examples_z_qbar_expansion}
z
& = &
\qbar - 8 \qbar^2 + 36 \qbar^3 - 168 \qbar^4 + 514 \qbar^5 - 2760 \qbar^6 
+ {\mathcal O}\left(\qbar^7\right).
\eq
The differential operator $L^{(0)}$ is a Calabi--Yau operator in the sense of \cite{2013arXiv1304.5434B} and has one non-trivial $\Yinvariant$-invariant, given by
\bq
\label{def_Y_invariant}
\Yinvariant_2 & = & 
\frac{d^2}{d\tau^2} \frac{\Frobeniusbasis_2}{\Frobeniusbasis_0}.
\eq
We may write $\Yinvariant_2$ in the form
\bq
\label{def_coeff_n_k}
\Yinvariant_2
& = & 
\frac{1}{24} \left( \qbar \frac{d}{d\qbar} \right)^3  
\left[ 4 \ln^3\qbar + \sum\limits_{k=1}^\infty n_k \mathrm{Li}_3\left(\qbar^k\right) \right].
\eq
The $n_k$ are integral invariants (so-called instanton numbers \cite{Candelas:1990rm} or Gopakumar--Vafa invariants \cite{Gopakumar:1998ii,Gopakumar:1998jq}) that have an enumerative interpretation as, roughly speaking, counting rational curves of degree $k$ on the mirror manifold $Y^{\text{mirror}}_\tau$. 
\begin{table}
	\begin{center}
		\begin{tabular}{c|rrrrrrrr}
			$k$ & $1$ & $2$ & $3$ & $4$ & $5$ & $6$ & $7$ & $8$ \\
			\hline
			$n_k$ & $24$ & $48$ & $224$ & $1248$ & $8400$ & $62816$ & $516336$ & $4539696$ \\
		\end{tabular}
	\end{center}
	\caption{
		The first few instanton numbers $n_k$. A more extensive table can be found in ref.~\cite{Candelas:2019llw}.
	}
	\label{table_n_k}
\end{table}
The first few entries of this sequence are recalled in table~\ref{table_n_k}.


\subsection{Topological data}
\label{sect:topological_data}

We will need some of the topological data of the manifold $Y_z$ and its mirror $\mirrorY$.
This information can be easily obtained by using the description of $\mirrorY$ as a quotient of a complete intersection (see ref.~\cite{Candelas:2021lkc} for details).

The integral cohomology $H^2(\mirrorY,\mathbb{Z})$ is one-dimensional, as the complete intersection $\text{H}\Lambda_{(\tau_1,\dots,\tau_5)}$ has five independent K\"ahler classes corresponding to the pull-backs of the K\"ahler classes of the ambient $\mathbb{P}^1$ factors. Taking the $\mathbb{Z}_5$-quotient identifies these K\"ahler classes, leaving us with one generator for $H^2(\mirrorY,{\mathbb Z})$. Let us denote this generator by $\omega^{\textrm{K\"ahler}}$. The triple intersection number of the mirror Calabi--Yau threefold $\mirrorY$ is given by
\bq
\label{eq_kappa}
\kappa & := & 
\int\limits_{\mirrorY} \omega^{\textrm{K\"ahler}} \wedge \omega^{\textrm{K\"ahler}} \wedge \omega^{\textrm{K\"ahler}} 
\; = \; 24~.
\eq
We define $\sigma$ to be
\bq
\sigma
& := &
\left\{ \begin{array}{ll}
	0, & \kappa \; \mbox{even}, \\ 
	\frac{1}{2}, & \kappa \; \mbox{odd}. \\
\end{array} \right.
\eq
For the case at hand we have
\bq
\label{eq_sigma}
\sigma & = & 0~.
\eq
We denote by $c_2(\mirrorY)$ the second Chern class of $\mirrorY$ and by $C_2$ 
the integrated second Chern class of $\mirrorY$. We have
\bq
\label{eq_C2}
C_2 & = &
\int\limits_{\mirrorY} c_2 \wedge \omega^{\textrm{K\"ahler}} 
\; = \;
24~.
\eq
Finally, we denote by $\chi(\mirrorY)$ the Euler characteristic of $\mirrorY$, which is obtained from the third Chern class 
\begin{align} \label{eq_chi}
	\chi(\mirrorY) = \int_{\mirrorY} c_3(\mirrorY) = -16~.
\end{align}
Using the relation of $\chi(\mirrorY)$ to the dimensions $h^{p,q}(\mirrorY)$ of $H^{(p,q)}(\mirrorY,\mathbb{C})$ 
\begin{align}
	\chi(\mirrorY) = \sum_{p,q} (-1)^{p+q} h^{p,q}(\mirrorY) = -16~,
\end{align}
we can solve for the remaining independent Hodge number $h^{1,2}(\mirrorY)=9$. As a consequence of mirror symmetry, $h^{1,2}(Y_z)=h^{1,1}(\mirrorY)$, so we can write down the full Hodge diamonds for both $Y_z$ and $\mirrorY$ in figure~\ref{fig_hodge_numbers}. 
\begin{figure}
	\begin{center}
		\bq
		\begin{array}{ccc}
			\begin{array}{ccccccc}
				& & & 1 & & & \\
				& & 0 & & 0 & & \\
				& 0 & & 9 & & 0 & \\
				1 & & 1 & & 1 & & 1 \\
				& 0 & & 9 & & 0 & \\
				& & 0 & & 0 & & \\
				& & & 1 & & & \\
			\end{array}
			& \hspace*{5mm} &
			\begin{array}{ccccccc}
				& & & 1 & & & \\
				& & 0 & & 0 & & \\
				& 0 & & 1 & & 0 & \\
				1 & & 9 & & 9 & & 1 \\
				& 0 & & 1 & & 0 & \\
				& & 0 & & 0 & & \\
				& & & 1 & & & \\
			\end{array}
			\\
			&& \\
			\mbox{Calabi--Yau manifold $Y$} && \mbox{Mirror manifold $Y^{\mathrm{mirror}}$} \\
		\end{array}
		\nonumber 
		\eq
	\end{center}
	\caption{
		The Hodge numbers $h^{p,q}$ for the Calabi--Yau manifolds $Y$ and $Y^{\mathrm{mirror}}$.
	}
	\label{fig_hodge_numbers}
\end{figure}



\subsection{From the Frobenius basis to the integral basis}
\label{sect:integral_basis}
We will now use the topological data derived in the previous section to obtain the basis of integral periods $\Pi_i(z)$, which is equivalent to finding a basis of $H^3(Y_z,\mathbb{Z})$, which is needed for the construction of the Griffiths and Weil intermediate Jacobians.

The identification of the basis of integral periods begins by noting that, analogously to the case of the complex structure moduli space $\mathcal{M}_{cs}(Y_z)$ of $Y_z$ discussed in section \ref{sect:Griffiths_Jacobian_general}, it is possible to define a prepotential also on the space $\mathcal{M}_{cK}(Y^{\text{mirror}}_\tau)$ of the complexified K\"ahler moduli space of the mirror manifold $Y^{\text{mirror}}_\tau$ \cite{Candelas:1989bb}:
\begin{align}
	G(\tau) = -\frac{1}{6} Y_{111} \tau^3 - \frac{1}{2} Y_{110} \tau^2 - \frac{1}{2} Y_{100}\tau - \frac{1}{6} Y_{000} - \frac{1}{(2\pi i)^3} \sum_{p=0}^\infty n_{p} \, \text{Li}_3(q^{p})~,
\end{align}
where, instead of using the projective coordinates $w^0,w^1$ of $\mathcal{M}_{cK}(\mirrorY)$, we have already written the prepotential in terms of the affine coordinate $\tau$ and set $w^0=1$. The projective coordinates can be restored by setting $\tau=w^1/w^0$ and homogenising. The coefficients of the polynomial part appearing here are given in terms of the topological data
\begin{align}
	\begin{split}
		Y_{111} &= \int_{\mirrorY} \omega^{\text{K\"ahler}} \wedge \omega^{\text{K\"ahler}} \wedge \omega^{\text{K\"ahler}} = \kappa~, \qquad Y_{110} = \sigma \in \left\{0, \frac{1}{2} \right\}~,\\
		Y_{100} &= - \frac{1}{12} \int_{\mirrorY} c_2 \wedge \omega^{\text{K\"ahler}} = -\frac{C_2}{12}~, \qquad Y_{000} = - 3 \chi(\mirrorY) \frac{\zeta(3)}{(2\pi i)^3}.
	\end{split}
\end{align}
The relation between $H^3(Y_z,\mathbb{C})$ and $H^{2*}(\mirrorY,\mathbb{C})$ predicted by mirror symmetry implies that there exists a symplectic integral basis  $(B_J, A^I)$ of $H_3(Y_z,\mathbb{Z})$ (see eq.~\ref{eq:AB_intersections}) such that the periods $X^I(z^0,z^1)$ and $F_J(z^0,z^1)$ (see eq.~\ref{eq:XIFJ}) in this basis are related to the coordinates $w^I$ and the derivatives $G_J(w^0,w^1)=\partial_{w^J}G(w^0,w^1)$ by
\begin{align} \label{eq:mirror_identifications}
	X^I = w^I~, \qquad G_J = F_J~.
\end{align}
In fact, it is easy to see that this relation uniquely fixes such a basis. Starting from the Frobenius basis and taking into account the mirror map \eqref{mirror_map}, we see that we can take
\begin{align}
	w^0 = \Frobeniusbasis^0, \qquad w^1 = \Frobeniusbasis^1~.
\end{align}
Let us denote by $\Frobeniusbasis$ the vector $(\Frobeniusbasis_3,\Frobeniusbasis_2,\Frobeniusbasis_1,\Frobeniusbasis_0)^T$ and
by $\Ip$ the vector $(G_1, G_0, w^1, w^0)^T$.
There exists a matrix $U$ such that
\begin{align} \label{eq:basis_change_Frob_integral}
	\Ip = U \Frobeniusbasis = \Pi~,
\end{align}
when the identification $\tau=\Frobeniusbasis^1/\Frobeniusbasis^0$ is made. Here the last equality follows from the fact that, by the discussion above, the corresponding basis of periods is integral.

It is easy to find the matrix $U$ explicitly by considering the above equation in the limit $z \to 0$. Note that the mirror map is given by
\begin{align}
	\tau = \frac{1}{2\pi i} \log z + \mathcal{O}(z)~,
\end{align}
and in the limit $z \to 0$, corresponding to $\tau \to \infty$, we can ignore the contributions of order $z \log^n z$ in the periods $\Frobeniusbasis^i$, as well as the contributions of order $q$ in the prepotential $G(\tau)$. Then to obtain the matrix $U$ in eq.~\eqref{eq:basis_change_Frob_integral}, we simply need to match the polynomial terms in $G_I(\tau)$ to the logarithmic terms of $\Frobeniusbasis^2$ and $\Frobeniusbasis^3$. A straightforward computation then gives
\bq \label{eq:change-of-basis_U}
U
& = &
\left( \begin{array}{cccc}
	0 & -\kappa & \sigma & \frac{C_2}{24} \\
	\kappa & 0 & \frac{C_2}{24} & \frac{\chi \zeta(3)}{\left(2 \pi i \right)^3} \\
	0 & 0 & 1 & 0 \\
	0 & 0 & 0 & 1 \\
\end{array} \right).
\eq

\subsection{The integral basis for the family \texorpdfstring{$Y_z$}{Y}}
For the case of the family $\mirrorY$, the topological data is given in  eq.~(\ref{eq_kappa}), eq.~(\ref{eq_sigma}), eq.~(\ref{eq_C2}), and eq.~(\ref{eq_chi}):
\begin{align}
	\kappa & = 24~,
	&
	\sigma & = 0~,
	&
	C_2 & = 24~,
	&
	\chi & = -16~.
\end{align}
With these values the matrix $U$ is given by
\bq
U
& = &
\left( \begin{array}{cccc}
	0 & -24 & 0 & 1 \\
	24 & 0 & 1 & - 16 \frac{\zeta(3)}{\left(2 \pi i \right)^3} \\
	0 & 0 & 1 & 0 \\
	0 & 0 & 0 & 1 \\
\end{array} \right).
\eq
The integral basis of periods of $Y_z$ is given explicitly by
\bq
\label{integral_basis_v1}
\Period
& = &
\left( \begin{array}{c}
	\Period_{B_1} \\ 
	\Period_{B_0} \\
	\Period_{A^1} \\
	\Period_{A^0} \\
\end{array} \right)
\; = \; 
\left( \begin{array}{c}
	\Frobeniusbasis_0 -24 \Frobeniusbasis_2 \\
	24 \Frobeniusbasis_3 + \Frobeniusbasis_1 - 16 \frac{\zeta(3)}{\left(2\pi i\right)^3} \Frobeniusbasis_0 \\
	\Frobeniusbasis_1 \\
	\Frobeniusbasis_0 \\	
\end{array} \right).
\eq
In the following it will be convenient to normalise the integral periods such that the last 
normalised integral period equals one. We write
\bq
\normalisedPeriod
& = &
(\normalisedPeriod_{B_1},\normalisedPeriod_{B_0},\normalisedPeriod_{A^1},\normalisedPeriod_{A^0})^T
\; = \; 
\frac{1}{\Frobeniusbasis_0}
(\Period_{B_1},\Period_{B_0},\Period_{A^1},\Period_{A^0})^T.
\eq
For the remainder of the article, due to the relation \eqref{eq:mirror_identifications}, we will use the mirror map implicitly and identify $G(\tau)$ and $F(\tau)$, making no distinction between them. With the help of the prepotential $F(\tau)$ we may express the normalised integral periods as
\bq
\label{integral_basis_v2}
\normalisedPeriod
& = &
\left( \begin{array}{c}
	\normalisedPeriod_{B_1} \\ 
	\normalisedPeriod_{B_0} \\
	\normalisedPeriod_{A^1} \\
	\normalisedPeriod_{A^0} \\
\end{array} \right)
\; = \; 
\left( \begin{array}{c}
	\partial_\tau F \\
	2 F - \tau \partial_\tau F \\
	\tau \\
	1 \\
\end{array} \right).
\eq
The prepotential $F(\tau)$ is given by
\bq
\label{def_prepotential}
F & = &
\frac{1}{2} \left( \normalisedPeriod_{B_0} + \tau \normalisedPeriod_{B_1} \right)
\; = \;
12 \frac{\left(\Frobeniusbasis_0 \Frobeniusbasis_3 - \Frobeniusbasis_1 \Frobeniusbasis_2 \right)}{\Frobeniusbasis_0^2} + \tau - 8 \frac{\zeta(3)}{\left(2\pi i\right)^3}~.
\eq
Its $q$-expansion reads\footnote{Recall that the coefficients $n_k$ appearing are here are the instanton numbers, the first few of which are given in table~\ref{table_n_k}.}
\bq
F
& = &
- 4 \tau^3 + \tau - 8 \frac{\zeta(3)}{\left(2\pi i\right)^3}
- \frac{1}{\left(2\pi i\right)^3} \sum\limits_{k=1}^\infty n_k \; \mathrm{Li}_3\left(\qbar^k\right).
\eq
We also need the first and second derivative of $F$ with respect to $\tau$. One finds
\bq
\partial_\tau F
& = &
- 12 \tau^2
+ 1
- \frac{1}{\left(2\pi i\right)^2} \sum\limits_{k=1}^\infty k n_k \; \mathrm{Li}_2\left(\qbar^k\right),
\nonumber \\
\partial_\tau^2 F
& = &
- 24 \tau
- \frac{1}{\left(2\pi i\right)} \sum\limits_{k=1}^\infty k^2 n_k \; \mathrm{Li}_1\left(\qbar^k\right).
\eq
We remark that the third derivative is related to the $\Yinvariant_2$-invariant defined in eq.~(\ref{def_Y_invariant}):
\bq
\partial_\tau^3 F
& = &
- 24 \Yinvariant_2~.
\eq
The $q$-expansions of $\normalisedPeriod_{B_1}$ and $\normalisedPeriod_{B_0}$ are given by
\bq
\normalisedPeriod_{B_1}
& = &
- 12 \tau^2
+ 1
- \frac{1}{\left(2\pi i\right)^2} \sum\limits_{k=1}^\infty k n_k \; \mathrm{Li}_2\left(\qbar^k\right),
\nonumber \\
\normalisedPeriod_{B_0}
& = &
4 \tau^3
+ \tau 
- 16 \frac{\zeta(3)}{\left(2\pi i\right)^3}
+ \frac{\tau}{\left(2\pi i\right)^2} \sum\limits_{k=1}^\infty k n_k \; \mathrm{Li}_2\left(\qbar^k\right)
- \frac{2}{\left(2\pi i\right)^3} \sum\limits_{k=1}^\infty n_k \; \mathrm{Li}_3\left(\qbar^k\right).
\eq

\section{The Correspondence for the Banana Integral}
\label{sect:jacobian}
In this section we spell out explicitly  the Calabi--Yau-to-curve correspondence for the family of equal-mass four-loop banana integrals~\eqref{def_four_loop_banana_integral}. As discussed in detail in section~\ref{sect:calabi_yau} these integrals relate to the one-parameter family of $\mathbb{Z}_5$ quotients $Y_z$ of Hulek--Verrill Calabi--Yau threefolds $\text{HV}_{(z,\dots,z)}$. Following the general logic presented in section~\ref{sect:Calabi-Yau_to_Curve}, we first compute in this section various intermediate Jacobians for the one-parameter family of Calabi--Yau threefolds $Y_z$, which allow us then to explicitly spell out a Calabi--Yau-to-curve correspondence to a family of genus-two curves.

As this family of Calabi--Yau threefolds depends on a single complex structure modulus, all families of second intermediate Jacobians are complex tori of complex dimension two of the form
\begin{equation}
	J_2(z)
	= 
	{\mathbb C}^2 / \left( {\mathbb Z}^2 +  {\mathbb Z}^2\bm{M}(z)\right) \ ,
\end{equation}
where the complex $2\times 2$-matrix $\bm{M}(z)$ varies with the complex structure modulus $z$ of the one-parameter family of Calabi--Yau threefolds $Y_z$. The precise dependence on the modulus $z$ depends on the type of second intermediate Jacobian to be analysed. 

We start with the calculation of the Griffiths intermediate Jacobian, defined by a symmetric $2 \times 2$-matrix $\bm{F}(z)$ that varies holomorphically with respect to the complex structure parameter $z$. However, for the Griffiths intermediate Jacobian there is no Calabi--Yau-to-curve correspondence as the matrices $\bm{F}(z)$ are not in the Siegel upper half-space $\mathcal{H}_2$. 

Then we introduce the Weil intermediate Jacobian in terms of the symmetric $2{\times}2$-matrix $\bm{N}(z)$, which varies analytically but not holomorphically with respect to the complex structure parameter $z$. The Weil intermediate Jacobian allows us to spell out the analytic Calabi--Yau-to-curve correspondence because the matrices $\bm{N}(z)$ do reside in the Siegel upper half-space $\mathcal{H}_2$. However, for computational purposes this correspondence is not very useful because of the lack of holomorphicity. 

Finally, we introduce the polarised holomorphic intermediate Jacobian that is suitable to describe the equal-mass banana integrals in a specific range of the kinematic variable $z$. This intermediate Jacobian is described by the $2\times 2$-matrix $\bm{H}(z)$ that varies holomorphically in a tubular neighbourhood in the complex structure moduli space of the one-parameter family of Calabi--Yau threefolds $Y_z$ that contains the considered range of the kinematic variable~$z$ as a Lagrangian sublocus. As the symmetric $2\times 2$-matrix $\bm{H}(z)$ varies holomorphically with respect to kinematic variable $z$ and as $\bm{H}(z)$ resides in the Siegel upper half-space $\mathcal{H}_2$, the polarised holomorphic intermediate Jacobian of $\bm{H}(z)$ parametrises a holomorphic family of genus-two curves via the holomorphic Calabi--Yau-to-curve correspondence proposed in section~\ref{sect:hol_corr}.

We describe and analyse the resulting family of curves of genus two in detail and relate them to the original Picard--Fuchs differential equation governing the complex structure moduli space of the one-parameter family of Calabi--Yau threefolds $Y_z$.

In this section we use both the projective prepotential $F(X^1,X^0)$ in terms of the two projective coordinates $X^1$ and $X^0$, which are (a choice of) A-periods of the $\mathbb{Z}_5$-quotient $Y_z$ of the Hulek--Verrill Calabi--Yau threefold, and the affine prepotential $F(\tau)$, which depends on the single affine coordinate $\tau = \frac{X^1}{X^0}$. Since the prepotential $F(X^1,X^0)$ is of homogeneous degree two, these two notions of the prepotential are related as
\begin{equation}
	F\left(X^1,X^0\right) = \left(X^0\right)^2F\left(\tau, 1 \right) =  \left(X^0\right)^2 F(\tau) \ ,
\end{equation}
with
\begin{equation}
	\tau = \frac{X^1}{X^0} \ , \qquad F(\tau) := F(\tau,1) \ .
\end{equation} 
Note that $(\tau,1)$ are the normalised A-periods in eq.~\eqref{integral_basis_v2}, whereas --- as a consequence of eqs.~\eqref{eq:XIFJ} and \eqref{eq:FI} --- the normalised B-periods in eq.~\eqref{integral_basis_v2} are given by 
\bq
\normalisedPeriod_{B_I}(\tau) & = & \left. \frac{\partial F(X^1,X^0)}{\partial X^I} \right|_{\left(X^1,X^0\right)=\left(\tau,1\right)} \ .
\eq

\subsection{The Griffiths intermediate Jacobian of the Calabi--Yau threefold} \label{sect:Griffiths_Jacobian}
According to eq.~\eqref{eq:FIJ} the Griffiths intermediate Jacobian $J_2^G$ is given by
\bq
J^G_2(\tau)
& = &
{\mathbb C}^2 / \left( {\mathbb Z}^2 +  {\mathbb Z}^2 \, \bm{F}(\tau) \right)~,
\eq
where the symmetric $2\times 2$-matrix~$\bm{F}(\tau)$ varies holomorphically with respect to the affine coordinate $\tau$ and it is given by 
\bq \label{eq:F_HV}
\bm{F}(\tau)
= 
\begin{pmatrix}
	F_{11}(\tau) & F_{01}(\tau) \\
	F_{01}(\tau) & F_{00}(\tau) 
\end{pmatrix} \ ,
& &
F_{IJ}(\tau) = \left. \frac{\partial^2 F(X^1,X^0)}{\partial X^I \partial X^J} \right|_{\left(X^1,X^0\right)=\left(\tau,1\right)} \ .
\eq
Explicitly, we compute from the expressions \eqref{eq:DefABPeriods} and \eqref{eq:FI} for the entries of the matrix $\bm{F}(\tau)$
\bq
\label{def_tau_G_explicit}
F_{00}(\tau)
& = &
2 F(\tau) - 2 \tau \partial_\tau F(\tau) + \tau^2 \partial_\tau^2 F(\tau)
\nonumber \\
& = &
- 8 \tau^3 - 16 \frac{\zeta(3)}{\left(2\pi i\right)^3}
- \frac{\tau^2}{\left(2\pi i\right)} \sum\limits_{k=1}^\infty k^2 n_k \; \mathrm{Li}_1\left(\qbar^k\right)
+ \frac{2\tau}{\left(2\pi i\right)^2} \sum\limits_{k=1}^\infty k n_k \; \mathrm{Li}_2\left(\qbar^k\right)
\nonumber \\
& &
- \frac{2}{\left(2\pi i\right)^3} \sum\limits_{k=1}^\infty n_k \; \mathrm{Li}_3\left(\qbar^k\right)\ ,
\nonumber \\
F_{01}(\tau)
& = &
\partial_\tau F(\tau) - \tau \partial_\tau^2 F(\tau)
\nonumber \\
& = &
12 \tau^2
+ 1
+ \frac{\tau}{\left(2\pi i\right)} \sum\limits_{k=1}^\infty k^2 n_k \; \mathrm{Li}_1\left(\qbar^k\right)
- \frac{1}{\left(2\pi i\right)^2} \sum\limits_{k=1}^\infty k n_k \; \mathrm{Li}_2\left(\qbar^k\right),
\nonumber \\
F_{11}(\tau)
& = &
\partial_\tau^2 F(\tau)
\nonumber \\
& = &
- 24 \tau
- \frac{1}{\left(2\pi i\right)} \sum\limits_{k=1}^\infty k^2 n_k \; \mathrm{Li}_1\left(\qbar^k\right)\ ,
\eq
with $\qbar = e^{2\pi i \tau}$.

From the general structure of the second Griffith intermediate Jacobian of one-parameter Calabi--Yau threefolds (c.f., section~\ref{sect:Griffiths_Jacobian_general}), we know that the imaginary matrix $\bm{F}$ has signature $(1,1)$. Let us demonstrate this explicitly for the matrix $\bm{F}(\tau)$ of the one-parameter family of Calabi--Yau threefolds $Y_z$ for a particular range of the kinematic parameter $z$. Namely,  
we consider $z \in ]0,z_{\mathrm{max}}[$ for $z_{\mathrm{max}}>0$ and sufficiently small, then the affine coordinate $\tau$ is purely imaginary and $\tau = i t$ for some $t > 0$. For small $z \in ]0,z_{\mathrm{max}}[$ the parameter $t$ gets large and we find  
\bq
\mathrm{Im}(\bm{F})
& = &
\begin{pmatrix}
	8 t^3 - \frac{2 \zeta(3)}{\pi^3} & 0 \\
	0 & - 24 t 
\end{pmatrix}
+
{\mathcal O}\left(\qbar\right)~.
\eq
Thus, we see explicitly that for $t$ large --- for which the performed approximation is reliable --- the matrix $\mathrm{Im}(\bm{F})$ has indefinite signature $(1,1)$.  Thus, neither $\bm{F}$ nor $-\bm{F}$ reside in the Siegel upper half-space~$\mathcal{H}_2$. Therefore --- as expected from the general discussion in section~\ref{sect:an_corr} --- the computed Griffiths intermediate Jacobians cannot give rise to a Calabi--Yau-to-curve correspondence.

\subsection{The Weil intermediate Jacobian}
\label{sect:Weil_Jacobian}
Next we determine the Weil intermediate Jacobian $J^W_2$ of the $\mathbb{Z}_5$-quotient $Y_z$ of the Hulek--Verrill Calabi--Yau threefold, which is given by
\begin{equation}
	J^W_2(\tau)
	= 
	{\mathbb C}^2 / \left( {\mathbb Z}^2 + {\mathbb Z}^2 \, \bm{N}(\tau) \right) \ .
\end{equation}
The symmetric $2\times 2$ matrix $\bm{N}(\tau)$ of one-parameter Calabi--Yau manifolds --- such as the one-parameter Calabi--Yau threefold $Y_z$ --- becomes according to eq.~\eqref{eq:DefN}
\bq
\label{def_tau_Weil}
\bm{N}(\tau)
= 
\begin{pmatrix}
	N_{11}(\tau) & N_{01}(\tau) \\
	N_{01}(\tau) & N_{00}(\tau) 
\end{pmatrix} \ , & &
N_{IJ}(\tau)
=
-\overline{F_{IJ}(\tau)}
- 2 i \left. \frac{I_{IK} X^K I_{JL} X^L}{X^M I_{MN}X^N} \right|_{\left(X^1,X^0\right)=\left(\tau,1\right)}\ ,
\eq
where $\overline{F_{IJ}(\tau)}$ is the complex conjugate of $F_{IJ}(\tau)$ defined in eq.~\eqref{eq:F_HV} and $I_{IJ}$ abbreviates the imaginary part of $F_{IJ}$, i.e. 
\begin{equation}
	I_{IJ}(X^1,X^0) = \mathrm{Im} \; F_{IJ}(X^1,X^0) \ .
\end{equation}  
With the relation $\overline{F_{IJ}(\tau)} = F_{IJ}(\tau) - 2i I_{IJ}(\tau)$, we rewrite the expression for $\bm{N}(\tau)$ given in eq.~\eqref{def_tau_Weil} to
\begin{equation} \label{def_tau_Weil_v2}
	N_{IJ}(\tau)
	= 
	-F_{IJ}(\tau)
	+ 2 i \left( I_{IJ}(\tau) - \left. \frac{I_{IK} X^K I_{JL} X^L}{X^M I_{MN} X^N} \right|_{\left(X^1,X^0\right)=\left(\tau,1\right)} \right) \ .
\end{equation}
The imaginary part of the $2\times 2$-matrix $\bm{N}(\tau)$ is positive definite \cite{deWit:1984rvr,deWit:1984wbb,Andrianopoli:1996cm}.\footnote{Note again the opposite sign convention for the matrix $\bm{N}(\tau)$ used in refs.~\cite{deWit:1984rvr,deWit:1984wbb,Andrianopoli:1996cm}.}  That is to say, the matrix $\bm{N}(\tau)$ resides in the Siegel upper half-space
\begin{equation}
	\bm{N}(\tau) \in \mathcal{H}_2 \ .
\end{equation} 
Moreover, it is also clear that $\bm{N}(\tau)$ does not vary holomorphically with $\tau$, the imaginary part $I_{IJ}(\tau)$ of the holomorphic matrix $F_{IJ}(\tau)$ is certainly not holomorphic in the complex structure parameter $\tau$. However, while the matrix $\bm{N}(\tau)$ is not holomorphic in $\tau$, its entries are still real analytic functions in $z$. 

As a result the second Weil intermediate Jacobian of the Calabi--Yau threefolds $Y_z$ gives rise to an analytic Calabi--Yau-to-curve correspondence along the lines presented in section~\ref{sect:an_corr}, where the $2\times 2$ matrix $\bm{N}(\tau)$ encodes the period matrix $\bm P$ (see eq.~\eqref{eq:genus-g_period_matrix_definition}) of the holomorphic one forms $\omega_0$ and $\omega_1$ of a genus-two curve $C$
\begin{equation}
	\bm{P} \mathbfcal{X}^{-1}  = \begin{pmatrix}  N_{11} & N_{01}  \\  N_{01} & N_{00} \\ 1 & 0 \\  0 & 1 \end{pmatrix} \ .
\end{equation}
We observe the following relation between the periods of the genus-two curve and the Calabi--Yau threefold 
\bq \label{eq_holomorphic_one_form}
 \bm{P} \mathbfcal{X}^{-1}  \begin{pmatrix}  \tau \\ 1 \end{pmatrix}=
 \begin{pmatrix}  N_{11} & N_{01}  \\  N_{01} & N_{00} \\ 1 & 0 \\  0 & 1 \end{pmatrix} \begin{pmatrix}  \tau \\ 1 \end{pmatrix}
=
\begin{pmatrix} -\normalisedPeriod_{B_1} \\ -\normalisedPeriod_{B_0} \\ 
\phantom{-}\normalisedPeriod_{A^1} \\ \phantom{-}\normalisedPeriod_{A^0} \end{pmatrix} \ .
\eq
The right hand side of eq.~\eqref{eq_holomorphic_one_form} is manifest holomorphic in $\tau$ and thus the piece $I_{IJ}(\tau)$ that is non-holomorphic in $\tau$ drops out in the multiplication on the left hand side. This demonstrates that on the level of cohomology the linear combination of holomorphic one-forms $\tau \,\omega_1 + \omega_0$ varies holomorphically in $\tau$. Furthermore, the projectively rescaled holomorphic cohomology class $X^1(z) \,\omega_1 +  X^0(z) \omega_0$ is annihilated by the Picard--Fuchs operator $L^{(0)}$ given in eq.~\eqref{def_Picard_Fuchs_eps_0}. However, as the period matrix $\bm{N}(\tau)$ varies non-holomorphically with respect to $\tau$ (and hence also with respect to $z$) the holomorphic one-forms linear independent to $\tau\,\omega_1 + \omega_0$ cannot vary holomorphically in~$\tau$.

To establish the holomorphic Calabi--Yau-to-curve correspondence, we focus on the real range $z \in ]0,\frac{1}{25}[$ for the kinetic variable $z$. 
This range realises a one-dimensional Lagrangian sublocus $\Delta_\mathbb{R} :=  ]0,\frac{1}{25}[$ in the complex structure moduli space of the $\mathbb{Z}_5$-quotient of the Hulek--Verrill Calabi--Yau threefold. On this Lagrangian sublocus both the prepotential $F(X^1(z),X^0(z))$ and the affine coordinate $\tau$ are purely imaginary. As a result we find that
\bq \label{eq:FLagLocus}
\left. \mathrm{Re} 
\begin{pmatrix} 
	F_{11} & F_{01} \\
	F_{01} & F_{00} 
\end{pmatrix}  \right|_{\Delta_\mathbb{R}}
= \left.
\begin{pmatrix}
	0 & F_{01} \\
	F_{01} & 0 
\end{pmatrix}\right|_{\Delta_\mathbb{R}} \ , \qquad
\left. \mathrm{Im} 
\begin{pmatrix} 
	F_{11} & F_{01} \\
	F_{01} & F_{00} 
\end{pmatrix}  \right|_{\Delta_\mathbb{R}}
= \left.
\frac{1}{i}
\begin{pmatrix}
	F_{11} & 0 \\
	0 & F_{00} 
\end{pmatrix}\right|_{\Delta_\mathbb{R}} \ .
\eq
Therefore, the entries of $\left.\bm{N}(\tau)\right|_{\Delta_\mathbb{R}}$ given in eq.~\eqref{def_tau_Weil_v2} are explicitly given by
\begin{equation}
\label{entries_tau_W_real_segment}
	\begin{alignedat}{3}
			\left.N_{00}\right|_{\Delta_\mathbb{R}}
	& = &
	-F_{00} \frac{F_{00}-\tau^2 F_{11}}{F_{00}+\tau^2 F_{11}}
	& = &&
	-\frac{\left(F-\tau \partial_\tau F\right)\left(2 F - 2 \tau \partial_\tau F + \tau^2 \partial_\tau^2 F\right)}{F-\tau \partial_\tau F + \tau^2 \partial_\tau^2 F} \ ,
	 \\
	\left.N_{01}\right|_{\Delta_\mathbb{R}}
	& = &
	\ -F_{01} - 2 \frac{\tau\, F_{00}F_{11}}{F_{00}+\tau^2 F_{11}}
	& = &&
	-\frac{F \partial_\tau F - \tau \left(\partial_\tau F\right)^2 + \tau F \partial_\tau^2 F}{F-\tau \partial_\tau F + \tau^2 \partial_\tau^2 F}\ ,
	 \\
	\left.N_{11}\right|_{\Delta_\mathbb{R}}
	& = &
	F_{11} \frac{F_{00}-\tau^2 F_{11}}{F_{00}+\tau^2 F_{11}}
	& = &&
	\frac{\left(F-\tau \partial_\tau F\right) \partial_\tau^2 F}{F-\tau \partial_\tau F + \tau^2 \partial_\tau^2 F}\ .
	\end{alignedat}
\end{equation}
Due to the structure~\eqref{eq:FLagLocus} of the symmetric $(2\times 2)$ matrix $\left.\bm{F}\right|_{\Delta_\mathbb{R}}$, we find that for the symmetric $(2\times 2)$-matrix $\left.\bm{N}(\tau)\right|_{\Delta_\mathbb{R}}$ the diagonal entries $\left.N_{00}\right|_{\Delta_\mathbb{R}}$ and $\left.N_{11}\right|_{\Delta_\mathbb{R}}$ are purely imaginary whereas the off-diagonal entries $\left.N_{01}\right|_{\Delta_\mathbb{R}}$ are real.

Setting $\tau = it$ with $t$ real, we find for large $t$ up to corrections of the order ${\mathcal O}(\qbar)$ with $\qbar = e^{-2\pi t}$ the explicit expressions 
\bq
\left.N_{00}\right|_{\Delta_\mathbb{R}}
& = &
4i t^3 
- 4i \frac{\zeta(3)}{16\pi^3} 
- 36 i \left(\frac{\zeta(3)}{16\pi^3}\right)^2 \frac{1}{t^3 - \frac{\zeta(3)}{16\pi^3}}
+ {\mathcal O}\left(\qbar\right)\ ,
\nonumber \\
\left.N_{01}\right|_{\Delta_\mathbb{R}}
& = &
-1 
+ 36  \frac{\zeta(3)}{16\pi^3} \frac{t^2}{t^3- \frac{\zeta(3)}{16\pi^3}}
+ {\mathcal O}\left(\qbar\right)\ ,
\nonumber \\
\left.N_{11}\right|_{\Delta_\mathbb{R}}
& = &
 12 i t
+ 36i  \frac{\zeta(3)}{16\pi^3} \frac{t}{t^3 -  \frac{\zeta(3)}{16\pi^3}}
+ {\mathcal O}\left(\qbar\right)\ .
\eq
The leading terms of $\left.N_{00}\right|_{\Delta_\mathbb{R}}$ and $\left.N_{11}\right|_{\Delta_\mathbb{R}}$ in $t$ are
\bq
\left.N_{00}\right|_{\Delta_\mathbb{R}}
= 
 4i  t^3
+ {\mathcal O}\left(t^2\right)\ ,
& &\left.N_{11}\right|_{\Delta_\mathbb{R}} 
= 
12it 
+ {\mathcal O}\left(t^{0}\right) \ .
\eq
Thus, we confirm that the imaginary part $\mathrm{Im}(\bm{N}(\tau))$ of the symmetric $2\times 2$-matrix $\bm{N}(\tau)$ (restricted to $\Delta_\mathbb{R}$ for large $t$) is indeed positive definite, i.e. $\bm{N}(\tau)\in \mathcal{H}_2$.

\subsection{The polarised holomorphic Jacobian for the banana integral}
\label{sect:holomorphic_Jacobian}

Let us construct from the Weil intermediate Jacobian of the $\mathbb{Z}_5$-quotient $Y_z$ of the Hulek--Verrill Calabi--Yau threefold restricted to the Lagrangian sublocus $\Delta_\mathbb{R} \, \equiv\,  ]0,\frac{1}{25}[$ (corresponding to the real range $z\in  ]0,\frac{1}{25}[$ of the kinematic variable $z$) the polarised holomorphic intermediate Jacobian introduced in section~\ref{sect:polholJac} for a tubular neighbourhood of $\Delta_\mathbb{R}$ via holomorphic continuation from $\Delta_\mathbb{R}$. That is to say, we construct the intermediate Jacobian 
\begin{equation}
	J^H_2(\tau)
	= 
	{\mathbb C}^2 / \left( {\mathbb Z}^2 + {\mathbb Z}^2 \, \bm{H}(\tau) \right) \ .
\end{equation}
where the entries of the matrix $\bm{H}(\tau)$ arise from the entries \eqref{entries_tau_W_real_segment} via holomorphic continuation
\begin{equation}
\label{entries_tau_H}
	\begin{alignedat}{3}
	H_{00}
	& = &
	-F_{00} \frac{F_{00}-\tau^2 F_{11}}{F_{00}+\tau^2 F_{11}}
	& = &&
	-\frac{\left(F-\tau \partial_\tau F\right)\left(2 F - 2 \tau \partial_\tau F + \tau^2 \partial_\tau^2 F\right)}{F-\tau \partial_\tau F + \tau^2 \partial_\tau^2 F} \ ,
	 \\
	H_{01}
	& = &
	\ -F_{01} - 2 \frac{\tau\, F_{00}F_{11}}{F_{00}+\tau^2 F_{11}}
	& = &&
	-\frac{F \partial_\tau F - \tau \left(\partial_\tau F\right)^2 + \tau F \partial_\tau^2 F}{F-\tau \partial_\tau F + \tau^2 \partial_\tau^2 F}\ ,
	 \\
	H_{11}
	& = &
	F_{11} \frac{F_{00}-\tau^2 F_{11}}{F_{00}+\tau^2 F_{11}}
	& = &&
	\frac{\left(F-\tau \partial_\tau F\right) \partial_\tau^2 F}{F-\tau \partial_\tau F + \tau^2 \partial_\tau^2 F}\ .
	\end{alignedat}
\end{equation}
By construction $\bm{H}(\tau)$ agrees with $\bm{N}(\tau)$ on $z \in \Delta_{\mathbb{R}}$ and varies holomorphically in the tubular neighbourhood of $\Delta_{\mathbb{R}}$. As for the Weil intermediate Jacobian, we have the identity
\bq
\label{eq_holomorphic_one_form_v2}
\left( \begin{array}{cccc}
	 H_{11} & H_{01}  \\
	 H_{01} & H_{00} \\
	 1 & 0 \\
	 0 & 1
\end{array} \right)
\left(\begin{array}{cc}  \tau \\ 1 \end{array} \right)
=
\begin{pmatrix} -\normalisedPeriod_{B_1} \\ -\normalisedPeriod_{B_0} \\ \phantom{-}\normalisedPeriod_{A^1} \\ \phantom{-}\normalisedPeriod_{A^0}
\end{pmatrix} \ .
\eq
This shows that the genus-two curve will have one holomorphic one-form, whose periods are annihilated by the Picard--Fuchs operator $L^{(0)}$.

\begin{figure}
	\centering
	\includegraphics[width=0.49\textwidth,valign=m]{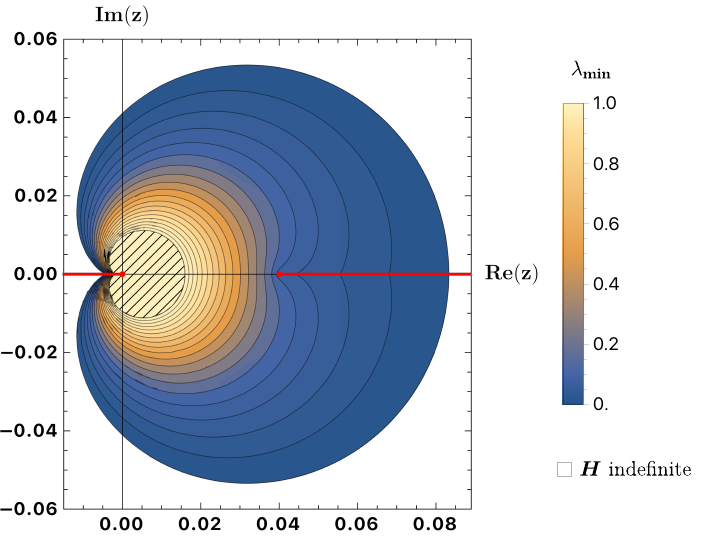}\hspace{1em}
	\includegraphics[width=0.45\textwidth,valign=m]{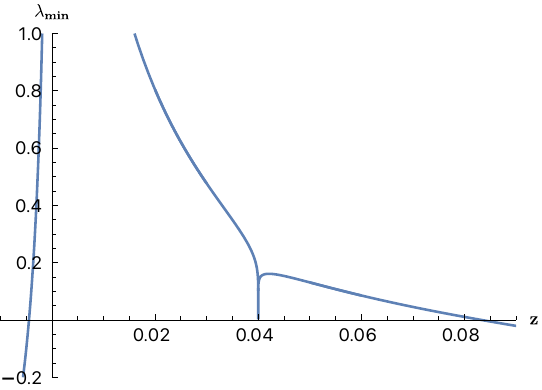}
	\caption{Plots showing the smallest eigenvalue $\lambda_{\text{min}}$ of $\text{Im}(\bm{H}(z))$, visualising the range of values of $z$, for which the polarised holomorphic intermediate Jacobian is positive definite and therefore in $\mathcal{H}_2$. Left: Contour plot over the complex $z$-plane. The red points mark the physical singularities $z=0$ and $z=\frac{1}{25}$, and the hatching corresponds to the region $\lambda_{\text{min}}>1$. The red lines denote branch cuts of $\bm{H}(z)$. In the uncoloured region we have $\lambda_{\text{min}}< 0$ and $\text{Im}(\bm{H}(z))$ is indefinite. The coloured area thus shows the region $U$ where the second polarised holomorphic intermediate Jacobian $J_2^{\Delta_{\mathbb{R}}}(Y_z)$ is defined. Right: Values of $\lambda_{\text{min}}$ along the real $z$-axis. At the physical singularity $z=\tfrac{1}{25}$, the eigenvalue $\lambda_{\text{min}}$ vanishes, corresponding to the boundary of $\mathcal{H}_2$.}
	\label{fig:HolomorphicJacobian}
\end{figure}

More importantly, we have in a sufficiently small neighbourhood of $\Delta_{\mathbb{R}}$ that the symmetric $2\times 2$-matrix $\bm{H}(\tau)$ is in the Siegel upper half-space $\mathcal{H}_2$. A plot of this neighborhood is shown in fig. \ref{fig:HolomorphicJacobian}.
Therefore, the constructed polarised holomorphic intermediate Jacobian $J^H_2$ is suitable to construct 
a local holomorphic Calabi--Yau-to-curve correspondence along the lines of section~\ref{sect:hol_corr} upon setting
\begin{equation}
	\bm{\tau}(\tau) = \bm{H}(\tau) \ .
\end{equation}
In the remaining parts of this section we construct the corresponding families of genus-two curves explicitly. To set the stage for this construction we first review some properties of theta functions of higher genus curves in subsection~\ref{sect:theta_functions}, which allows us then to spell out in detail the holomorphic Calabi--Yau-to-curve correspondence starting in subsection~\ref{sect:construction}.

Let us stress that the proposed construction is not limited to the kinematic range $z\in  ]0,\frac{1}{25}[$. Choosing a different kinematic range requires the choice of a different Lagrangian sublocus, which can be achieved by analytically continuing the Calabi--Yau periods for the kinematic range  $z\in  ]0,\frac{1}{25}[$ to the newly chosen Lagrangian sublocus. 

\subsection{Riemann theta functions}
\label{sect:theta_functions}

For $\bm{\tau} \in \mathcal H_g$ and $z \in {\mathbb C}^g$ the Riemann theta function is defined by
\bq
\vartheta\left(z,\bm{\tau}\right)
& = &
\sum\limits_{n \in {\mathbb Z}^g} e^{i \pi \left( n^T \, \bm{\tau} \, n + 2 n^T z \right)}~.
\eq
Theta functions with characteristic are defined for $a,b \in {\mathbb Q}^g$ by
\bq
\vartheta\left[\begin{array}{c} a \\ b \\ \end{array}\right]\left(z,\bm{\tau}\right)
& = &
\sum\limits_{n \in {\mathbb Z}^g} e^{i \pi \left( \left(n+a\right)^T \, \bm{\tau} \, \left(n+a\right) + 2 \left(n+a\right)^T \left(z+b\right) \right)}
\nonumber \\
& = &
e^{i \pi \left( a^T \, \bm{\tau} \, a + 2 a^T \left(z+b\right) \right)}
\vartheta\left(z+ \bm{\tau} \, a + b,\bm{\tau}\right).
\eq
Of particular importance is the case, where $a,b \in ({\mathbb Z}/2)^g$.
In this case $4a^T b$ is an integer, and the characteristic is called even (respectively odd) if this integer is even (respectively odd).

Let us now specialise to $g=2$.
In this case we have ten even characteristics and six odd characteristics.
We introduce the following short-hand notation for the even theta constants (the notation follows ref.~\cite{shaska:2012theta}):
\begin{align}
	\label{def_even_theta_constants}
	\theta_1(\bm \tau) & =  \vartheta\left[\begin{array}{cc} 0 & 0 \\ 0 & 0 \\ \end{array}\right]\left(0,\bm{\tau}\right)~,
	&
	\theta_2(\bm \tau) & =  \vartheta\left[\begin{array}{cc} 0 & 0 \\ \frac{1}{2} & \frac{1}{2} \\ \end{array}\right]\left(0,\bm{\tau}\right)~,
	\nonumber \\
	\theta_3(\bm \tau) & =  \vartheta\left[\begin{array}{cc} 0 & 0 \\ \frac{1}{2} & 0 \\ \end{array}\right]\left(0,\bm{\tau}\right)~,
	&
	\theta_4(\bm \tau) & =  \vartheta\left[\begin{array}{cc} 0 & 0 \\ 0 & \frac{1}{2} \\ \end{array}\right]\left(0,\bm{\tau}\right)~,
	\nonumber \\
	\theta_5(\bm \tau) & =  \vartheta\left[\begin{array}{cc} \frac{1}{2} & 0  \\ 0 & 0 \\ \end{array}\right]\left(0,\bm{\tau}\right)~,
	&
	\theta_6(\bm \tau) & =  \vartheta\left[\begin{array}{cc} \frac{1}{2} & 0  \\ 0 & \frac{1}{2} \\ \end{array}\right]\left(0,\bm{\tau}\right)~,
	\nonumber \\
	\theta_7(\bm \tau) & =  \vartheta\left[\begin{array}{cc} 0 & \frac{1}{2}  \\ 0 & 0 \\ \end{array}\right]\left(0,\bm{\tau}\right)~,
	&
	\theta_8(\bm \tau) & =  \vartheta\left[\begin{array}{cc} \frac{1}{2} & \frac{1}{2}  \\ 0 & 0 \\ \end{array}\right]\left(0,\bm{\tau}\right)~,
	\nonumber \\
	\theta_9(\bm \tau) & =  \vartheta\left[\begin{array}{cc} 0 & \frac{1}{2}  \\ \frac{1}{2} & 0 \\ \end{array}\right]\left(0,\bm{\tau}\right)~,
	&
	\theta_{10} & =  \vartheta\left[\begin{array}{cc} \frac{1}{2} & \frac{1}{2}  \\ \frac{1}{2} & \frac{1}{2} \\ \end{array}\right]\left(0,\bm{\tau}\right)~.
\end{align}
Let $z=(z_0,z_1)$.
For the odd theta constants we use the notation ($i \in \{0,1\}$)
\begin{align}
	\label{def_odd_theta_constants}
	\partial_i \theta_{11}(\bm \tau) & =  \left. \frac{\partial}{\partial z_i} \vartheta\left[\begin{array}{cc} 0 & \frac{1}{2} \\ 0 & \frac{1}{2} \\ \end{array}\right]\left(z,\bm{\tau}\right) \right|_{z=0}~,
	&
	\partial_i \theta_{12}(\bm \tau) & =  \left. \frac{\partial}{\partial z_i} \vartheta\left[\begin{array}{cc} 0 & \frac{1}{2} \\ \frac{1}{2} & \frac{1}{2} \\ \end{array}\right]\left(z,\bm{\tau}\right) \right|_{z=0}~,
	\nonumber \\
	\partial_i \theta_{13}(\bm \tau) & =  \left. \frac{\partial}{\partial z_i} \vartheta\left[\begin{array}{cc} \frac{1}{2} & 0 \\ \frac{1}{2} & 0 \\ \end{array}\right]\left(z,\bm{\tau}\right) \right|_{z=0}~,
	&
	\partial_i \theta_{14}(\bm \tau) & =  \left. \frac{\partial}{\partial z_i} \vartheta\left[\begin{array}{cc} \frac{1}{2} & \frac{1}{2} \\ \frac{1}{2} & 0 \\ \end{array}\right]\left(z,\bm{\tau}\right) \right|_{z=0}~,
	\nonumber \\
	\partial_i \theta_{15}(\bm \tau) & =  \left. \frac{\partial}{\partial z_i} \vartheta\left[\begin{array}{cc} \frac{1}{2} & 0 \\ \frac{1}{2} & \frac{1}{2} \\ \end{array}\right]\left(z,\bm{\tau}\right) \right|_{z=0}~,
	&
	\partial_i \theta_{16}(\bm \tau) & =  \left. \frac{\partial}{\partial z_i} \vartheta\left[\begin{array}{cc} \frac{1}{2} & \frac{1}{2} \\ 0 & \frac{1}{2} \\ \end{array}\right]\left(z,\bm{\tau}\right) \right|_{z=0}~.
\end{align}


\subsection{A family of genus-two curves from the polarised holomorphic Jacobian}
\label{sect:construction}
For the $\mathbb{Z}_5$-quotients $Y_z$ of Hulek--Verrill Calabi--Yau threefolds, we have constructed the family $\mathcal{J}_2^{\Delta_{\mathbb{R}}}$ of polarised holomorphic intermediate Jacobians explicitly in section \ref{sect:holomorphic_Jacobian}. There we gave the Jacobians as the quotients of $\mathbb{C}^2$ by the lattice $\left( {\mathbb Z}^2 + {\mathbb Z}^2 \, \bm{H}(\tau) \right)$ defined by the family $\bm{H}(\tau)$ of $2{\times}2$ matrices depending holomorphically on $\tau$. In this section, we construct a family $\mathcal{C}_U$ of genus-two curves $C_z$, $z \in U$, whose intermediate Jacobians agree with those of the family $\mathcal{J}_2^{\Delta_{\mathbb{R}}}$. In effect, we are explicitly constructing the inverse map \eqref{eq:inverse_Jacobian_map} taking us from the Schottky locus $S_g$ of intermediate Jacobians to the moduli space $\overline{\mathcal{M}}_2^\mathcal{A}$ of genus-two curves.

\begin{figure}
	\begin{center}
		\includegraphics[scale=0.8]{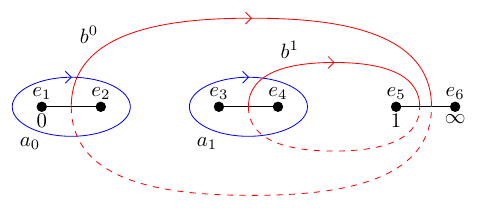}
	\end{center}
	\caption{
		The branch points $\branchpoint_1$ through $\branchpoint_6$ of the genus-two curve $C_z$ together with the cycles $a^0$, $a^1$, $b_0$ and $b_1$.
	}
	\label{fig_genus_2_cuts}
\end{figure}

To start, recall that every genus-two curve $C_z$ is a hyperelliptic curve so that, working over $\mathbb{C}$, we may write the defining equation as
\bq \label{eq:genus-2_hyperelliptic}
v^2 = P_5\left(u\right)~, 
& &
P_5 = u \left(u-\branchpoint_2(z)\right) \left(u-\branchpoint_3(z)\right) \left(u-\branchpoint_4(z)\right) \left(u-1\right)~,
\eq
where we have, without loss of generality, used the automorphisms of the complex plane to fix three branch points of $C_z$ as
\begin{align}
	\branchpoint_1 \; = \; 0~, \qquad  \branchpoint_5 \; = \; 1~, \qquad \branchpoint_6 \; = \; \infty~. 
\end{align}
We may also take the branch cuts to be between $\branchpoint_1$ and $\branchpoint_2$, between $\branchpoint_3$ and $\branchpoint_4$ and between $\branchpoint_5$ and $\branchpoint_6=\infty$.
For this curve, a choice of a symplectic basis of the homology group $H_1(C_z,\mathbb{Z})$, as defined in eq.~\eqref{eq:symplectic_homology_basis_C}, is displayed in fig.~\ref{fig_genus_2_cuts}.

Given a matrix $\bm{H}(\tau) \in {\mathcal H}_2$, a lemma of Picard (see ref.~\cite{shaska:2012theta}, Lemma 8) can be used to find a set of branch points $\branchpoint_2,\branchpoint_3$, and $\branchpoint_4$ so that the Jacobian $J_1(C_z)$ of the corresponding curve \eqref{eq:genus-2_hyperelliptic} is equal to the polarised holomorphic intermediate Jacobian $J_2^H(Y_z,V_z)$ which corresponds to the matrix $\bm{H}(\tau)$.  The branch points are given in terms of theta constants $\theta_i := \theta_i(\bm H(\tau))$ as
\begin{align}
	\label{def_branch_points}
	\branchpoint_2(\bm H(\tau)) & = \frac{\theta_5^2 \theta_6^2}{\theta_1^2 \theta_4^2}~,
	&
	\branchpoint_3(\bm H(\tau)) & = \frac{\theta_6^2 \theta_7^2}{\theta_4^2 \theta_8^2}~,
	&
	\branchpoint_4(\bm H(\tau)) & = \frac{\theta_5^2 \theta_7^2}{\theta_1^2 \theta_8^2}~.
\end{align}
On the level of period matrices this implies that there exists a symplectic integral basis of $H_1(C_z,\mathbb{Z})$ such that the matrix $\bm \tau$ is equal to the matrix $\bm{H}(\tau)$ on which the construction of $J_2^H(Y_z,V_z)$ is based.\footnote{Note that the matrix $\bm \tau$ is by construction invariant under the change of basis of $H^1(X,\mathbb{C})$, so we need not worry about possible change of basis in this space.} Concretely, this means that there exists a symplectic change of basis $4 \times 4$-matrix
\begin{align} \label{eq:HV_periods_symplectic_transformation}
		\left(\begin{array}{c}
		b \\
		a \\
	\end{array} \right) 
\mapsto 
\bm \Gamma 	\left(\begin{array}{c}
	b \\
	a \\
\end{array} \right)~, 
\qquad \text{where} \qquad	
\bm{\Gamma}	=
	\left(\begin{array}{cc}
		\bm{A} & \bm{B} \\
		\bm{C} & \bm{D} \\
	\end{array} \right)
	\in 
	\mathrm{Sp}(4,{\mathbb Z})~,
\end{align}
such that under this change of basis, the matrix $\bm \tau$ transforms so as to agree with $\bm H(\tau)$:
\bq
\label{def_Gamma_trafo}
\bm{H}(\tau)
& = &
\left( \bm{A} \bm \tau  + \bm{B} \right) \left( \bm{C} \bm \tau + \bm{D} \right)^{-1}~.
\eq
In fact, this condition can be used to solve for the matrix $\bm \Gamma$.
In this way, eq.~\eqref{eq:genus-2_hyperelliptic} gives us an algebraic equation for the genus-two curve $J_1^{-1}(J_2^H)$ corresponding to the polarised holomorphic intermediate Jacobian $J_2^H$ constructed in subsection~\ref{sect:holomorphic_Jacobian}.


\subsection{Periods and holomorphic forms}
\label{sect:holomorphic_forms}
Having obtained an algebraic equation for the polarised holomorphic intermediate Jacobians in the family $\mathcal{J}^{\Delta_{\mathbb{R}}}_2(\mathcal{Y}_U,\mathcal{V}_U)$, we now wish to write down the period matrix $\bm{P}$ from eq.~\eqref{eq:genus-g_period_matrix_definition} explicitly in terms of the canonical choice of one-forms in $H^{(1,0)}(C_z,\mathbb{Z})$ and the symplectic basis $(b_j, a^i)$ displayed in fig.~\ref{fig_genus_2_cuts}. After this, we will explicitly spell out the relation of these periods to the periods of the Calabi--Yau threefold $Y_z$, and in this way show that they are annihilated by the Picard--Fuchs operator $L^{(0)}$ \eqref{def_Picard_Fuchs_eps_0}.

For the genus-two curve $C_z$ the space $H^{(1,0)}(C_z,\mathbb{C})$ is two-dimensional, and we may take as the basis the canonical holomorphic one-forms
\bq
\label{def_standard_holomorphic_one_forms}
\omega_0 \; = \; \frac{du}{\sqrt{P_5\left(u\right)}}~,
& &
\omega_1 \; = \; \frac{u du}{\sqrt{P_5\left(u\right)}}~.
\eq
Recall from section \ref{sect:genus-g_curves} the definition \eqref{eq:genus-g_period_matrix_definition} of the $4{\times}2$ period matrix $\bm P$, and the $2{\times}2$ matrices $\mathbfcal{X}$, $\mathbfcal{T}$, and $\bm \tau$:\footnote{It is convenient to order the indices backwards as $(1,0)$ as in section \ref{sect:genus-g_curves}.}
\begin{align}
\label{eq:genus-g_period_matrices_recalled}
 \mathbfcal{X} =  
 \left( \begin{array}{cc}
  \int\limits_{a^1} \omega_1 & \int\limits_{a^1} \omega_0 \\
  \int\limits_{a^0} \omega_1 & \int\limits_{a^0} \omega_0 \\
 \end{array} \right),
 \quad
 \mathbfcal{T} = 
 \left( \begin{array}{cc}
  \int\limits_{b_1} \omega_1 & \int\limits_{b_1} \omega_0 \\
  \int\limits_{b_0} \omega_1 & \int\limits_{b_0} \omega_0 \\
 \end{array} \right),
 \quad
 \bm{\tau} =\mathbfcal{T} \ \mathbfcal{X}^{-1} ~, \quad
\bm P = \begin{pmatrix}
		\mathbfcal{T} \\ \mathbfcal{X}
	\end{pmatrix} ~.
\end{align}
The $a$-cycle periods can also be expressed in terms of theta constants of the matrix $\bm H(\tau)$ using Thomae's formulae (see ref.~ \cite{Enolski:2008aaa}): 
\bq
\label{P_a_in_terms_of_theta}
\mathbfcal{X}
& = &
2 \frac{\theta_1 \theta_4 \theta_8}{\theta_2 \theta_3 \theta_9 \theta_{10}}
\left( \begin{array}{cc}
	\partial_1 \theta_{11} & \frac{\theta_1 \theta_4 \theta_8 \partial_1 \theta_{16}}{\theta_5 \theta_6 \theta_7} \\
        \partial_0 \theta_{11} & \frac{\theta_1 \theta_4 \theta_8 \partial_0 \theta_{16}}{\theta_5 \theta_6 \theta_7} \\
\end{array} \right),
\eq
and the $b$-periods can be computed from this if we know the matrix $\bm \tau$ --- which is ultimately expressed in terms of the matrix $\bm H(\tau)$ via the relations \eqref{def_branch_points} --- using the definitions \eqref{eq:genus-g_period_matrices_recalled} recalled above:
\bq
\mathbfcal{T} & = & \bm{\tau} \mathbfcal{X} ~.
\eq
The family of curves $C_z$ was constructed in the previous section so that the Jacobian varieties $J_1(C_z)$ of $C_z$ coincide with the polarised holomorphic intermediate Jacobians $J_2^H(Y_z,V_z)$ of the Calabi--Yau manifolds $Y_z$ when $z$ belongs to the tubular neighbourhood $U$ defined in \ref{sect:polholJac}. By construction, the period matrices $\bm \tau$ of this family of curves must agree, up to a symplectic change of basis \eqref{def_Gamma_trafo}, with the matrices $\bm H(\tau)$ on which the construction of $J_2^H(Y_z,V_z)$ is based. Therefore, we obtain that
\begin{align}
	\begin{pmatrix} \bm H(\tau) \\ \bm{1} \end{pmatrix} =
	\begin{pmatrix}    \left( \bm{A} \bm \tau  + \bm{B} \right) \left( \bm{C} \bm \tau + \bm{D} \right)^{-1} \\ \bm 1  \end{pmatrix} \ ,
\end{align}
where the matrices $\bm A$, $\bm B$, $\bm C$, and $\bm D$ can be solved from \eqref{def_Gamma_trafo}. 
From eq.~(\ref{eq_holomorphic_one_form_v2}) we have
\bq
\label{eq_relation_genus_two_to_CY}
 \Frobeniusbasis_0
 \begin{pmatrix}    \left( \bm{A} \bm \tau  + \bm{B} \right) \left( \bm{C} \bm \tau + \bm{D} \right)^{-1} \\ \bm 1  \end{pmatrix}
 \begin{pmatrix} \tau \\ 1 \end{pmatrix}
 & = &
\begin{pmatrix} -\Period_{B_1} \\ -\Period_{B_0} \\ \phantom{-}\Period_{A^1} \\ \phantom{-}\Period_{A^0}
\end{pmatrix} \ .
\eq
The periods on the right-hand side of eq.~(\ref{eq_relation_genus_two_to_CY}) are manifestly annihilated by the Picard--Fuchs operator $L^{(0)}$. Therefore, we define a holomorphic one-form $\omega \in H^{(1,0)}(C_z,\mathbb{Z})$ by
\bq
\label{def_omega_one_form}
\omega 
& = &
\Frobeniusbasis_0 \;
\begin{pmatrix} \omega_1, \omega_0 \end{pmatrix}
 {\mathbfcal{X}}^{-1}
 \left( \bm{C} \bm{\tau} + \bm{D} \right)^{-1}
\begin{pmatrix} \tau \\ 1  \end{pmatrix}
\ ,
\eq
and denote the corresponding vector of periods by $\pi = (\pi_{b_1},\pi_{b_0},\pi_{a^1},\pi_{a^0})^T$. 
The periods of $\omega$ are related by a $\mathrm{Sp}(4,{\mathbb Z})$-transformation to the ones in eq.~(\ref{eq_relation_genus_two_to_CY}), hence they 
give a basis of solutions of the Picard--Fuchs equation $L^{(0)}f = 0$.

Let us close this section by noting that if we choose the cycles $a^0, a^1, b_0, b_1$ such that $\bm{\Gamma}=\bm{1}$,\footnote{We have checked that the choice of the cycles as in fig.~(\ref{fig_genus_2_cuts}) or fig.~(\ref{fig_genus_2_cuts_v2}) has this property.} then eq.~(\ref{def_omega_one_form}) simplifies and we obtain
\bq
\omega & = & c_0 \omega_0 + c_1 \omega_1
\eq
with
\begin{equation}
	\label{def_c0_c1}
	\begin{alignedat}{3}
		c_0
		& = &
		\frac{\Frobeniusbasis_0 {\mathcal{X}^1}_{1} - \Frobeniusbasis_1 {\mathcal{X}^0}_{1}}{\det{\mathbfcal{X}}}
		& = &&
		\frac{\theta_2 \theta_3 \theta_5 \theta_6 \theta_7 \theta_9 \theta_{10}}{2 \theta_1^2 \theta_4^2 \theta_8^2} 
		\frac{\Frobeniusbasis_0 \partial_1\theta_{11} - \Frobeniusbasis_1 \partial_0\theta_{11}}{\partial_0 \theta_{16} \partial_1 \theta_{11} - \partial_0 \theta_{11} \partial_1 \theta_{16}}~,
		\\
		c_1
		& = &
		\frac{\Frobeniusbasis_1 {\mathcal{X}^0}_{0} - \Frobeniusbasis_0 {\mathcal{X}^1}_{0} }{\det{\mathbfcal{X}}}
		& = &&
		\frac{\theta_2 \theta_3 \theta_9 \theta_{10}}{2 \theta_1 \theta_4 \theta_8} \frac{\Frobeniusbasis_0 \partial_1\theta_{16} - \Frobeniusbasis_1 \partial_0\theta_{16}}{\partial_0 \theta_{16} \partial_1 \theta_{11} - \partial_0 \theta_{11} \partial_1 \theta_{16}}~.
	\end{alignedat}
\end{equation}
The relation between the periods of $\omega$ and the periods of the holomorphic $(3,0)$-form $\Omega$ of the Calabi--Yau threefold $Y_z$ is then simply given by
\begin{align}
	(\pi_{b_1},\pi_{b_0},\pi_{a^1},\pi_{a^0}) = (-\Pi_{B_1},-\Pi_{B_0},\Pi_{A^1},\Pi_{A^0})~.
\end{align}
This finishes the explicit construction of a family $\mathcal{C}_U$ of genus-two curves $C_z$ together
with holomorphic one-forms $\omega(z) \in H^{(1,0)}(C_z,\mathbb{C})$ whose periods are annihilated by~$L^{(0)}$. In the following sections, we first study the behaviour of the family near the MUM-point of $L^{(0)}$ at $z = 0$ in section \ref{sect:expansions}, after which we briefly review the Igusa invariants in section \ref{sect:igusa}. We use these to show in section \ref{eq:transcendentality} that the family $C_z$ is transcendental, in the sense that at least one of the branch points $\branchpoint_2(z)$, $\branchpoint_3(z)$, and $\branchpoint_4(z)$ defining the curve $C_z$ via eq.~\eqref{eq:genus-2_hyperelliptic} is necessarily a transcendental function of $z$.
																
\subsection{Expansions near the MUM-point}
\label{sect:expansions}

In this section we consider the case where the variable $z$ is small.
From eq.~(\ref{examples_qbar_z_expansion}) we have
\bq
\qbar & = & z + {\mathcal O}\left(z^2\right).
\eq
We set
\begin{align}
	\qbar_{00,2} & = \exp\left(2 \pi i \frac{N_{00}}{2}\right),
	& 
	\qbar_{11,2} & = \exp\left(2 \pi i \frac{N_{11}}{2}\right),
	&
	\qbar_{01,2} & = \exp\left(2 \pi i \frac{N_{01}}{2}\right).
\end{align}
For small $z$ we have approximately
\begin{align}
	\qbar_{00,2} & \approx \exp\left(\frac{\ln^3z}{2 \pi^2}\right)~,
	& 
	\qbar_{11,2} & \approx z^6~,
	&
	\frac{i}{2}\left(1+\qbar_{01,2}\right)  & \approx -\frac{9 \zeta(3)}{4 \pi \ln z}~.
\end{align}
All three expressions on the right-hand sides go to zero as $z \rightarrow 0$, albeit at different rates.
\begin{table}
	\begin{center}
		\renewcommand{\arraystretch}{2}
		\begin{tabular}{c|cccccc}
			$z$ & $10^{-7}$ & $10^{-6}$ & $10^{-5}$ & $10^{-4}$ & $10^{-3}$ & $10^{-2}$ \\
			\hline
			\hline
			$e^{\frac{\ln^3z}{2 \pi^2}}$ & $7.4 \cdot 10^{-93}$ & $9.6 \cdot 10^{-59}$ & $2.7\cdot 10^{-34}$ & $6.5 \cdot 10^{-18}$ & $5.6 \cdot 10^{-8}$ & $7\cdot 10^{-3}$ \\
			\hline
			$z^6$ & $10^{-42}$ & $10^{-36}$ & $10^{-30}$ & $10^{-24}$ & $10^{-18}$ & $10^{-12}$ \\
			\hline
			$-\frac{9 \zeta(3)}{4 \pi \ln z}$ & $0.053$ & $0.062$ & $0.075$ & $0.093$ & $0.12$ & $0.19$ \\
		\end{tabular}
	\end{center}
	\caption{
		The hierarchy of small parameters.
	}
	\label{table_scales}
\end{table}
For sufficiently small values of $z$ we have 
the hierarchy
\bq
e^{\frac{\ln^3z}{2\pi^2}} \; \ll \; z^6 \; \ll \; -\frac{9 \zeta(3)}{4 \pi \ln z}~.
\eq
A few values are shown in table~\ref{table_scales}.
For $z \lesssim 10^{-5}$ the parameter $\qbar_{00,2}$ is the smallest quantity among
the set $\{\qbar_{00,2},\qbar_{11,2},\frac{i}{2}(1+\qbar_{01,2})\}$, followed by $\qbar_{11,2}$ as the second smallest quantity.
We may therefore first expand in $\qbar_{00,2}$. If a second expansion is needed, we expand in $\qbar_{11,2}$.

For the branch points we find
\bq
\label{expansion_branch_points}
\branchpoint_2
& = &
16 \qbar_{00,2}
+ {\mathcal O}\left(\qbar_{00,2}^2\right)
+ {\mathcal O}\left(\qbar_{11,2}^2\right),
\nonumber \\
\branchpoint_3
& = &
\frac{1}{\left(1+\qbar_{01,2}\right)^2} 
\left[ 4 \qbar_{01,2} - 8 \left(1-\qbar_{01,2}\right)^2 \qbar_{11,2} 
\right] 
+ {\mathcal O}\left(\qbar_{00,2}^1\right)
+ {\mathcal O}\left(\qbar_{11,2}^2\right),
\nonumber \\
\branchpoint_4
& = &
\frac{1}{\left(1+\qbar_{01,2}\right)^2} 
\left[ 4 \qbar_{01,2} + 8 \left(1-\qbar_{01,2}\right)^2 \qbar_{11,2} 
\right] 
+ {\mathcal O}\left(\qbar_{00,2}^1\right)
+ {\mathcal O}\left(\qbar_{11,2}^2\right).
\eq
For $z \in ]0,\frac{1}{25}[$ we have that $H_{01}$ is real, $\qbar_{01,2}$ a phase and
\bq
\frac{4 \qbar_{01,2}}{\left(1+\qbar_{01,2}\right)^2}
\; = \; 
\frac{2}{1+\cos\left(\pi \,H_{01} \right)}~,
& &
\frac{\left(1-\qbar_{01,2}\right)^2}{\left(1+\qbar_{01,2}\right)^2}
\; = \; 
-\frac{1-\cos\left(\pi \,H_{01}\right)}{1+\cos\left(\pi \,H_{01}\right)}~.
\eq
As $\qbar_{00,2}$ and $\qbar_{11,2}$ are real as well, this shows that for $z \in ]0,\frac{1}{25}[$
all branch points 
are on the real line (at least to the approximation of eq.~(\ref{expansion_branch_points})).
We have checked numerically that this holds also independently of the expansion used in eq.~(\ref{expansion_branch_points})).
We find that they are ordered as
\bq
0 \; < \; \branchpoint_2 \; < \; 1 \; < \; \branchpoint_4 \; < \; \branchpoint_3 \; < \; \infty~.
\eq
The location of the branch points 
\begin{figure}
	\begin{center}
		\includegraphics[scale=0.8]{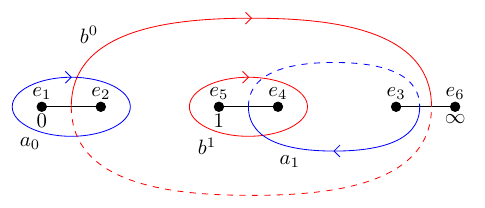}
	\end{center}
	\caption{
		The location of the branch points $\branchpoint_1-\branchpoint_6$ and the location of the cycles $a_0$, $a_1$, $b^0$ and $b^1$ for $z \in ]0,\frac{1}{25}[$.
	}
	\label{fig_genus_2_cuts_v2}
\end{figure}
and the location of the cycles are shown in fig.~\ref{fig_genus_2_cuts_v2}.
Numerically, we find for $z=0.002$
\bq
\branchpoint_2 \; = \; 9.58 \cdot 10^{-5}~,
\;\;\;
\branchpoint_4 \; = \; 47.5~,
\;\;\;
\branchpoint_3 \; = \; \branchpoint_4 + 3.9 \cdot 10^{-14}~.
\eq
For $z$ sufficiently small we have
\bq
\branchpoint_2 - \branchpoint_1 \; = \; {\mathcal O}\left(\qbar_{00,2}\right)~,
& &
\branchpoint_3 - \branchpoint_4 \; = \; {\mathcal O}\left(\qbar_{11,2}\right)~.
\eq
In the limit $z\rightarrow 0$ the genus-two curve degenerates to a genus zero curve, as both $\qbar_{00,2}$ and $\qbar_{11,2}$ go
to zero.


\subsection{The Igusa invariants}
\label{sect:igusa}

Let
\bq
v^2 & = & c \prod\limits_{i=1}^{6} \left( u - \branchpoint_i \right)
\eq
be a hyperelliptic curve of genus $2$.
We write $(ij)$ for $\branchpoint_{i}-\branchpoint_{j}$.
The Igusa invariants $I_2, I_4, I_6$ and $I_{10}$ are defined by \cite{Igusa_1960}
\bq
I_2
& = &
c^2 \sum\limits_{15} \left(12\right)^2 \left(34\right)^2 \left(56\right)^2~,
\nonumber \\
I_4
& = &
c^4 \sum\limits_{10} \left(12\right)^2 \left(23\right)^2 \left(31\right)^2 \left(45\right)^2 \left(56\right)^2 \left(64\right)^2~,
\nonumber \\
I_6
& = &
c^6 \sum\limits_{60} \left(12\right)^2 \left(23\right)^2 \left(31\right)^2 \left(45\right)^2 \left(56\right)^2 \left(64\right)^2 \left(14\right)^2 \left(25\right)^2 \left(36\right)^2~,
\nonumber \\
I_{10}
& = &
c^{10} \prod\limits_{i<j} \left(ij\right)^2~.
\eq
The subscripts on the sums give the number of permutations to be summed over.
The absolute Igusa invariants are defined by
\bq
j_1 \; = \; \frac{I_2^5}{I_{10}}~,
\;\;\;\;\;\;
j_2 \; = \; \frac{I_2^3 I_4}{I_{10}}~,
\;\;\;\;\;\;
j_3 \; = \; \frac{I_2^2 I_6}{I_{10}}~.
\eq
The Clebsch invariants $a,b,c,d$ \cite{Clebsch_1866}
are related to the Igusa invariants $I_2, I_4, I_6, I_{10}$ by
\bq
I_2 & = &  - 120 a~,
\\
I_4 & = & -720a^2+6750b~,
\nonumber \\
I_6 & = & 8640a^3-108000ab+202500c~,
\nonumber \\
I_{10} & = & -62208a^5+972000a^3b+1620000a^2c-3037500ab^2-6075000bc-4556250d~.
\nonumber
\eq
We may express the absolute Igusa invariants in terms of theta constants.
One finds
\begin{align}
	j_1 & = \frac{32 N_1^5}{p_0^6 p_1^3 p_2^3 p_3^3}~,
	&
	j_2 & = \frac{32N_1^3N_2}{p_0^4 p_1^2 p_2^2 p_3^2}~,
	&
	j_3 & = \frac{4N_1^2 N_3}{p_0^4 p_1^2 p_2^2 p_3^2}~,
\end{align}
where
\bq
N_1
& = &
3 p_1 p_2 p_3
+ 2 p_0 \left(p_1p_2+p_2p_3+p_3p_1\right)
- 2 p_0^2 \left(p_1+p_2+p_3\right)
+ p_0^2 p_4
- 4 p_0^3~,
\nonumber \\
N_2 & = & -3p_0-p_1-p_2-p_3+p_4~,
\nonumber \\
N_3 & = &
- 10 p_1 p_2 p_3 \left(p_1+p_2+p_3\right)
+ \left(p_1p_2+p_2p_3+p_3p_1\right)^2
- 18 p_0 p_1 p_2 p_3
\nonumber \\
& &
- 8 p_0 \left(p_1+p_2+p_3\right) \left(p_1p_2+p_2p_3+p_3p_1\right)
+ 6 p_0 \left(p_1p_2+p_2p_3+p_3p_1\right) p_4
\nonumber \\
& &
- 18 p_0^2 \left(p_1p_2+p_2p_3+p_3p_1\right)
- 4 p_0^2 \left(p_1+p_2+p_3\right)^2
+ p_0^2 p_4^2
- 4 p_0^3 \left(p_1+p_2+p_3\right)
\nonumber \\
& &
+ 8 p_0^4
+ 7 p_1 p_2 p_3 p_4
- 6 p_0^3 p_4~,
\eq
and
\bq
p_0 
& = & 
\theta_1^2 \theta_4^2 \theta_5^2 \theta_6^2~,
\nonumber \\
p_1
& = & 
\left( \theta_1^2 \theta_5^2 - \theta_4^2 \theta_6^2 \right)^2~,
\nonumber \\
p_2
& = & 
\left( \theta_1^2 \theta_4^2 - \theta_5^2 \theta_6^2 \right)^2~,
\nonumber \\
p_3
& = & 
\left( \theta_1^2 \theta_6^2 - \theta_4^2 \theta_5^2 \right)^2~,
\nonumber \\
p_4
& = & 
\theta_1^8 + \theta_4^8 + \theta_5^8 + \theta_6^8~.
\eq
\subsection{Transcendentality of the family of genus-two curves} \label{eq:transcendentality}
The coefficients $\branchpoint_2(z), \branchpoint_3(z), \branchpoint_4(z)$ defining the family $\mathcal{C}_U$ of genus-two curves $C_z$ via eq.~(\ref{def_branch_points}), corresponding to the locations of the branch points, are a priori transcendental functions of the variable $z$. We may ask if there is a family $\mathcal{C}'_U$ of genus-two curves $C'_z$ which are isomorphic to the curves $C_z$, such that the branch points of the curve $C'_z$ are algebraic functions of $z$. In this section, we show that this is impossible.

Let us first assume that the branch points $\branchpoint_2(z), \branchpoint_3(z), \branchpoint_4(z)$ of $C'_z$ are algebraic functions of $z$.
The absolute Igusa invariants are then algebraic functions of $z$ as well. We have shown above that as $z\rightarrow 0$, the genus-two curve degenerates, which implies that some absolute Igusa invariants must go to zero or infinity. If we assume that all absolute Igusa invariants are algebraic functions of $z$, they must behave in the limit $z \rightarrow 0$ as
\bq
z^r~,
\eq
where $r \in {\mathbb Q}$ is a rational number.

If the curves $C_z$ and $C'_z$ are isomorphic, they have the same absolute Igusa invariants.
We now investigate the behaviour of the absolute Igusa invariants of the curves $C_z$ in the limit $z \rightarrow 0$.
We find for the leading term of $j_1$
\bq
j_1
& = &
8 \qbar_{00,2}^{-2} \qbar_{11,2}^{-2}
\frac{q_{01,2}^2\left(1+10q_{01,2}^2+q_{01,2}^4\right)}{\left(1-q_{01,2}^2\right)^{12}}
+ \dots~,
\eq
where the dots stand for less singular terms.
The dominant behaviour comes from $\qbar_{00,2}^{-2}$.
We have
\bq
\qbar_{00,2}^{-2}
& \approx &
\exp\left(-\frac{\ln^3z}{\pi^2}\right)
\; = \;
z^{-\frac{\ln^2z}{\pi^2}}~,
\eq
which diverges stronger than any $z^r$, with $r\in {\mathbb Q}$ fixed.
This shows that $j_1$ is not an algebraic function of $z$ and so a family $\mathcal{C}'$ of curves $C'_z$, where all branch points would be algebraic functions of $z$, cannot exist.


\subsection{Summary of the explicit construction}
\label{sect:summary_genus:two}
Let us summarise the explicit construction of the family $\mathcal{C}_U$ of genus-two curves $C_z$ whose Jacobians $J_1(C_z)$ agree with the Jacobians $J^H_2(Y_z,V_z)$ of the holomorphic family $\mathcal{J}_2^{\Delta_{\mathbb{R}}}(\mathcal{Y}_U,\mathcal{V}_U)$ introduced in section \ref{sect:polholJac}: We start from the Picard--Fuchs operator $L^{(0)}$, which can be defined via 
\bq
\lefteqn{
	z^4 \left(1-z\right)\left(1-9z\right)\left(1-25z\right) \; L^{(0)} 
	= } & &
\nonumber \\
& &
\theta_z^4
- z \left(35\theta_z^4+70\theta_z^3+63\theta_z^2+28\theta_z+5\right)
+ z^2\left(\theta_z+1\right)^2\left(259\theta_z^2+518\theta_z+285\right)
\nonumber \\
& &
- 225z^3\left(\theta_z+1\right)^2\left(\theta_z+2\right)^2~,
\eq
and which is the differential operator which annihilates the maximal cut of the Feynman integral $I_{11111}$ in $D=2$ space-time dimensions. The differential operator has a point of maximal unipotent monodromy at $z=0$, to which the closest singular point is at $z=1/25$.

Near $z=0$ we first obtain the Frobenius solutions  $\Frobeniusbasis_0$ through $\Frobeniusbasis_3$ of the fourth-order differential equation $L^{(0)} \Frobeniusbasis = 0$. The explicit form of these solutions is given by eqs.~(\ref{Frobenius_series})-(\ref{Frobenius_coeffs}).

We then set\footnote{The quantities $\tau$ and $F$ have geometrical interpretations in terms of the geometry of the $\mathbb{Z}_5$-quotients $Y_z$ of Hulek--Verrill Calabi--Yau threefolds, in which case the equations below become non-trivial results rather than definitions  (see eq.~(\ref{mirror_map}) and eq.~(\ref{def_prepotential})). However, if one takes a purely instrumentalist approach that does not make reference to the Calabi--Yau geometry, one can view these as definitions. It should be borne in mind, however, that the correspondence which shows the family of curves $\mathcal{C}$ corresponds to the differential operator $L^{(0)}$ proceeds via using the Calabi--Yau geometry.}
\bq
\tau \; = \; \frac{\Frobeniusbasis_1}{\Frobeniusbasis_0}~,
& &
F 
\; = \; 
12 \frac{\left(\Frobeniusbasis_0 \Frobeniusbasis_3 - \Frobeniusbasis_1 \Frobeniusbasis_2 \right)}{\Frobeniusbasis_0^2} + \frac{\Frobeniusbasis_1}{\Frobeniusbasis_0} - 8 \frac{\zeta(3)}{\left(2\pi i\right)^3}~.
\eq
With $\tau$ and $F$ at hand, we define a symmetric $(2 \times 2)$-matrix $\bm H(\tau)$ (see eq.~(\ref{entries_tau_H})) by
\bq
\bm H(\tau) 
& = &
\begin{pmatrix}
	H_{11} & H_{01} \\
	H_{01} & H_{00} 
\end{pmatrix} \ ,
\eq
where
\begin{alignat}{2}
	H_{00}
	& = &&
	- \frac{\left(F-\tau \partial_\tau F\right)\left(2 F - 2 \tau \partial_\tau F + \tau^2 \partial_\tau^2 F\right)}{F-\tau \partial_\tau F + \tau^2 \partial_\tau^2 F}~,
	\nonumber \\
	H_{01}
	& = &&
	- \frac{F \partial_\tau F - \tau \left(\partial_\tau F\right)^2 + \tau F \partial_\tau^2 F}{F-\tau \partial_\tau F + \tau^2 \partial_\tau^2 F}~,
	\nonumber \\
	H_{11} & = &&
	\frac{\left(F-\tau \partial_\tau F\right) \partial_\tau^2 F}{F-\tau \partial_\tau F + \tau^2 \partial_\tau^2 F}~.
\end{alignat}
This matrix has the important property that it varies holomorphically as a function of $z$, and in some neighbourhood $U$ of the open interval $\Delta_{\mathbb{R}} := ]0,1/25[$, its imaginary part is positive definite so that $\bm H(\tau) \in \mathcal{H}_2$ (see section \ref{sect:polholJac}).

Given such a matrix $\bm H(\tau)$, we obtain the even theta constants $\theta_1,\dots,\theta_{10}$ and the odd theta constants $\partial_i \theta_{11}, \dots, \partial_i \theta_{16}$ as defined in eqs.~(\ref{def_even_theta_constants})-(\ref{def_odd_theta_constants}).
The genus-two curve $C_z$ is then given by
\bq
C_z & : & 
v^2 \; = \; u \left(u-\branchpoint_2(z)\right) \left(u-\branchpoint_3(z)\right) \left(u-\branchpoint_4(z)\right) \left(u-1\right)~,
\eq
with the branch points $\branchpoint_2, \branchpoint_3$ and $\branchpoint_4$ given by
(see eq.~(\ref{def_branch_points}))
\begin{align}
	\branchpoint_2(z) & = \frac{\theta_5^2 \theta_6^2}{\theta_1^2 \theta_4^2}~,
	&
	\branchpoint_3(z) & = \frac{\theta_6^2 \theta_7^2}{\theta_4^2 \theta_8^2}~,
	&
	\branchpoint_4(z) & = \frac{\theta_5^2 \theta_7^2}{\theta_1^2 \theta_8^2}~.
\end{align}
Furthermore, it can be shown (see section \ref{sect:polholJac}) that the genus-two curve $C_z$ varies holomorphically with $z$.

One can then show that in a suitable integral symplectic homology basis of $H^1(C_z,\mathbb{Z})$, the components of the four-component vector 
\begin{align}
	\psi_0 \; \begin{pmatrix} \bm \tau \\ \bm 1 \end{pmatrix} \begin{pmatrix} \tau  \\ 1  \end{pmatrix} \ ,
\end{align}
with $\bm \tau$ defined in \eqref{eq:period_matrix_tau_definition},
give a basis of solutions of the Picard--Fuchs equation $L^{(0)} f  = 0$. 

An alternative way of viewing this result is as follows: The genus-two curve $C_z$ has two independent holomorphic $(1,0)$-forms $\omega_i \in H^1(C_z,\mathbb{C})$, which we may take as
(see eq.~(\ref{def_standard_holomorphic_one_forms}))
\bq
\omega_0 \; = \; \frac{du}{v}~,
& &
\omega_1 \; = \; \frac{u \, du}{v}~.
\eq
We define an holomorphic $(1,0)$-form $\omega$ by
\bq
\omega & = & c_0 \omega_0 + c_1 \omega_1~,
\eq
where the coefficients $c_0$ and $c_1$ are defined, for a suitable choice of the integral symplectic homology basis of $H^1(C_z,\mathbb{Z})$, by eq.~(\ref{def_c0_c1}). Then it can be shown (see section \ref{sect:holomorphic_forms}) that the four periods of $\omega$ form a basis of solutions of the Picard--Fuchs equation $L^{(0)} f = 0$.

\section{Conclusions}
\label{sect:conclusions}

The maximal cut of the equal-mass four-loop banana integral has long been understood to be associated with a Calabi--Yau threefold~\cite{Bloch:2014qca,Bloch:2016izu}. In this paper, we have shown that it is also possible to interpret this maximal cut as a period of a genus-two curve. In fact, we have constructed this curve explicitly, by identifying its Jacobian variety with what we refer to as the holomorphic intermediate Jacobian, which can be constructed out of data from the Calabi--Yau threefold.  The periods of this genus-two curve depend holomorphically on $z$, and one column of the period matrix can be chosen such that it is annihilated by the same Picard--Fuchs operator that annihilates the periods of the Calabi--Yau threefold. 

While we have only studied the maximal cut of the banana integral in this paper, 
our expectation is that it should also be possible to express the full banana integral in terms of iterated integrals 
defined on this genus-two curve in $D=2$. 
In fact, our construction generalises known results from lower loop orders: The two-loop equal-mass banana integral
is known to be related to an elliptic curve and evaluates to iterated integrals of modular forms. 
The modular forms are naturally defined on the intermediate Jacobian $J_1$ of the elliptic curve.
The intermediate Jacobian $J_1$ has complex dimension one.
In this paper we related the four-loop equal-mass banana integrals to intermediate Jacobians of complex dimension two. 

It is natural to wonder how special the four-loop banana integral is in being expressible not only in terms of iterated integrals whose kernels involve the periods of a Calabi--Yau threefold, but also in terms of iterated integrals whose kernels involve periods of a genus-two curve. As developed in this work, any Feynman integral that involves periods of a family of Calabi--Yau threefolds enjoys a description in terms of the associated families of intermediate Jacobians. In particular, in suitable kinematic regions, we can locally determine the family of polarised intermediate Jacobians that realise these Feynman integrals in terms of a holomorphic family of Abelian varieties. For Feynman integrals that depend only on one or two kinematic variables, the Calabi--Yau-to-curve correspondence that we have introduced thereby yields either a one-parameter family of genus-two curves or a two-parameter family of genus-three curves. The equal-mass banana integral gives us an explicit example of the former type. It would be interesting to work out an example of the latter type as well. 

For Feynman integrals that involve Calabi--Yau threefolds that depend on more than two parameters, the associated family of Abelian varieties does not generically arise from a family of higher genus curves. This is a consequence of the Schottky problem, which states that not all Abelian varieties arise as the intermediate Jacobian of a (higher genus) curve.
On na\"ive dimensional grounds, we expect for Feynman integrals with $m$ kinematic variables (with $m\ge 1$) attributed to an $m$-parameter family of Calabi--Yau threefolds the following: Such Feynman integrals map to a family of polarised holomorphic Abelian varieties of complex dimension $m+1$. This family describes a codimension $\frac12(m^2 + m +2)$ locus in the moduli space of Abelian varieties $\mathcal{A}_{m+1}$. Moreover, the Schottky locus $\mathcal{S}_{m+1} \subset \mathcal{A}_{m+1}$ of genus-($m+1$) curves is of codimension $\frac12(m-1)(m-2)$ in $\mathcal{A}_{m+1}$.  Assuming that the Schottky locus intersects the family of Abelian varieties arising from the Feynman integral transversely, we expect that such a Feynman integral with $m$ kinematic variables possesses a $\left(\frac{1}{2}m(5-m) - 1\right)$-dimensional slice in its $m$-dimensional parameter space, for which the polarised holomorphic Abelian variety maps to the Schottky locus $\mathcal{S}_{m+1}$ of a genus-$(m+1)$ curve. Hence, along such a slice, we can formulate a Calabi--Yau-to-curve correspondence as introduced in this text. With the transversality assumption (which may be violated), we can only hope for a Calabi--Yau-to-curve correspondence for (sub-)families of Calabi--Yau threefolds if $m(5-m) \ge 2$, i.e. for $1\le m \le 4$.\footnote{The five-parameter Hulek--Verrill Calabi--Yau threefold family that describes the four-loop banana integral for generic masses does not obey this transversality assumption with the Schottky locus, as it possesses the one-parameter subfamily of Calabi--Yau threefolds, which yields the Calabi--Yau-to-curve correspondence in the equal-mass case as discussed in depth in this work.} It would be interesting to construct examples of such more general Calabi--Yau-to-curve correspondences explicitly and study their properties systematically.

It may also be possible to formulate a Calabi--Yau-to-curve correspondence that can be applied to Feynman integrals in which families of Calabi--Yau manifolds arise that are of odd complex dimension greater than three. The middle-dimensional cohomology classes of these higher-dimensional Calabi--Yau manifolds are again equipped with a symplectic structure, allowing us to define various types of intermediate Jacobians.\footnote{As even-dimensional Calabi--Yau manifolds do not possess a canonical symplectic structure  on the middle-dimensional cohomology, our construction cannot be generalised in an obvious way to even-dimensional Calabi--Yau manifolds.} When these intermediate Jacobians give rise to a family of Abelian varieties, there may again exist a Calabi--Yau-to-curve correspondence along the Schottky locus in the moduli space of Abelian varieties.

On the whole, we believe that the Calabi--Yau-to-curve correspondence we have presented here offers an interesting contrast to the standard view of how geometries can be associated with Feynman integrals. However, we leave the task of working out the deeper implications of this correspondence to future work.

\subsection*{Acknowledgments}
We would like to thank Per Berglund, Lance Dixon, Mohamed Elmi, 
Vasily Golyshev, Alexandros Kanargias, Joseph McGovern,
Manfred Lehn and Duco van Straten
for discussions and correspondences.
This work has been supported by the Cluster of Excellence Precision Physics, Fundamental Interactions, and Structure of
Matter (PRISMA EXC 2118/1) funded by the German Research Foundation (DFG) within
the German Excellence Strategy (Project ID 390831469), by the Royal Society grant URF$\backslash$R1$\backslash$221233, and by the Excellence Cluster ORIGINS funded by the DFG under Grant No.~EXC-2094-390783311.


\begin{appendix}
\section{The Complex Structure of Intermediate Jacobians}
\label{appendix:intermediate_Jacobians}
In the main part of this paper, we discuss various versions of intermediate Jacobians on middle dimensional cohomologies of (higher genus) curves and of odd-dimensional Calabi--Yau geometries. From a mathematical perspective these different choices of intermediate Jacobians arise from picking distinct complex structures on the vector space of middle dimensional cohomology classes. While this perspective is not explicitly used in the main text, the aim of this appendix is to briefly explain this perspective.

Let us consider a compact K\"ahler manifold $X$ of complex dimension $n$ and its odd cohomology group $H^{2k-1}(X,\mathbb{C})$ for some $k=1,\ldots,n$. These odd cohomology groups are always even dimensional, say $\dim_\mathbb{C} H^{2k-1}(X,\mathbb{C}) = 2g$ for some non-negative integer $g$. Then for any complex subvector space $V$ of $H^{2k-1}(X,\mathbb{C})$ of dimension $g$ with $V \oplus \overline V = H^{2k-1}(X,\mathbb{C})$, we define the $k$th intermediate Jacobian with respect to $V$ as the quotient
\begin{equation} \label{eq:kInt}
  J_k(X,V) = H^{2k-1}\left(X,{\mathbb C}\right) / \left( V \oplus H^{2k-1}\left(X,{\mathbb Z}\right) \right) \ ,
\end{equation}
which is a complex torus of complex dimension~$g$.\footnote{In this context a torus $T^{2g}$ of real dimension $2g$ with a complex structure $I$ is called a complex torus of complex dimension $g$. It can always be realised as the quotient $\mathbb{C}^g /\Lambda$ for some lattice $\Lambda \subset \mathbb{C}^g$, where the complex structure is induced from its universal cover $\mathbb{C}^g$.}

An alternative useful point of view for the $k$th intermediate Jacobian is the following. Let us consider the real cohomology group $H^{2k-1}(X,\mathbb{R})$ of real dimension $2g$. We make $H^{2k-1}(X,\mathbb{R})$ into a complex vector space of dimension $g$ by specifying a complex structure $I$ on $H^{2k-1}(X,\mathbb{R})$. That is to say, the complex structure $I$ is an automorphism 
\begin{equation}
   I : \ H^{2k-1}(X,\mathbb{R}) \to H^{2k-1}(X,\mathbb{R}) \quad \text{with}\quad I^2 = - \bm{1} \ .
\end{equation}
We denote the cohomology group $H^{2k-1}(X,\mathbb{R})$ with the complex structure $I$ by $H^{2k-1}_I(X,\mathbb{R})$.\footnote{The cohomology group $H^{2k-1}_I(X,\mathbb{R})$ of complex dimension $g$ should not be confused with the complexification $H^{2k-1}(X,\mathbb{R}) \otimes \mathbb{C} =  H^{2k-1}(X,\mathbb{C})$, which has complex dimension $2g$ and which plays a role shortly as well.} The scalar multiplication of any element $\theta \in H^{2k-1}(X,\mathbb{R})$ with a complex number $x+i y \in \mathbb{C}$, $x,y \in\mathbb{R}$, is well-defined by
\begin{equation}
    (x+i y) \cdot \theta := x\,\theta + y \, I(\theta) \ .
\end{equation}
On the complexification $H^{2k-1}(X,\mathbb{R}) \otimes \mathbb{C} = H^{2k-1}(X,\mathbb{C})$ the automorphism $I$ of $H^{2k-1}(X,\mathbb{R})$ extends linearly and splits the vector space $H^{2k-1}(X,\mathbb{C})$ into the direct sum of eigenspaces
\begin{equation}
  H^{2k-1}(X,\mathbb{C}) = V \oplus \overline V \ .
\end{equation}  
Here $V$ is the $g$-dimensional eigenspace of $I$ with eigenvalue $-i$ and $\overline V$ its complex conjugate eigenspace with eigenvalue $+i$. The cohomology group $H^{2k-1}_I(X,\mathbb{R})$ with complex structure $I$ is isomorphic to the quotient $H^{2k-1}(X,\mathbb{C}) / V \simeq \overline V$ via the map
\begin{equation}
  \varphi: \ H^{2k-1}(X,\mathbb{R}) \to  \overline V \,, \ 
    \theta \mapsto \chi = \frac12\big( \theta - i \, I(\theta) \big) \ ,
\end{equation}
and the inverse map
\begin{equation}
  \varphi^{-1}: \ \overline V \to H^{2k-1}(X,\mathbb{R}) \,, \
   \chi \mapsto \frac12 \big( \chi + \overline\chi \big) \ .
\end{equation}
Note that the map $\varphi$ is an isomorphism as complex vector spaces because the scalar multiplication with a complex number $x+iy \in \mathbb{C}$, $x,y\in\mathbb{R}$, commutes with the automorphism $\varphi$, i.e. 
\begin{equation}
  \varphi( (x+i y) \cdot \theta) = \varphi( x\,\theta + y \,I(\theta) )= x \,\varphi(\theta) + y\, \varphi(I(\theta)) =(x + i y) \chi \ .
\end{equation}  
As a consequence $\varphi$ establishes an isomorphism of the $k$th intermediate Jacobians~\eqref{eq:kInt} to the complex torus
\begin{equation}
  J_k(X,V) \simeq H^{2k-1}_I(X,\mathbb{R}) /H^{2k-1}(X,\mathbb{Z}) \ .
\end{equation}

In order to determine in this equivalent formulation the lattice $\Lambda \subset \mathbb{C}^g$ that describes the $k$th intermediate Jacobian~$J_k(X,V)$ as the quotient $\mathbb{C}^g/\Lambda$, we consider an integral basis of $(2k-1)$-form classes $\Gamma_a$, $a=1,\ldots,2g$, which generates (the free part of) the integral cohomology group $H^{2k-1}(X,\mathbb{Z})$. Without loss of generality, we assume that this integral basis is chosen and ordered in such a way that the real $2g$-dimensional vector space $H^{2k-1}(X,\mathbb{R}) \simeq \mathbb{R}^{2g}$ is generated as
\begin{equation}
  \langle\!\langle \Gamma_1,\ldots,\Gamma_g,I(\Gamma_1),\ldots,I(\Gamma_g) \rangle\!\rangle = H^{2k-1}(X,\mathbb{R}) \ .
\end{equation}
Then a complex basis for the complex vector space $H^{2k-1}_I(X,\mathbb{R}) \simeq \mathbb{C}^g$ is given by the integral vectors $\langle\!\langle \Gamma_1,\ldots,\Gamma_g\rangle\!\rangle$. In this basis the remaining integral vectors $\Gamma_{\tilde a}$, $\tilde a=g+1,\ldots,2g$, enjoy the expansion
\begin{equation}
  \Gamma_{\tilde a} = \sum_{b=1}^g \left(  {R_{\tilde a}}^b \Gamma_ b  + {S_{\tilde a}}^b I(\Gamma_ b) \right)
  = \sum_{b=1}^g \left( {R_{\tilde a}}^b + i {S_{\tilde a}}^b \right) \cdot \Gamma_b \ , \quad \tilde a = g+1,\ldots,2g \ ,
\end{equation}
that determines the complex $g\times g$-matrix
\begin{equation}
   \bm{T} = \left(  {R_{\tilde a}}^b + i {S_{\tilde a}}^b \right) \ , \qquad \tilde a=g+1,\ldots, 2g, \quad  b=1,\ldots,g \ ,
\end{equation}
which realizes the $k$th intermediate Jacobian $J_k(X,V)$ in terms of the lattice quotient
\begin{equation}
   J_k(X,V) \simeq \mathbb{C}^g / (\mathbb{Z}^g + \mathbb{Z}^g \, \bm{T}) \ .
\end{equation}

Finally, let us close this discussion by spelling out the complex structure $I^G$ and $I^W$ for the Griffiths and Weil intermediate Jacobian, respectively. Let us introduce the Hodge decomposition of $H^{2k-1}(X,\mathbb{C})$ in terms of Dolbeault cohomology groups 
\begin{equation}
  H^{2k-1}(X,\mathbb{C}) = \bigoplus_{p+q = 2k-1} H^{(p,q)}(X) \ .
\end{equation}  
Then the complex structure automorphisms $I^G$ and $I^W$ for the Griffiths and the Weil intermediate Jacobian on the complexified vector space $H^{2k-1}(X,\mathbb{C})$ are respectively defined by their actions on $(p,q)$-form classes $\theta^{(p,q)} \in H^{(p,q)}(X)$, which are respectively given as
\begin{equation}
    I^G \theta^{(p,q)}  = i^{\, -\text{sign}(p-q)}\theta^{(p,q)}  \ , \qquad  I^W \theta^{(p,q)}  =  i^{(p-q)} \theta^{(p,q)}  \ .
\end{equation}    

Note that the Griffiths and Weil complex structures on $H^1(X,\mathbb{C})$ coincide, such that their first Griffith and Weil intermediate Jacobians become identical. This is for instance the case for the first Griffiths and Weil intermediate Jacobians of complex curves of genus $g$. By Poincar\'e duality the same identification is true for the $n$-th Griffiths and Weil intermediate Jacobian of a complex $n$-dimensional K\"ahler manifold $X$. However, for the $k$th Griffiths and Weil intermediate Jacobians with $k\ne1$ and $k\ne n$ of a complex $n$-dimensional K\"ahler manifold this is not true if the Hodge decomposition of the corresponding cohomology group $H^{2k-1}(X,\mathbb{C})$ is non-trivial.  

For the Hodge decomposition of the middle dimensional cohomology $H^3(Y,\mathbb{C})$ of a Calabi--Yau threefold $Y$ --- which is of particular relevance to this work --- we explicitly spell out the eigenspaces with respect to the Griffiths and Weil complex structures $I^G$ and $I^W$ in table~\ref{table_eigenvalues_I_G_I_W}. The former complex structure gives rise to the Griffiths intermediate Jacobian defined in eq.~\eqref{eq:DefJ2G}, whereas the latter complex structures yield the Weil intermediate Jacobian introduced in eq.~\eqref{eq:DefJ2W}. 

\begin{table}
\begin{center}
\begin{tabular}{c|cccc}
 \strut Dolbeault cohomologies & ~$H^{(3,0)}(X)$~ & ~$H^{(2,1)}(X)$~ & ~$H^{(1,2)}(X)$~ & ~$H^{(0,3)}(X)$~ \\
 \hline
\strut Eigenvalues of $I^G$ & $-i$ & $-i$ & $+i$ & $+i$ \\
\strut Eigenvalues of $I^W$ & $-i$ & $+i$ & $-i$ & $+i$ \\
 \hline
\end{tabular}
\end{center}
\caption{
The eigenspaces of the Griffiths and Weil complex structures $I^G$ and $I^W$ for the Hodge decomposition of the three-form cohomology group $H^3(X,\mathbb{C})$.}
\label{table_eigenvalues_I_G_I_W}
\end{table}


\section{Gymnastics with Theta Constants}
\label{appendix:theta_constants}
In this appendix, we give a summary of relations among theta constants.
The main references are refs.~\cite{Mumford:theta_I,Mumford:theta_II,Mumford:theta_III,Enolski:2008aaa,shaska:2012theta}.
We first review G\"opel groups for any genus $g$. From the G\"opel groups one may derive relations among theta constants.
We then consider hyperelliptic curves and recall the Thomae formulae, which relate certain theta constants 
to the branch points of the hyperelliptic curve.
Finally we specialise to the genus two case and express the $a$-cycle periods in term of theta constants.

\subsection{G\"opel groups}

We consider half-integer characteristics.
For two half-integer characteristics 
\bq
 \epsilon_1
 = 
 \left[ \begin{array}{c}
    a_{1} \\
    b_{1} \\
 \end{array} \right]
 = 
 \left[ \begin{array}{cccc}
    a_{11} & a_{12} & \dots & a_{1g} \\
    b_{11} & b_{12} & \dots & b_{1g} \\
 \end{array} \right],
 & &
 \epsilon_2
 = 
 \left[ \begin{array}{c}
    a_{2} \\
    b_{2} \\
 \end{array} \right]
 =  
 \left[ \begin{array}{cccc}
    a_{21} & a_{22} & \dots & a_{2g} \\
    b_{21} & b_{22} & \dots & b_{2g} \\
 \end{array} \right],
 \;\;\;
\eq
we define an addition by
\bq
 \epsilon_1 + \epsilon_2
 & = &
 \left[ \begin{array}{cccc}
    a_{11}+a_{21} & a_{12}+a_{22} & \dots & a_{1g}+a_{2g} \\
    b_{11}+b_{21} & b_{12}+b_{22} & \dots & b_{1g}+b_{2g} \\
 \end{array} \right] \bmod 1~.
\eq
We define
\bq
 \left\langle \epsilon_1, \epsilon_2 \right\rangle
 & = &
 4 \left(a_1 b_2 - b_1 a_2 \right).
\eq
Two half-integer characteristics $\epsilon_1$ and $\epsilon_2$ are called syzygetic, if
\bq
 \left\langle \epsilon_1, \epsilon_2 \right\rangle & = & 0 \bmod 2~.
\eq
A G\"opel group $G$ is a set of $2^k$ half-integer characteristics (with $k\le g$) 
together with the addition defined above, such that 
any two $\epsilon_1, \epsilon_2, \in G$ are syzygetic.
We may view a G\"opel group as a subgroup of the group of all $2^{2g}$ characteristics.
In the following we assume that $\epsilon_1$ and $\epsilon_2$ are even characteristics and $\epsilon_1 \neq \epsilon_2$.
By considering cosets of G\"opel groups, one derives two relations:
The first one reads
\bq
 \theta^2\left[\epsilon_1\right] \theta^2\left[\epsilon_2\right]
 & = &
 \frac{1}{2^{g-1}}
 \sum\limits_{\epsilon_3}
 \left(-1\right)^{4 \left(a_2+a_3\right) \cdot \left(b_1+b_3\right)}
 \theta^2\left[\epsilon_3\right] \theta^2\left[\epsilon_1+\epsilon_2+\epsilon_3\right]~,
\eq
where the sum is over all even characteristics $\epsilon_3$ with 
$\epsilon_3 \neq \epsilon_1,\epsilon_2$
and such that $\epsilon_1+\epsilon_2+\epsilon_3$ is even as well.
An example for genus two is given by
\bq
 \theta_7^2 \theta_8^2 \; = \; \theta_1^2 \theta_5^2 - \theta_4^2 \theta_6^2~.
\eq
The second relation reads
\bq
\lefteqn{
 \theta^4\left[\epsilon_1\right] 
 + \left(-1\right)^{\left\langle \epsilon_1, \epsilon_2 \right\rangle} \theta^4\left[\epsilon_2\right]
 = } & &
 \nonumber \\
 & &
 \frac{1}{2^{g-1}}
 \sum\limits_{\epsilon_3}
 \left(-1\right)^{4 \left(a_1+a_3\right) \cdot \left(b_1+b_3\right)}
 \left(
  \theta^4\left[\epsilon_3\right] 
  + \left(-1\right)^{\left\langle \epsilon_1, \epsilon_2 \right\rangle} \theta^4\left[\epsilon_1+\epsilon_2+\epsilon_3\right]
 \right)~,
\eq
where the sum is again over all even characteristics $\epsilon_3$ with 
$\epsilon_3 \neq \epsilon_1,\epsilon_2$
and such that $\epsilon_1+\epsilon_2+\epsilon_3$ is even as well.
An example for genus two is given by
\bq
 \theta_7^4 + \theta_8^4 \; = \; \theta_1^4 + \theta_5^4 - \theta_4^4 - \theta_6^4~,
\eq

\subsection{Hyperelliptic curves and Thomae formulae}

Let us now specialise to theta constants related to hyperelliptic curves, 
i.e. we assume that $\bm{\tau}$ corresponds to a
hyperelliptic curve.
In this case we may express the branch points and the $a$-cycle periods in terms of theta constants.
In this appendix we use the standard (increasing) order for the indices: $0,1,\dots,g-1$ and $1,2,\dots,g$. 
Note that in section~\ref{sect:holomorphic_forms} it was more convenient to use the decreasing order for the indices.

We consider a hyperelliptic curve of genus $g$ defined by the equation
\bq
 v^2 & = & \prod\limits_{i=1}^{2g+1} \left( u - \branchpoint_i\right)~,
\eq
There are $(2g+1)$ finite branch points, one branch point is at infinity $\branchpoint_{2g+2}=\infty$.
We assume that the branch cuts are between $\branchpoint_{2j+1}$ and $\branchpoint_{2j+2}$ for $0 \le j \le g$.
For $g=2$ this is shown in fig.~\ref{fig_genus_2_cuts}.
A curve of genus $g$ has $g$ holomorphic differentials $\vec{\omega}=(\omega_0,\dots,\omega_{g-1})$.
We may identify each branch point $\branchpoint_j$ with a vector $\vec{\eta}_j$ through
\bq
 \vec{\eta}_j & = & \int\limits_{\infty}^{\branchpoint_i} \vec{\omega}
 \; =\; 
 \vec{b} \mathbfcal{X} + \vec{a} \mathbfcal{T}~.
\eq
As these are half-periods, all entries of $\vec{a}$ and $\vec{b}$ modulo $1$ are either $0$ or $\frac{1}{2}$.
They define a characteristic through
\bq
 \eta_j
 & = &
 \left[ \begin{array}{cccc}
  a_1 & a_2 & \dots & a_g \\
  b_1 & b_2 & \dots & b_g \\
 \end{array} \right]~.
\eq
Working this out for the finite branch points one finds
\bq
 \eta_{2j+1}
 & = &
 \frac{1}{2}
 \left[ \begin{array}{cccccccc}
  0 & 0 & \dots & 0 & 1 & 0 & \dots & 0 \\
  1 & 1 & \dots & 1 & 0 & 0 & \dots & 0 \\
 \end{array} \right]~,
 \;\;\;\;\;\;
 0 \; \le \; j \; \le \; g~,
 \nonumber \\
 \eta_{2j+2}
 & = &
 \frac{1}{2}
 \left[ \begin{array}{cccccccc}
  0 & 0 & \dots & 0 & 1 & 0 & \dots & 0 \\
  1 & 1 & \dots & 1 & 1 & 0 & \dots & 0 \\
 \end{array} \right]~,
 \;\;\;\;\;\;
 0 \; \le \; j \; \le \; g-1~,
\eq
where the $1$ in the first row occurs at position $(j+1)$.
$\eta_{2g+2}$ is (obviously) defined as
\bq
\label{def_eta_infinity}
 \eta_{2g+2}
 & = &
 \left[ \begin{array}{ccc}
  0 & \dots & 0 \\
  0 & \dots & 0 \\
 \end{array} \right]~.
\eq
It is an easy exercise to check that
\bq
\label{sum_basic_characteristics}
 \sum\limits_{j=1}^{2g+2} \eta_j
 & = &
 \left[ \begin{array}{ccc}
  0 & \dots & 0 \\
  0 & \dots & 0 \\
 \end{array} \right] \bmod 1~.
\eq
As an example we have for $g=2$:
\begin{align}
 \eta_{1}
 & = 
 \frac{1}{2}
 \left[ \begin{array}{cc}
  1 & 0 \\
  0 & 0 \\
 \end{array} \right]~,
 &
 \eta_{3}
 & = 
 \frac{1}{2}
 \left[ \begin{array}{cc}
  0 & 1 \\
  1 & 0 \\
 \end{array} \right]~,
 &
 \eta_{5}
 & = 
 \frac{1}{2}
 \left[ \begin{array}{cc}
  0 & 0 \\
  1 & 1 \\
 \end{array} \right]~,
 \nonumber \\
 \eta_{2}
 & = 
 \frac{1}{2}
 \left[ \begin{array}{cc}
  1 & 0 \\
  1 & 0 \\
 \end{array} \right]~,
 &
 \eta_{4}
 & = 
 \frac{1}{2}
 \left[ \begin{array}{cc}
  0 & 1 \\
  1 & 1 \\
 \end{array} \right]~,
 &
 \eta_{6}
 & = 
 \frac{1}{2}
 \left[ \begin{array}{cc}
  0 & 0 \\
  0 & 0 \\
 \end{array} \right]~.
\end{align}
Let $G=\{1,2,\dots,2g+2\}$ the set of all indices and $U=\{1,3,5,\dots,2g+1\}$ the set of the odd indices.
There is a one-to-one correspondence between partitions of the set
\bq
\label{def_subset_J}
 G & = & J \cup \bar{J},
 \;\;\; \mbox{with} \;\;\;
 \bar{J} \; = \; G \backslash J,
 \;\;\;
 \left| J \right| \; = \; g+1-2k,
 \;\;\;
 0 \; \le \; k \; \le \; \lfloor\frac{g+1}{2} \rfloor
\eq
and the $2^{2g}$ different characteristics.
The map between a subset $J$ and a characteristic $\epsilon$ is given by
\bq
\label{def_Enolskii}
 \epsilon
 & = &
 \left(
 \sum\limits_{k \in J} \eta_k
 +
 \sum\limits_{j=1}^g \eta_{2j}
 \right) \bmod 1~.
\eq
It can be shown that subsets with $k$ even correspond to even characteristics, while
subsets with $k$ odd correspond to odd characteristics.
Characteristics obtained from subsets $J$ with $k=0$ are called non-singular even characteristics,
characteristics obtained from subsets $J$ with $k=1$ are called non-singular odd characteristics,
all other characteristics are called singular characteristics.
Singular characteristics only exist for $g \ge 3$.

Due to eq.~(\ref{sum_basic_characteristics}) we may replace $J$ in eq.~(\ref{def_Enolskii}) 
by the complement $G \backslash J$, this will not change $\epsilon$.
Due to eq~(\ref{def_eta_infinity}) we may further omit (if present) the index $(2g+2)$ from set $J$, this will not
change $\epsilon$ either.
The correspondence between subsets $J$ of $G$ as in eq.~(\ref{def_subset_J})
and characteristics motivates the following definition:
\bq
 \theta_E\left\{J\right\}\left(z,\bm{\tau}\right)
 &= &
 \vartheta\left[\epsilon\right]\left(z,\bm{\tau}\right)~,
\eq
with $\epsilon$ defined by eq.~(\ref{def_Enolskii}).
This notation is used by Enolskii and Richter \cite{Enolski:2008aaa}.
 
Mumford uses a different notation for the non-singular even characteristics \cite{Mumford:theta_II}.
This notation eliminates the second term in eq.~(\ref{def_Enolskii})
at the expense of having a more involved definition of the subsets.
For two sets $A$ and $B$ define $\ominus$ by
\bq
 A \ominus B & = & \left(A \cup B \right) \backslash \left( A \cap B \right)~.
\eq
This operation satisfies
\bq
 \left( A \ominus B \right) \ominus B & = & A~.
\eq
Mumford considers subsets $S$ of $G\backslash\{2g+2\}$ with
\bq
\label{condition_Mumford}
 \left|S\right| \, \mbox{even}
 & \mbox{and} &
 \left| S \ominus U \right| \; = \; g+1~.
\eq
For such a set one defines the corresponding characteristics by
\bq
\label{def_Mumford}
 \epsilon
 & = &
 \sum\limits_{k \in S} \eta_k~,
\eq
and we set
\bq
 \theta_M\left\{S\right\}\left(z,\bm{\tau}\right)
 &= &
 \vartheta\left[\epsilon\right]\left(z,\bm{\tau}\right),
\eq
with $\epsilon$ defined by eq.~(\ref{def_Mumford}).

For $g=2$ we summarise the different conventions for the even characteristics 
in table~\ref{table_conventions_even_theta_constants} and for the odd characteristics in
in table~\ref{table_conventions_odd_theta_constants}.
\begin{table}
\bq
\begin{array}{cccc}
 \mbox{Mumford \cite{Mumford:theta_II}} & \mbox{Enolskii/Richter \cite{Enolski:2008aaa}} & \mbox{Shaska/Wijesiri \cite{shaska:2012theta}} & \mbox{characteristic} \\
 S & J & \theta_i & \epsilon \\  \hline \\[-2ex]
 \left\{\right\} & \left\{2,4\right\} & \theta_{1} & \frac{1}{2} \left[\begin{array}{cc} 0 & 0  \\ 0 & 0 \\ \end{array}\right] \\[2ex]
 \left\{1,2\right\} & \left\{1,4\right\} & \theta_{3} & \frac{1}{2} \left[\begin{array}{cc} 0 & 0  \\ 1 & 0 \\ \end{array}\right] \\[2ex]
 \left\{1,4\right\} & \left\{1,2\right\} & \theta_{10} & \frac{1}{2} \left[\begin{array}{cc} 1 & 1  \\ 1 & 1 \\ \end{array}\right] \\[2ex]
 \left\{2,3\right\} & \left\{3,4\right\} & \theta_{8} & \frac{1}{2} \left[\begin{array}{cc} 1 & 1  \\ 0 & 0 \\ \end{array}\right] \\[2ex]
 \left\{3,4\right\} & \left\{2,3\right\} & \theta_{4} & \frac{1}{2} \left[\begin{array}{cc} 0 & 0  \\ 0 & 1 \\ \end{array}\right] \\[2ex]
 \left\{2,5\right\} & \left\{4,5\right\} & \theta_{6} & \frac{1}{2} \left[\begin{array}{cc} 1 & 0  \\ 0 & 1 \\ \end{array}\right] \\[2ex]
 \left\{4,5\right\} & \left\{2,5\right\} & \theta_{7} & \frac{1}{2} \left[\begin{array}{cc} 0 & 1  \\ 0 & 0 \\ \end{array}\right] \\[2ex]
 \left\{1,2,3,4\right\} & \left\{1,3\right\} & \theta_{2} & \frac{1}{2} \left[\begin{array}{cc} 0 & 0  \\ 1 & 1 \\ \end{array}\right] \\[2ex]
 \left\{1,2,4,5\right\} & \left\{1,5\right\} & \theta_{9} & \frac{1}{2} \left[\begin{array}{cc} 0 & 1  \\ 1 & 0 \\ \end{array}\right] \\[2ex]
 \left\{2,3,4,5\right\} & \left\{3,5\right\} & \theta_{5} & \frac{1}{2} \left[\begin{array}{cc} 1 & 0  \\ 0 & 0 \\ \end{array}\right]
\end{array}
\nonumber
\eq
\caption{
The different conventions of labelling the even theta constants for genus two. 
}
\label{table_conventions_even_theta_constants}
\end{table}
\begin{table}
\bq
\begin{array}{ccc}
 \mbox{Enolskii/Richter \cite{Enolski:2008aaa}} & \mbox{Shaska/Wijesiri \cite{shaska:2012theta}} & \mbox{characteristic} \\
 J & \theta_i & \epsilon \\  \hline \\[-2ex]
 \left\{1\right\} & \theta_{11} & \frac{1}{2} \left[\begin{array}{cc} 0 & 1  \\ 0 & 1 \\ \end{array}\right] \\[2ex]
 \left\{2\right\} & \theta_{12} & \frac{1}{2} \left[\begin{array}{cc} 0 & 1  \\ 1 & 1 \\ \end{array}\right] \\[2ex]
 \left\{3\right\} & \theta_{15} & \frac{1}{2} \left[\begin{array}{cc} 1 & 0  \\ 1 & 1 \\ \end{array}\right] \\[2ex]
 \left\{4\right\} & \theta_{13} & \frac{1}{2} \left[\begin{array}{cc} 1 & 0  \\ 1 & 0 \\ \end{array}\right] \\[2ex]
 \left\{5\right\} & \theta_{14} & \frac{1}{2} \left[\begin{array}{cc} 1 & 1  \\ 1 & 0 \\ \end{array}\right] \\[2ex]
 \left\{6\right\} & \theta_{16} & \frac{1}{2} \left[\begin{array}{cc} 1 & 1  \\ 0 & 1 \\ \end{array}\right]
\end{array}
\nonumber
\eq
\caption{
Conventions of labelling the odd theta constants for genus two. 
}
\label{table_conventions_odd_theta_constants}
\end{table}

The Thomae formula relates the fourth power of the non-singular even theta constants to the finite branch points
\bq
 \theta_E\left\{J\right\}^4
 & = &
 \frac{\left| \mathbfcal{X} \right|^2}{\left(2 \pi i\right)^{2g}}
 \prod\limits_{\stackrel{i<j<2g+2}{i,j \in J}} \left(\branchpoint_i-\branchpoint_j\right)
 \prod\limits_{\stackrel{i<j<2g+2}{i,j \notin J}} \left(\branchpoint_i-\branchpoint_j\right)~.
\eq
The Thomae formula leads to eq.~(\ref{def_branch_points}),
which gives the branch points of the genus-two hyperelliptic curve in terms of theta constants.

\subsection{Genus two}

Let us now specialise to genus two. Any genus-two curve is hyperelliptic.
We may express the $a$-cycle periods in term of theta constants.
Let $k,l,p,q,r \in \{1,2,3,4,5\}$ such that they are pairwise distinct.
We have \cite{Enolski:2008aaa}
\bq
 \mathbfcal{X}
 & = &
 \frac{2 T_{pqr}}{\theta_E\left\{k,l\right\}\left(\branchpoint_k-\branchpoint_l\right)^{\frac{3}{2}}}
 \left( \begin{array}{cc}
  \partial_0 \theta_E\left\{l\right\} & \partial_0 \theta_E\left\{k\right\} \\
  \partial_1 \theta_E\left\{l\right\} & \partial_1 \theta_E\left\{k\right\} \\
 \end{array} \right)
 \left( \begin{array}{cc}
  \frac{1}{T_l} & 0 \\
  0 & \frac{1}{T_k} \\
 \end{array} \right)
 \left( \begin{array}{rr}
  1 & \branchpoint_k \\
  -1 & -\branchpoint_l \\
 \end{array} \right),
\eq
where
\bq
 T_k & = & \theta_E\left\{p,k\right\} \theta_E\left\{q,k\right\} \theta_E\left\{r,k\right\}~,
 \nonumber \\
 T_l & = & \theta_E\left\{p,l\right\} \theta_E\left\{q,l\right\} \theta_E\left\{r,l\right\}~,
 \nonumber \\
 T_{pqr} & = & \theta_E\left\{p,q\right\} \theta_E\left\{q,r\right\} \theta_E\left\{p,r\right\}~.
\eq
Using this formula for $\branchpoint_1=0$ and $\branchpoint_5=1$ with $l=1$ and $k=5$ yields eq.~(\ref{P_a_in_terms_of_theta}).
For eq. ~(\ref{P_a_in_terms_of_theta}) we used in addition the relation
\bq
 \theta_5 \theta_6 \theta_7 \partial_i \theta_{11}  
 & = & \theta_2 \theta_3 \theta_{10} \partial_i \theta_{14} 
 + \theta_1 \theta_4 \theta_8 \partial_i \theta_{16}~,
 \;\;\;\;\;\;
 i \; \in \; \{0,1\}~,
\eq
for simplification.

\newpage

\end{appendix}

{\footnotesize
\bibliography{biblio}
\bibliographystyle{JHEP.bst}
}

\end{document}